\documentclass[]{aastex63}
\graphicspath{{./}{Figures/}}
\shorttitle{APOGEE-2N Targeting}
\shortauthors{Beaton, Oelkers, Hayes, Covey, et al.}
\usepackage{amsmath}
\usepackage[framemethod=tikz]{mdframed}
\newcommand{\tess}{{\it TESS}}

\newcommand{\kepler}{{\it Kepler}}
\newcommand{\gaia}{{\it Gaia}}
\newcommand{\ktwo}{{K2}}
\newcommand{\spitzer}{{Spitzer}}
 
\newcommand{\vizier}{VizieR}

\newcommand{\logg}{$\log{g}$}
\newcommand{\teff}{$T_{\rm eff}$}
\newcommand{\ejk}{$E(J-K)$}
\newcommand{\jk}{$(J-K)_{0}$}

\newcommand{\degs}{^{\circ}}
\newcommand{\kms}{km s$^{-1}$}

\newcommand{\inprep}[1]{\textbf{#1}}

\newcommand{\santanatext}{F.~Santana et al. (submitted)}
\newcommand{\santanapar}{(F.~Santana et al. submitted)}
\newcommand{\holtzmanprep}{(J.~Holtzman et al., in prep.)}
\newcommand{\revise}[1]{#1}
\begin{document}
\title{Final Targeting Strategy for the SDSS-IV APOGEE-2N Survey}
\correspondingauthor{Rachael~L.~Beaton}
\email{rbeaton@princeton.edu}
%%
% All authors and affiliations are set in the file:
%%%%%%%%%%%%%%%%%%%%%%%%%%%%%%%%%%%%%%%%%%%%%%%%%%%%%%%%%%%%%%%%%%%%%%%%%%%%%%%%%%%%
% People who wrote things or wrangled people to write things:
%%%%%%%%%%%%%%%%%%%%%%%%%%%%%%%%%%%%%%%%%%%%%%%%%%%%%%%%%%%%%%%%%%%%%%%%%%%%%%%%%%%%
\author[0000-0002-1691-8217]{Rachael L. Beaton}
\altaffiliation{Much of this work was completed while this author was a NASA Hubble Fellow at Princeton University.}
\altaffiliation{Carnegie-Princeton Fellow}
\affiliation{Department of Astrophysical Sciences, Princeton University, 4 Ivy Lane, Princeton, NJ~08544}
\affiliation{The Observatories of the Carnegie Institution for Science, 813 Santa Barbara St., Pasadena, CA~91101}
\author{Ryan~J.~Oelkers} 
\affiliation{Department of Physics \& Astronomy, Vanderbilt University, Nashville, TN, 37235, USA}
\author[0000-0003-2969-2445]{Christian R.~Hayes} 
\affiliation{Department of Astronomy, University of Washington, Seattle, WA, 98195, USA}
\author[0000-0001-6914-7797]{Kevin~R.~Covey} 
\affiliation{Department of Physics \& Astronomy, Western Washington University, Bellingham, WA, 98225, USA}
%
% Northern Plate Designer
%
\author{S.~D.~Chojnowski}
\affiliation{Department of Astronomy, New Mexico State University, Las Cruces, NM, 88001, USA}
\affiliation{Department of Physics, Montana State University, P.O. Box 173840, Bozeman, MT 59717-3840, USA}
%
% The Irreplaceable Nathan who has an official title that I can never remember: Northern Scheduler. 
%
\author{Nathan De Lee} 
\affiliation{Department of Physics, Geology, and Engineering Technology, Northern Kentucky University, Highland Heights, KY 41099, USA}
\affiliation{Department of Physics \& Astronomy, Vanderbilt University, Nashville, TN, 37235, USA}
%%%%
% Jen & Steve
%%%
\author{Jennifer~S.~Sobeck} 
\affiliation{Department of Astronomy, University of Washington, Seattle, WA, 98195, USA}
\author{Steven~R.~Majewski} 
\affiliation{Department of Astronomy, University of Virginia, Charlottesville, VA, 22903, USA}
% 
%%%%%%%%%%%%%%%%%%%%%%%%%%%%%%%%%%%%%%%%%%%%%%%%%%%%%%%%%%%%%%%%%%%%%%%%%%%%%%%%%%%%
% Other Targeting Personnel; e.g., Regular Telecon Attendees, Targeting Reviews, Other Tasks. 
%%%%%%%%%%%%%%%%%%%%%%%%%%%%%%%%%%%%%%%%%%%%%%%%%%%%%%%%%%%%%%%%%%%%%%%%%%%%%%%%%%%%
%
\author{Roger Cohen} 
\affiliation{Space Telescope Science Institute, Baltimore, MD, 21218, USA}
\author{Jos\'e Fern\'andez-Trincado} 
\affiliation{Instituto de Astronom\'ia y Ciencias Planetarias, Universidad de Atacama, Copayapu 485, Copiap\'o, Chile}
\author{Pen\'elope~Longa-Pe\~na} 
\affiliation{Unidad de Astronom\'ia, Universidad de Antofagasta, Avenida Angamos 601, Antofagasta 1270300, Chile}
\author[0000-0003-2321-950X]{Julia E. O\rq{}Connell}
\affiliation{Departamento de Astronom{\'{\i}}a, Universidad de Concepci{\'o}n, Casilla 160-C, Concepci{\'o}n, Chile}
\author{Felipe~A.~Santana} 
\affiliation{Universidad de Chile, Av. Libertador Bernardo O'Higgins 1058, Santiago De Chile}
\author[0000-0003-1479-3059]{Guy~S.~Stringfellow} 
\affiliation{Center for Astrophysics and Space Astronomy, Department of Astrophysical and Planetary Sciences, University of Colorado, Boulder, CO, 80309, USA}
\author{Gail~Zasowski} 
\affiliation{Department of Physics \& Astronomy, University of Utah, Salt Lake City, UT, 84112, USA}
%
%%%%%%%%%%%%%%%%%%%%%%%%%%%%%%%%%%%%%%%%%%%%%%%%%%%%%%%%%%%%%%%%%%%%%%%%%%%%%%%%%%%%
% Goal Program Targeting or Main Survey Contributions; Alphabetical:
%%%%%%%%%%%%%%%%%%%%%%%%%%%%%%%%%%%%%%%%%%%%%%%%%%%%%%%%%%%%%%%%%%%%%%%%%%%%%%%%%%%%
\author{Conny Aerts} 
\affiliation{Institute of Astronomy, KU Leuven, Celestijnenlaan 200D, B-3001 Leuven, Belgium}
\author{Borja Anguiano} 
\affiliation{Department of Astronomy, University of Virginia, Charlottesville, VA, 22903, USA}
\author{Chad~Bender} 
\affiliation{Steward Observatory, The University of Arizona, Tucson, AZ, 85719, USA}
\author{Caleb I.~Ca\~nas}
\affiliation{Department of Astronomy \& Astrophysics, The Pennsylvania State University, 525 Davey Lab, University Park, PA 16802, USA} 
\affiliation{Center for Exoplanets \& Habitable Worlds, University Park, PA 16802, USA} 
\affiliation{Penn State Astrobiology Research Center, University Park, PA 16802, USA}
\author{Katia Cunha} 
\affiliation{Steward Observatory, The University of Arizona, Tucson, AZ, 85719, USA} \affiliation{Observat\'{o}rio Nacional, 20921-400 So Crist\'{o}vao, Rio de Janeiro, RJ, Brazil}
\author{John~Donor} 
\affiliation{Department of Physics \& Astronomy, Texas Christian University, Fort Worth, TX, 76129, USA}
\author{Scott~W.~Fleming} 
\affiliation{Space Telescope Science Institute, Baltimore, MD, 21218, USA}
\author[0000-0002-0740-8346]{Peter~M.~Frinchaboy} 
\affiliation{Department of Physics \& Astronomy, Texas Christian University, Fort Worth, TX, 76129, USA}
\author{Diane~Feuillet} 
\affiliation{Department of Astronomy and Theoretical Physics, Lund Observatory, Lund University, Box 43, SE-221 00 Lund, Sweden} 
\affiliation{Max-Planck-Institut f\"{u}r Astronomie, K\"{o}nigstuhl 17, D-69117 Heidelberg, Germany}
\author{Paul~Harding} 
\affiliation{Department of Astronomy, Case Western Reserve University, Cleveland, OH, 44106, USA}
\author{Sten~Hasselquist}
\affiliation{Department of Physics \& Astronomy, University of Utah, Salt Lake City, UT, 84112, USA} 
\affiliation{NSF Astronomy and Astrophysics Postdoctoral Fellow} 
\author{Jon~Holtzman} 
\affiliation{Department of Astronomy, New Mexico State University, Las Cruces, NM, 88001, USA}
\author{Jennifer~A.~Johnson} 
\affiliation{Department of Astronomy, The Ohio State University, Columbus, OH, 43210, USA}
\author{Juna~A.~Kollmeier}
\affiliation{The Observatories of the Carnegie Institution for Science, 813 Santa Barbara St., Pasadena, CA~91101}
\author{Marina Kounkel}
\affiliation{Department of Physics \& Astronomy, Western Washington University, Bellingham, WA, 98225, USA}
\affiliation{Department of Physics \& Astronomy, Vanderbilt University, Nashville, TN, 37235, USA}
\author{Suvrath Mahadevan}
\affiliation{Department of Astronomy \& Astrophysics, The Pennsylvania State University, 525 Davey Lab, University Park, PA 16802, USA}
\affiliation{Center for Exoplanets \& Habitable Worlds, University Park, PA 16802, USA} 
\affiliation{Penn State Astrobiology Research Center, University Park, PA 16802, USA}
\author[0000-0003-0872-7098]{Adrian.~M.~Price-Whelan}
\affiliation{Center for Computational Astrophysics, Flatiron Institute, 162 Fifth Avenue, New York, NY 10010, USA}
\author{Alvaro Rojas-Arriagada}
\affiliation{Instituto de Astrof\'{i}sica, Facultad de F\'{i}sica, Pontificia Universidad Cat\'{o}lica de Chile, Av. Vicu\~{n}a Mackenna 4860, Santiago, Chile}
\affiliation{Millennium Institute of Astrophysics, Av. Vicu\~{n}a Mackenna 4860, 782-0436, Macul, Santiago, Chile}
\author[0000-0001-8600-4798]{Carlos Rom\'an-Z\'u\~niga}
\affiliation{Instituto de Astronom\'ia, UNAM, Ensenada, C.P. 22860, Baja California, M\'exico}
\author{Edward F.~Schlafly}
\affiliation{Lawrence Berkeley National Laboratory, One Cyclotron Road, Berkeley, CA, 94720, USA}
\author{Mathias Schultheis}
\affiliation{Observatoire de la Cote d'Azur, Lagrange Boulevard de l'Observatoire 06304 Nice, France}
\author{Matthew~Shetrone}
\affiliation{University of California Observatories, Santa Cruz, CA 95064, USA}
\author{Joshua~D.~Simon}
\affiliation{The Observatories of the Carnegie Institution for Science, 813 Santa Barbara St., Pasadena, CA~91101}
\author[0000-0002-3481-9052]{Keivan~G.~Stassun}
\affiliation{Department of Physics \& Astronomy, Vanderbilt University, Nashville, TN, 37235, USA}
\author{Amelia~M.~Stutz}
\affiliation{Departamento de Astronom{\'{\i}}a, Universidad de Concepci{\'o}n, Casilla 160-C, Concepci{\'o}n, Chile}
\author[0000-0002-4818-7885]{Jamie~Tayar} 
\altaffiliation{Hubble Fellow}
\affiliation{Institute for Astronomy, University of Hawai\rq{}i, 2680 Woodlawn Drive, Honolulu, HI, 96822, USA}
\author{Johanna~Teske} 
\affiliation{Earth and Planets Laboratory, Carnegie Institution for Science, 5241 Broad Branch Road, NW, Washington, DC 20015, USA}
\altaffiliation{Much of this work was completed while this author was a NASA Hubble Fellow at the Observatories of the Carnegie Institution for Science.}
\author{Andrew~Tkachenko} 
\affiliation{Institute of Astronomy, KU Leuven, Celestijnenlaan 200D, B-3001 Leuven, Belgium}
\author{Nick~Troup} 
\affiliation{Department of Physics, Salisbury University, Salisbury, MD, 21801, USA}
%%%%%%%%%%%%%%%%%%%%%%%%%%%%%%%%%%%%%%%%%%%%%%%%%%%%%%%%%%%%%%%%%%%%%%%%%%%%%%%%%%%%
% Ancillary PIs or Ancillary Co-Is with Significant Effort; Alphabetical:
%%%%%%%%%%%%%%%%%%%%%%%%%%%%%%%%%%%%%%%%%%%%%%%%%%%%%%%%%%%%%%%%%%%%%%%%%%%%%%%%%%%%
\author{Franco~D.~Albareti}
\affiliation{Instituto de F{\'i}sica Te{\'o}rica UAM/CSIC, Universidad Aut{\'o}noma de Madrid, Cantoblanco, E-28049 Madrid, Spain}
\affiliation{Campus of International Excellence UAM+CSIC, Cantoblanco, E-28049 Madrid, Spain}
\author{Dmitry Bizyaev}
\affiliation{Apache Point Observatory and New Mexico State University, Sunspot, NM 88349, USA}
\affiliation{Sternberg Astronomical Institute, Moscow State University, Moscow, Russia}
\affiliation{Special Astrophysical Observatory of the Russian AS, 369167, Nizhnij Arkhyz, Russia}
\author{Jo~Bovy}
\affiliation{David A.~Dunlap Department of Astronomy and Astrophysics, University of Toronto, 50 St. George Street, Toronto, ON M5S 3H4, Canada}
\affiliation{Dunlap Institute for Astronomy and Astrophysics, University of Toronto, 50 St. George Street, Toronto, ON M5S 3H4, Canada}
\author{Adam J.~Burgasser}
\affiliation{Center for Astrophysics and Space Science, University of California San Diego, La Jolla, CA 92093, USA}
\author{Johan~Comparat}
\affiliation{Max-Planck-Institut f\"{u}r extraterrestrische Physik (MPE), Giessenbachstrasse 1, D-85748 Garching bei M\"unchen, Germany}  
\author[0000-0001-6559-2578]{Juan Jos\'e~Downes}
\affiliation{Centro Universitario Regional del Este, Universidad de la Rep\'ublica, AP 264, Rocha 27000, Uruguay}
\author{Doug Geisler}
\affiliation{Departamento de Astronom{\'{\i}}a, Universidad de Concepci{\'o}n, Casilla 160-C, Concepci{\'o}n, Chile}
\affiliation{Instituto de Investigaci\'on Multidisciplinario en Ciencia y Tecnolog\'ia, Universidad de La Serena. Avenida Ra\'ul Bitr\'an S/N, La Serena, Chile}
\affiliation{Departamento de Astronom\'ia, Universidad de La Serena - Av. Juan Cisternas, 1200 North, La Serena, Chile}
\author{Laura Inno}
\affiliation{INAF Osservatorio Astrofisico di Arcetri, Largo E. Fermi 5, I-50125 Firenze, Italy}
\author{Arturo Manchado}
\affiliation{Instituto de Astrofísica de Canarias (IAC), La Laguna, E-38205 Tenerife, Spain}
\affiliation{Universidad de La Laguna (ULL), Departamento de Astrof\'isica, La Laguna, E-38206 Tenerife, Spain}
\affiliation{Instituto de Astrof\'\i sica de Andaluc\'\i a, CSIC, Glorieta de la Astronom\'\i a s/n, E-18008, Granada, Spain}  
\author{Melissa K.~Ness}
\affiliation{Department of Astronomy, Columbia University, Pupin Physics Laboratories, New York, NY 10027, USA}
\affiliation{Center for Computational Astrophysics, Flatiron Institute, 162 Fifth Avenue, New York, NY 10010, USA}
\author{Marc H.~Pinsonneault}
\affiliation{Department of Astronomy, The Ohio State University, Columbus, OH, 43210, USA}
\author{Francisco~Prada}
\affiliation{Instituto de Astrof\'\i sica de Andaluc\'\i a, CSIC, Glorieta de la Astronom\'\i a s/n, E-18008, Granada, Spain}  
\author[0000-0002-1379-4204]{Alexandre~Roman-Lopes}
\affiliation{Departamento de Astronom\'ia, Universidad de La Serena - Av. Juan Cisternas, 1200 North, La Serena, Chile}
\author{Gregory V.~A.~Simonian}
\affiliation{Department of Astronomy, The Ohio State University, Columbus, OH, 43210, USA}
\author{Verne V.~Smith}
\affiliation{National Optical Astronomy Observatory, 950 North Cherry Avenue, Tucson, AZ 85719, USA}
\author{Renbin~Yan}
\affiliation{Department of Physics and Astronomy, University of Kentucky, 505 Rose St., Lexington, KY 40506-0057, USA} 
\author{Olga Zamora}
\affiliation{Instituto de Astrofísica de Canarias (IAC), La Laguna, E-38205 Tenerife, Spain}
\affiliation{Universidad de La Laguna (ULL), Departamento de Astrof\'isica, La Laguna, E-38206 Tenerife, Spain}
%
%%

%%%%%%%%%%%%%%%%%%%%%%%%%%%%%%%%%%%%%%%%%%%%%%%%%%%%%%%%%%%%%%%%%%%%
\begin{abstract}
APOGEE-2 is a dual-hemisphere, near-infrared (NIR), spectroscopic survey with the goal of producing a chemo-dynamical mapping of the Milky Way Galaxy. 
The targeting for APOGEE-2 is complex and has evolved with time.  
In this paper, we present the updates and additions to the initial targeting strategy for APOGEE-2N presented in \citet{zasowski_2017}.
These modifications come in two implementation modes: 
    (i) ``Ancillary Science Programs'' competitively awarded to SDSS-IV PIs through proposal calls in 2015 and 2017 for the pursuit of new scientific avenues outside the main survey, and 
    (ii) an effective 1.5-year expansion of the survey, known as the Bright Time Extension, made possible through accrued efficiency gains over the first years of the APOGEE-2N project. 
For the 23 distinct ancillary programs, we provide descriptions of the scientific aims, target selection, and how to identify these targets within the APOGEE-2 sample.
The Bright Time Extension permitted changes to the main survey strategy, the inclusion of new programs in response to scientific discoveries or to exploit major new datasets not available at the outset of the survey design, and expansions of existing programs to enhance their scientific success and reach.
After describing the motivations, implementation, and assessment of these programs, we also leave a summary of lessons learned from nearly a decade of APOGEE-1 and APOGEE-2 survey operations.
A companion paper, \santanatext, provides a complementary presentation of targeting modifications relevant to APOGEE-2 operations in the Southern Hemisphere.
\end{abstract}
%%%%%%%%%%%%%%%%%%%%%%%%%%%%%%%%%%%%%%%%%%%%%%%%%%%%%%%%%%%%%%%%%%%%

%%
%%%%%%%%%%%%%%%%%%%%%%%%%%%%%%%%%%%%%%%%%%%%%%%%%%%%%%%%%%%%%%%%%%%%%%%%%%%%%%%%%%%%
\section{Introduction} \label{sec:intro}

The Apache Point Observatory Galactic Evolution Experiment 2 (APOGEE-2; \inprep{S.~Majewski et al.~in prep}) is one of the programs in the Sloan Digital Sky Survey IV \citep[SDSS-IV;][]{blanton_2017} and the successor survey to APOGEE \citep[referred to as APOGEE-1, hereafter, for clarity;][]{majewski_2017} in SDSS-III \citep{eisenstein_2011}. 
APOGEE-2 operates in both the Northern and Southern hemispheres, with the original APOGEE spectrograph on the Sloan Foundation Telescope at Apache Point Observatory \citep[APO;][]{gunn_2006} and an almost clone spectrograph on the Ir\'en\'ee DuPont telescope at Las Campanas Observatory \citep[LCO;][]{Bowen_1973}; both instruments are described in \citet{wilson_2019}.
When referring to the SDSS-III program, we will use ``APOGEE-1'' and, for the combination of APOGEE-1 and APOGEE-2, we will use ``APOGEE'' to encompass the joint dataset.\footnote{Because SDSS data release are cumulative, the user will find the distinction between APOGEE-1 and APOGEE-2 to be academic. Yet, in terms of targeting, targeting-flags, and some elements of operations, there are many notable distinctions.}
When necessary to specify a specific APOGEE-2 survey component, ``APOGEE-2N'' (for APOGEE-2 North) refers to the survey, data, or instrument associated with APO, and ``APOGEE-2S'' (for APOGEE-2 South) to those at LCO. 
If we need to refer to the spectrograph specifically, APOGEE-N is the spectrograph at APO and APOGEE-S is the spectrograph at LCO.

\defcitealias{zasowski_2017}{Z17}
\defcitealias{zasowski_2013}{Z13}

The scientific goals of the APOGEE-1 survey \revise{in SDSS-III \citep{eisenstein_2011}}, and how those scientific goals were mapped into hardware, software, and data processing requirements, are given in \citet{majewski_2017}. 
The overall targeting strategy for APOGEE-1, including a description of APOGEE-1's ancillary programs, was given in \citet[][hereafter Z13]{zasowski_2013}. 
The cornerstone targeting strategy for APOGEE-1 was the use of a simple set of color and magnitude criteria in de-reddened color magnitude diagrams (CMDs) that permits precise modelling of the survey selection function; this targeting strategy forms the ``main red star sample'' that aims to target late-type giants based on their intrinsic colors in the near-infrared (NIR) and mid-infrared (MIR). 
There were, however, deviations from this strategy when deemed necessary to best achieve specific scientific goals---for example, the use of Washington$+DDO51$ photometry to pre-select likely red giant stars in the Milky Way halo among the dominant disk foreground of dwarf type stars \citep[the technique is described in][]{majewski_2000}. 
While some of these deviations require only a modification to the selection function, other programs demanded the selection of individual stars for explicit inclusion in the survey, such as confirmed member-stars in open clusters \citep[a detailed description is given in][]{frinchaboy_2013}, or the targets required for ancillary programs \citep[such as the M\,31 star clusters presented in][]{sakari_2016}, reflecting the specificity of their focused scientific goals.
Thus, \citetalias{zasowski_2013} not only included the methodology for target selection, but also how to identify the methodology for a given source using a set of targeting flags. 
\revise{APOGEE-1 observations occurred from September 2011 until July 2014, with Data Releases in 2013 \citep[DR10;][]{dr10_sdss,dr10_apogee}, 2015 \citep[DR12][]{dr12_sdss,holtzman_2015}, and the final final data release from SDSS-III in 2016 \citep[DR13][]{dr13_sdss,holtzman_2015}.}

\revise{In SDSS-IV \citep{blanton_2017}}, APOGEE-2 continues the large-scale goal of chemo-dynamical mapping of the key structural components of the Milky Way and its environment in both the Northern Hemisphere and the Southern Hemisphere, where it expands to new scientific areas due to its access to the full sky \revise{and a six-year operational timescale for APOGEE-2N} (\inprep{S.~Majewski et al.~in prep}).
APOGEE-2 largely adopted the same underlying targeting strategies described above for APOGEE, including its foundation ``main red star sample.''
\revise{However,} APOGEE-2 also elevated several ancillary projects from APOGEE-1 to key ``core'' programs, such that the science goals of these programs  became part of APOGEE-2's primary \revise{scientific goals}, while, at the same time, granted time to new ancillary programs that further expand the impact and legacy of APOGEE \revise{as a scientific project}.
\revise{APOGEE-2 also included new targeting classes, spanning from RR Lyrae in the inner Galaxy to red giants in the dwarf Spheroidal companions to the Milky Way. 
With its broader goals, longer timeline, and dual-instruments, APOGEE-2 presented a significant change from the overall targeting strategies of APOGEE-1.}

The initial targeting plans for APOGEE-2 were presented in \citet[][hereafter Z17]{zasowski_2017} and focused on the overall strategy for the \revise{joint APOGEE-2N and APOGEE-2S observing programs.}
The publication of \citetalias{zasowski_2017} was timed to accompany the first \revise{APOGEE-2} data release from SDSS-IV \citep[Data Release 14;][DR14]{dr14} \revise{that contained new observations with the APOGEE-N spectrograph from July 2014 to July 2016 \citep{holtzman_2018}.}
\revise{A subsequent APOGEE-2 Data Release occurred in 2019 with Data Release 16 \citep[DR16;][]{dr16,Jonsson_2020} that included observations from 2016 to 2018, including $\sim$1 year of observations from APOGEE-2S.
The final Data Release will occur in December 2021 (DR17) and will include the complete observational program from APOGEE-2 alongside new processing of APOGEE-1 observations.} 

When \citetalias{zasowski_2017} was written, modifications to this \revise{base plan} were anticipated, \revise{at a minimum} due to the then-incomplete commissioning of the APOGEE-2S instrument \citep[see][]{wilson_2019} and the on-going APOGEE-2N ancillary program allocation process \revise{(there were application cycles in 2015 and 2017, with implementation of programs often taking months, and observations taken over several years)}.
\revise{Since \citetalias{zasowski_2017}, additional changes in the targeting plan have also occurred through the Contributed Programs in APOGEE-2S and the Bright Time Extension (BTX) for APOGEE-2N, the latter served as an effective 1.5 year extension due to unanticipated gains in operational efficiency at APO.}

The objective of this paper is to present the summation of modifications to the field plan and targeting strategy for APOGEE-2N that was presented in \citepalias[][]{zasowski_2017} as it applies to the \revise{now complete} APOGEE-2N survey from APO (with observations from 2014 to 2020). 
A companion paper, \santanatext, provides a complementary presentation of the changes made in the APOGEE-2S observing program at LCO \revise{that began in February 2017 and completed in January 2021 as were required by on-sky performance and time allocations.} 
\revise{These papers are intended to be a formal presentation of survey strategy and motivations.}
\revise{The material is presented separately for APOGEE-2N and APOGEE-2S to make apparent the different contexts through which the survey plans from \citetalias{zasowski_2017} were modified in terms of nights available, target visibility, instrumental throughput, and operational procedures that required distinct planning and implementation strategies.}
\revise{The web documentation accompanying Data Release 17 contains a streamlined and, perhaps more practical, presentation of the total observational programs spanning APOGEE-1 and APOGEE-2.\footnote{The DR16 Documentation for Targeting is here: \url{https://www.sdss.org/dr16/irspec/targets/}, for Special Programs here: \url{https://www.sdss.org/dr16/irspec/targets/special-programs/}, and a discussion of Selection Biases and the Survey Selection Function is here: \url{https://www.sdss.org/dr16/irspec/targets/selection-biases/}. The URLs can be updated for ``dr17'' to point to the DR17 versions when DR17 becomes public in December 2021.} }

\revise{This paper for APOGEE-2N  and its companion for APOGEE-2S \santanapar\ are} intended to serve in supplement to the existing APOGEE-2 targeting paper, \citetalias{zasowski_2017}, \revise{and are intended to accompany Data Release 17 (DR17) planned for December 2021 that will release the cumulative APOGEE-1 and APOGEE-2 observations (\inprep{J.~Holtzman et al.~in prep.}).}
Thus, readers new to APOGEE-2 are strongly advised to review \citetalias{zasowski_2017} as well as \revise{the APOGEE-1 Targeting Paper \citetalias{zasowski_2013} that provides the base-line strategies}.
Only those programs \revise{in APOGEE-2N} that were modified relative to the descriptions in \citetalias{zasowski_2017} or programs that are completely new are discussed in this paper. 
However, certain keystone-information regarding targeting is repeated from \citetalias{zasowski_2017} to provide enough context that elements of this paper can stand alone \revise{(predominantly in \autoref{sec:prelims})}.
Because of the broad scientific scope of the programs in APOGEE-2N, descriptions of prior work from APOGEE-1 along with other scientific background is required to explain how that prior work has influenced the targeting and implementation of programs within APOGEE-2N. 

This paper is organized as follows. 
A summary of general information regarding APOGEE-2N targeting and the general motivation for the modifications of its strategy are given in \autoref{sec:prelims}.
The paper is organized by the \revise{significance of the} modification from \revise{the original survey strategies and plans for APOGEE-2N described in} \citetalias{zasowski_2017} from the most significant to the least significant.
\autoref{sec:strategy} details \revise{changes to the targeting strategies or selection algorithms described in \citetalias{zasowski_2017}.}
\autoref{sec:new_programs} describes the newly added programs that were not described in \citetalias{zasowski_2017} \revise{because they were added after that publication because survey observations beyond the initial six year plan became feasible}.
\autoref{sec:expansion} describes the extensions of some programs in \citetalias{zasowski_2017} during the BTX \revise{to include additional fields or span longer time baselines}.
Finally, \autoref{sec:wrapup} provides a summary that reflects on the overall process of modifying the combined APOGEE-1 and APOGEE-2 survey over its decade of operations \revise{at APO}, with a focus on highlighting intentional choices in our survey-planning \revise{specific to its targeting operations} that enabled the project to grow and adapt. 
\revise{As discussed in \santanatext, these strategies were pivotal in the successful completion of the APOGEE-2S scientific program.}

%%%%%%%%%%%%%%%%%%%%%%%%%%%%%%%%%%%%%%%%%%%%%%%%%%%%%%%%%%%%%%%%%%%%%%%%%%%%%%%%%%%%%%%%%%%%%%%%%%%%%%%%%%%%%%%%%%%%%%%%%%%%%%%%%%%%%%%
\section{Preliminaries} \label{sec:prelims}
%%%%%%%%%%%%%%%%%%%%%%%%%%%%%%%%%%%%%%%%%%%%%%%%%%%%%%%%%%%%%%%%%%%%%%%%%%%%%%%%%%%%%%%%%%%%%%%%%%%%%%%%%%%%%%%%%%%%%%%%%%%%%%%%%%%%%%%

To aid the reader, in the subsections that follow, we briefly review observing and targeting concepts from \citetalias{zasowski_2013} that continue to be used in APOGEE-2 (\autoref{sec:obsoverview} and \autoref{sec:targoverview}).
A number of SDSS- and APOGEE-specific terms and acronyms will be introduced and, as an additional aid, a glossary of these terms is given in \autoref{sec:glossary} (following that of  \citetalias{zasowski_2013} and  \citetalias{zasowski_2017}, but also including new terms used for this present paper). 
After these summaries, we provide a broad overview of the updates to the APOGEE-2N survey that motivated modifications to its targeting (\autoref{ssec:ancprograms_overview} and \autoref{ssec:btx_overview}).
We close by reviewing the final field plan for the APOGEE-2 Survey (\autoref{ssec:field_plan}) and summarizing the datasets used in this paper (\autoref{ssec:datasets}).

%%%%%%%%%%%%%%%%%%%%%%%%%%%%%%%%%%%%%%%%%%%%%
\subsection{Observational Framework} \label{sec:obsoverview}
%%%%%%%%%%%%%%%%%%%%%%%%%%%%%%%%%%%%%%%%%%%%%

As described in depth by \citet{wilson_2019}, both APOGEE spectrographs are fed by fibers that are held to a target position using plug-plates; thus, the specifics of the plate and fiber positioning places fundamental constraints on the targeting for the survey.
Plates designed for the APOGEE-N spectrograph have a 3$\degs$ diameter field-of-view; each fiber is approximately 3\arcsec\ in diameter on the sky, and two adjacent fibers cannot be placed closer than $\sim$72\arcsec\ apart, the ``fiber collision radius''. 
\revise{In addition, no targets can be placed within 96$''$ of the field center due to the central post that supports the plate \citep{Owen_1994}.}
Each unique APOGEE pointing on the sky is referred to as a field, and each unique set of targets selected for that field is called a design; even a difference of a single target will create a unique design that will be indicated by a unique design ID (an integer assigned to each design).
Designs are then drilled onto plates, but the exact locations of target holes on given plate is set by the intended hour angle at observation, to minimize the impacts of  differential refraction across the field-of-view and during an integration \citep[see discussion in][]{majewski_2017,wilson_2019}. 
Thus, a given field can have multiple designs, and any given design can have multiple associated plates. 
The final APOGEE-2 field plan is given in \autoref{fig:fieldmap}, where each circle represents a single field, but the number of targets per field depends on the number of designs and the number of stars common to those designs.

Each observation of a given plate is known as a ``visit''; a typical visit consists of about one hour of integration that is broken into eight exposures separated into two sets of spectral dithering sequences (each an ``ABBA'' sequence).\footnote{Half pixel dithering in the spectral dimension is employed by APOGEE to recover Nyquist sampling of the intrinsic spectrograph resolution \citep[see][]{majewski_2017,wilson_2019}.}
Deviations from this procedure occur when a plate has been designed specifically to focus on faint targets, and the changes are in two forms, intended to improve the overall signal-to-noise ($S/N$) in the visit spectra: 
\begin{enumerate} \itemsep -2pt
    \item beginning in late 2017, when a given plate has more than five faint targets (defined as $H$\textgreater~13.5), only a single dither sequence is used over the same time frame, with each individual exposure being doubled in length  %(effectively, receiving double length exposures; 
    (such a visit is referred to as ``DAB''); and 
    \item beginning in late 2019, when a given plate is dominated by faint stars, the plate receives an extra dither pair in a single DAB-style visit, such that its visit would sum to an hour and a half of exposure time (referred to as ``TDAB''). 
\end{enumerate}
A given field is planned to have a specific number of visits determined by its target magnitude depth, and these visits are implemented according to specific temporal spacing requirements, referred to as the ``cadence rule''.
\revise{Cadence rules vary by the scientific program. 
In the main red star sample, the cadence rules were designed with a spacing optimized to the anticipated radial velocity variation from close binaries on the red giant branch with the goal of, at a minimum, removing such stars as a potential source of uncertainty in studies of detailed Galactic dynamics and, more optimally, to determine the true systematic velocity of the binary system for use in such studies.
We note that programs specifically aiming to characterize and not just detect such variations have more complicated cadence rules.}

Generally, APOGEE aims for spectra with a minimum $S/N$ of 100 per pixel to ensure the highest quality stellar parameters and chemical abundances. 
This $S/N$ target is a fundamental constraint on the targeting and, along with the intended magnitude limit, determines the number of visits a plate will receive. 
Because the APOGEE reduction pipeline performance has proven to be similarly reliable from $S/N$ = 70 to 100 \citep[for a detailed performance assessment, see][]{Jonsson_2020}, for some programs the targeting-imposed magnitude limits have been altered; when this is the case, it will is noted.\footnote{For the purposes of this paper, ``reliable'' from $S/N$ = 70 to 100 refers to the observation that the stellar parameter and chemical abundance uncertainties, generally, over this $S/N$-range are similar. In contrast, for $S/N<70$, the uncertainties increase rapidly. 
By this reasoning, the \texttt{STARFLAG} bit for \texttt{SN\_WARN} is automatically set for all stars with $S/N<70$ \citep{holtzman_2015,holtzman_2018,Jonsson_2020}.} 
    %RLB: KEEPING THIS DISCUSSION TO MAKE SURE WE THINK THIS POINT IS CLEAR IN THE TEXT:
    %SRM:  I think better to say "to converge to stable stellar parameters for spectra with $S/N > 70%"??  Not sure what you are going for here, and I think the sentence will be confusing to the reader, but also I would prefer not to convey a thought of "unreliability" to the reader here.  Actually, unless this is a common situation, I recommend removing this whole "sometimes 70" sentence -- or simply just say "though sometimes the S/N limit was lowered to 70, when 100 was deemed unreachable in the available observing time"?
    %CRH: The S/N > 70 is not a point at which stellar parameters converge to stability, you still get stable parameters at far lower S/N, however, the random uncertainties and scatter go up quickly once you go below S/N=70, whereas over the range from 70 to 100, the reliability, i.e., uncertainties and spread don't change too much. 
\autoref{tab:maglims} summarizes the approximate $S/N$ attained for a typical star at the intended faint $H$ magnitude limit on some common design configurations (we note that the brightest target allowed is $H\sim7$ due to instrument detector saturation concerns). 
The visit number needed to reach $S/N = 100$ for a given plate magnitude limit is shown in \autoref{tab:maglims} and provides the guideline for survey planning and scheduling.  
Plates are generally categorized as ``3-visit plates'', ``6-visit plates'', etc., according to the prescriptions shown. 
Obviously, the actual on-sky $S/N$ performance achieved during visits varies due to observing conditions.

However, as can be seen, any $H=11$ star that might happen to be in a 24-visit plate, would obtain an estimated $S/N\sim490$, greatly exceeding what is required to achieve our scientific goals. 
For this reason, stars in targeted fields are often grouped by magnitude bins to form ``cohorts''; cohorts are normally only observed for the number of visits required to obtain $S/N\sim$100 for the faintest star in the cohort.  
Many fields have more than one cohort per design and this strategy combines each faint cohort with several brighter cohorts, so that the same faint stars are included in multiple designs while bright targets are switched out; this strategy increases the overall number of stars in a field that are included in the survey.
As a rule, a design will have no more than three cohorts, generally referred to as the `short,' `medium', and `long' cohorts, where the short cohorts require the fewest visits, and the long cohorts are included on the full complement of field designs to achieve the maximum exposure time.
The visits per cohort and relative number of stars in each cohorts are specific to a scientific program \citepalias{zasowski_2013,zasowski_2017}. 

%%%%%%%%%%%%%%%%%%%%%%%%%%%%%%%%%%%%%%%%%%%%%
%%
% Table
%%
\begin{table}[h]
    \centering
    \begin{mdframed}
    \caption{Signal-to-Noise Estimates for Common Faint-Magnitude and Visit Configurations}
    \label{tab:maglims}
    \begin{tabular}{c | ccccc}
        \hline \hline
         Target $H$   & \multicolumn{5}{c}{Number of 1-hour Visits} \\
         (mag)        &    1       &         3 &         6 &        12 &       24 \\
        \hline
        11.0          &  {\bf 100} &       173 &       245 &       346 &      490  \\ 
        12.2          &         53 & {\bf 100} &       141 &       200 &      283  \\ 
        12.8          &         41 &        71 & {\bf 100} &       141 &      200  \\ 
        13.3          &         29 &        50 &        71 & {\bf 100} &      141  \\ 
        13.8          &         20 &        35 &        50 &        71 & {\bf 100} \\ 
        \hline 
        \hline
        \multicolumn{6}{l}{Note: Real $S/N$ varies by observing conditions.}
    \end{tabular}
    \end{mdframed}
\end{table}
%%
% Table Spreadsheet link: https://docs.google.com/spreadsheets/d/19LH3uY87DaEcTAEUdGTmjtHWYm4dy8sNjifg2uOOl4M/edit?usp=sharing
%%
% GZ suggested we make this a figure, I don't disagree, but I am le tired.
%%
%%%%%%%%%%%%%%%%%%%%%%%%%%%%%%%%%%%%%%%%%%%%%

%%%%%%%%%%%%%%%%%%%%%%%%%%%%%%%%%%%%%%%%%%%%%%%%%%%%%%%%%%%%%%%%%%%%%%%%%%%%%%%%%%%%%%%%%%
\subsection{General Targeting Overview} \label{sec:targoverview}
%%%%%%%%%%%%%%%%%%%%%%%%%%%%%%%%%%%%%%%%%%%%%%%%%%%%%%%%%%%%%%%%%%%%%%%%%%%%%%%%%%%%%%%%%%

`Targeting' is our term for the implementation of the observational strategies of APOGEE to achieve its scientific goals through the assignment of fibers to targets. 
Both APOGEE spectrographs have 300 total fibers. 
For all plates designed in APOGEE-2, 15 fibers are assigned to hot (more ``featureless'') stars for the derivation of telluric absorption corrections, 35 fibers are assigned to ``blank sky'' positions distributed across the plate \citepalias[see][]{zasowski_2013}, and 250 fibers are available for science programs.

Because our data are only as good as the calibrations, the selection of suitable calibration fibers, both tellurics and blank sky, occur at high priority.
Telluric stars are selected to be the bluest stars in a given field that can also achieve $S/N>100$ in a single visit; because of their importance, telluric stars are selected and assigned first.
Candidate ``blank sky'' positions are selected as regions with no 2MASS point source \citep{Skrutskie_06_2mass} within 6$\arcsec$ of the position.
Though the selection of a large number of candidate sky regions occurs early in the design process, their fibers are not assigned until the science targets have been selected and \revise{then the 35 blank sky fibers are selected from the candidate positions and distributed uniformly across the plate. 
The design will not be drilled unless the 15 tellurics and 35 blank sky fibers are successfully allocated.}

Allocation of science fibers occurs in a two phase process. 
In the first phase, `special targets' are assigned following star-by-star priorities.
Special targets are generally stars that are unlikely to be picked from our standard algorithms; specific cases for the main survey include 
    (i) extremely rare or sparse targets (for example, a specific type of photometric variable star), or 
    (ii) known member stars of a substructure (for example, a star cluster or dwarf galaxy).
All targets for scientifically-focused Survey programs (for example, programs specifically targeting stars with \kepler\ observations) and targets from Ancillary Science Programs (\autoref{ssec:ancprograms_overview}) are considered special targets. 
Typically, special target lists are prepared and prioritized field-by-field by the relevant science working group or Ancillary Science Program principal investigator (PI) and submitted to the targeting team for plate design. 
In all cases, special targets are identified by special targeting flags (\autoref{tab:targeting_bits}).
If multiple sets of special targets exist for a given field, the special programs themselves are given a priority schema such that the special program with the smallest number of special targets is given the highest priority; the one exception to this rule is a field assigned to a specific program, like an open cluster, in which case the targets for that program come at highest priority. 

Once special targets are assigned, the remaining fibers are allocated following specific selection algorithms.
The core of APOGEE is the ``main red star sample'' \citepalias{zasowski_2013,zasowski_2017}, which is selected by fixed and simple color-magnitude criteria.
The color-limit and color-selection criteria adopted are set by the primary Galactic component targeted by the plate (effectively, disk, bulge, halo) and the magnitude range is set in accordance with the number of visits and cohorting scheme for a given field.
The main red star sample is well described in  \citetalias{zasowski_2013} and \citetalias{zasowski_2017}, with modifications for the APOGEE-2S survey given in \santanatext. 
The definitions of the plates designed to target specific Galactic components and their associated color cuts are given in \autoref{tab:colorcuts}. 

For fields that are devoted to specific science cases, the majority of fibers are assigned to that science case as special targets and the remaining targets are drawn from the appropriate ``main red star sample'' selection function for that part of the sky. 
For example, fibers not assigned to dwarf spheroidal galaxy members or candidate members would be assigned following the halo selection function and cohort scheme for a 24-visit field \citepalias[see \autoref{sec:dsph} in this work and also][]{zasowski_2017}.

%%%%%%%%%%%%%%%%%%%%%%%%%%%%%%%%%%%%%%%%%%%%%
\begin{table*} %%%%%%%%%%%%%%%%%%%%%%%%%%%%%%%%%%%%%%%%%%%%%%%%%%%%%%%%%%%%%%%%%%%%%%%%%%%%%%%%%%%%%%%%%%%%%%%%%%%%%%%%%%%%%%%%%%%%%%%%%%%%%%%%%%
    \centering
    \begin{mdframed}
    \caption{Color Cuts for Galactic Regions \label{tab:colorcuts}}
    \begin{tabular}{c cc l l l} 
    \hline \hline
    Galactic & $\ell$ & $b$   & Color Selection$^{a}$ &  \multicolumn{2}{c}{Targeting Flag in {\tt APOGEE2\_TARGET1}$^{b}$} \\
    Region   &  Range & Range & [mag]           &      bit    &  Description \\
    \hline 
    Bulge  & \textless~20$\degs$ or \textgreater~340$\degs$ & \textless~25$\degs$ & $0.5~\leq~(J-K_{\rm s})_0$         &  0 & APOGEE2\_ONEBIN\_GT\_0\_5 \\
    Disk   & $\geq$~20$\degs$ and $\leq$~340$\degs$         & \textless~25$\degs$ & $0.5~\leq~(J-K_s)_0$~\textless~0.8 &  1 & APOGEE2\_TWOBIN\_0\_5\_TO\_0\_8 \\
           &                                                &                     & $0.8~\leq~(J-K_{\rm s})_0$         &  2 & APOGEE2\_TWOBIN\_GT\_0\_8 \\ 
    Halo   & no $\ell$ limits                               & $\geq$~25$\degs$    & $0.3~\leq~(J-K_{\rm s})_0$         & 16 & APOGEE2\_ONEBIN\_GT\_0\_3 \\
    \hline \hline
    \multicolumn{6}{l}{ (a) The values for a star are coded in the {\tt MIN\_JK} and {\tt MAX\_JK} tags.} \\
    \multicolumn{6}{l}{ (b) The equivalent bit for APOGEE1 is {\tt APOGEE1\_TARGET1} and it follows the same definitions.}
    \end{tabular}
    \end{mdframed}
\end{table*} %%%%%%%%%%%%%%%%%%%%%%%%%%%%%%%%%%%%%%%%%%%%%%%%%%%%%%%%%%%%%%%%%%%%%%%%%%%%%%%%%%%%%%%%%%%%%%%%%%%%%%%%%%%%%%%%%%%%%%%%%%%%%%%%%%
%%%%%%%%%%%%%%%%%%%%%%%%%%%%%%%%%%%%%%%%%%%%%

%%%%%%%%%%%%%%%%%%%%%%%%%%%%%%%%%%%%%%%%%%%%%
\subsubsection{Targeting Bits} \label{sec:targbits}
%%%%%%%%%%%%%%%%%%%%%%%%%%%%%%%%%%%%%%%%%%%%%
APOGEE-2 uses bit flags to convey the targeting schema. 
The flags are not a comprehensive way of identifying stars that meet particular scientific criteria. 
Rather the flags serve to identify why a particular set of stars was targeted to enable study of (and correction for) the selection function of the survey. 
\autoref{tab:targeting_bits} provides a summary and description of the APOGEE-2 targeting bits that span four bit flags; three are used currently, \texttt{APOGEE2\_TARGET1}, \texttt{APOGEE2\_TARGET2}, and \texttt{APOGEE2\_TARGET3}, and the fourth, \texttt{APOGEE2\_TARGET4}, was added to the DR17 data model but is not currently in use. 
The specific bits that have been put into use since \citetalias{zasowski_2017} are shown in \autoref{tab:targeting_bits} in bold for ease of identification. 
The newly allocated bits largely indicate those targets selected under a specific schema, which will be described in the sections that follow. 
\santanatext\ provides a more comprehensive discussion of modifications to the targeting bits relative to that described in \citetalias{zasowski_2017}. 

%%%%%%%%%%%%%%%%%%%%%%%%%%%%%%%%%%%%%%%%%%%%%
\begin{table*}[h]
\centering
\movetabledown=1.7in
\begin{mdframed}
\begin{rotatetable}
%\begin{table*}
\caption{APOGEE-2 Targeting Bits \label{tab:targeting_bits}}
\footnotesize
%\begin{center}
\begin{tabular}{llllll} 
\hline \hline
\multicolumn{2}{c}{\texttt{APOGEE2\_TARGET1}} & \multicolumn{2}{c}{\texttt{APOGEE2\_TARGET2}} & \multicolumn{2}{c}{\texttt{APOGEE2\_TARGET3}} \\
{\it Bit} & {\it Criterion} & {\it Bit} & {\it Criterion} & {\it Bit} & {\it Criterion} \\
%\hline
\hline \hline
0  & Single $(J-K_s)_0 > 0.5$ bin         & 0  & {\bf \ktwo\ GAP Program}                    & 0  & KOI target \\
1  & ``Blue'' $0.5 < (J-K_s)_0 < 0.8$ bin & 1  & \textbf{California Cloud Target}    & 1  & Eclipsing binary \\
2  & ``Red'' $(J-K_s)_0 > 0.8$ bin        & 2  & Abundance/parameters standard               & 2  & KOI control target \\
3  & Dereddened with RJCE/IRAC            & 3  & RV standard                                 & 3  & M dwarf\\
4  & Dereddened with RJCE/WISE            & 4  & Sky fiber                                   & 4  & Substellar companion search target \\
5  & Dereddened with SFD $E(B-V)$         & 5  & External survey calibration                 & 5  & Young cluster target \\
6  & No dereddening                       & 6  & Internal survey calibration (APOGEE-1+2)    & 6  & {\bf K2 Star} \\
7  & Washington+DDO51 giant               & 7  & {\bf Outer Disk Substructure Member}        & 7  & {\bf APOGEE2 Target} \\
8  & Washington+DDO51 dwarf               & 8  & {\bf Outer Disk Substructure Candidate}     & 8  & Ancillary target \\
9  & Probable (open) cluster member       & 9  & Telluric calibrator                         & 9  & {\bf Massive Star} \\
10 & {\bf Globular Cluster Candidate}     & 10 & Calibration cluster member                  & 10 & -- {\it QSOs} \\
11 & Short cohort (1--3 visits)           & 11 & {\bf K2 Planet Host}                        & 11 & -- {\it Cepheids} \\
12 & Medium cohort (3--6 visits)          & 12 & -- {\it Kepler Synchronized Binaries}       & 12 & -- {\it The Distant Disk} \\
13 & Long cohort (12--24 visits)          & 13 & Literature calibration                      & 13 & -- {\it Emission Line Stars}\\
14 & Random sample member                 & 14 & Gaia-ESO overlap                            & 14 & -- {\it Moving Groups} \\
15 & MaNGA-led design                     & 15 & ARGOS overlap                               & 15 & -- {\it NGC 6791 Populations} \\
16 & Single $(J-K_s)_0 > 0.3$ bin         & 16 & {\it Gaia} overlap                          & 16 & -- {\it Cannon Calibrators} \\
17 & No Washington+DDO51 classification   & 17 & GALAH overlap                               & 17 & -- {\it Faint APOKASC Giants}\\
18 & Confirmed tidal stream member        & 18 & RAVE overlap                                & 18 & -- {\it W3-4-5 Star Forming Regions}\\
19 & Potential tidal stream member        & 19 & APOGEE-2S commissioning target              & 19 & -- {\it Massive Evolved Stars} \\
20 & Confirmed dSph member (non Sgr)      & 20 & {\bf Halo Member}                           & 20 & -- {\it Extinction Law} \\
21 & Potential dSph member (non Sgr)      & 21 & {\bf Halo Candidate}                        & 21 & -- {\it Kepler M Dwarfs} \\
22 & Confirmed Mag Cloud member 	      & 22 & 1-m target                                  & 22 & -- {\it AGB Stars} \\
23 & Potential Mag Cloud member 	      & 23 & Modified bright limit cohort ($H>10$)       & 23 & -- {\it M33 Clusters} \\
24 & RR Lyra star                         & 24 & Carnegie (CIS) program target               & 24 & -- {\it Ultracool Dwarfs} \\
25 & Potential bulge RC star              & 25 & Chilean (CNTAC) community target            & 25 & -- {\it SEGUE Giants} \\
26 & Sgr dSph member                      & 26 & Proprietary program target                  & 26 & -- {\it Cepheids} \\
27 & APOKASC ``giant'' sample 	          & 27 & {\bf N-CVZ OBAF stars}                      & 27 & -- {\it Kapteyn Field SA57} \\
28 & APOKASC ``dwarf'' sample 	          & 28 & {\bf N-CVZ GI Programs}                     & 28 & -- {\it K2 M Dwarfs} \\
29 & ``Faint'' target 			          & 29 & {\bf N-CVZ CTL star}                        & 29 & -- {\it RV Variables} \\
30 & APOKASC sample 			          & 30 & {\bf N-CVZ Giant with RPMJ}                 & 30 & -- {\it M31 Disk} \\
\hline 
\hline
\multicolumn{6}{l}{Note 1: A new bitmask, \texttt{APOGEE2\_TARGET4}, has been added to the data model for DR17 but is currently unpopulated.} \\ 
\multicolumn{6}{l}{Note 2: Flags that are new or different than what was presented in \citetalias{zasowski_2017} are highlighted in bold.}
\end{tabular}
%\end{center}
%\end{table*}
\end{rotatetable}
\end{mdframed}
\end{table*}

%%%%%%%%%%%%%%%%%%%%%%%%%%%%%%%%%%%%%%%%%%%%%

%%%%%%%%%%%%%%%%%%%%%%%%%%%%%%%%%%%%%%%%%%%%%
\subsection{Ancillary Science Programs} \label{ssec:ancprograms_overview}
%%%%%%%%%%%%%%%%%%%%%%%%%%%%%%%%%%%%%%%%%%%%%

In APOGEE-1 and in APOGEE-2N, a fiber reserve was intentionally budgeted into the survey plan for Ancillary Science Programs. 
This aspect of the survey design has been exceptionally beneficial, as Ancillary Science Programs from APOGEE-1 (e.g., the APOKASC and KOI programs, see \citetalias{zasowski_2013}) eventually became core components of APOGEE-2N \citepalias{zasowski_2017}.

In APOGEE-2N, approximately 5\% of the fiber hours \revise{for a six years of bright time operations} were reserved for Ancillary Science Programs. 
A fiber hour is defined as one visit for a single fiber, such that a single plate represents 265 fiber hours allocated for stars; \revise{allocation by fiber hour} allows for more flexibility in the implementation of Ancillary Science Programs. 
These were awarded by way of a competitive, internal review process that resulted in the selection of 23 programs \revise{over two application cycles}.  
The corresponding allocations could be through sparse fibers across many plates, a concentration of targets in dedicated (often new) fields or APOGEE-N observations via the fiber link to the NMSU 1-meter telescope \citep{holtzman_2010,holtzman_2015}.
\citetalias{zasowski_2017} described the general process of selecting and implementing Ancillary Science Programs, but, because the full implementation and even allocation of some of these programs was then still underway, could not include detailed descriptions of the programs, as had been done for APOGEE-1 Ancillary Science Programs in \citetalias{zasowski_2013}.

A \revise{description} of each program is given in \autoref{sec:ancillary2015} for the 2015 programs and \autoref{sec:ancillary2017} for the 2017 programs. 
\revise{These descriptions include the scientific motivations for the observations, the observations undertaken, and the specific goals of the program. 
The scope of the programs vary a great deal. 
All Ancillary Targets are input into the targeting procedure as ``special targets'' and are then drilled onto plates at high priority. 
Only in rare cases would the targets be modified (e.g., if the star was too bright and could compromise other observations). 
}

For ease, \autoref{tab:anc_sum} summarizes the 23 programs with their title and subsection reference, contact scientists, and appropriate targeting flag.
Dedicated fields for ancillary programs have \texttt{PROGRAMNAME} given as ``ancillary,'' \revise{such that whole fields could be identified and all} individual targets are flagged as summarized in \autoref{tab:anc_sum}.
We note that the timing of the 2017 call for Ancillary Science Programs relative to the Bright Time Extension (BTX; \autoref{ssec:btx_overview}) resulted in some programs from the former being absorbed into the latter; if this occurred for a particular program, it is noted both in the main text and in the appendices. 

%%%%%%%%%%%%%%%%%%%%%%%%%%%%%%
\begin{table*}[h]
    \centering
    \movetabledown=1.7in
    \begin{mdframed}
    \begin{rotatetable}
    \caption{Summary of APOGEE-2N Ancillary Programs}
    \small
    \label{tab:anc_sum}
    \begin{tabular}{l c c l}
         Program Name &  SubSection & Target Bit & Contact Scientists \\
         \hline \hline 
         Quasar Survey & \autoref{anc:quasar_survey} & {\tt APOGEE2\_TARGET3} = 10 & F.~Albareti, F.~Prada, J.~Comparat \\
         Cepheid Metallicity & \autoref{anc:cephmetals} & {\tt APOGEE2\_TARGET3} = 11 & R.~Beaton \\
         Far Disk in Low Extinction Windows & \autoref{anc:lowextwindow} & {\tt APOGEE2\_TARGET3} = 12 & J.~Bovy \\
         Hot Emission Line Stars & \autoref{anc:be_stars} & {\tt APOGEE2\_TARGET3} = 13 & D.~Chojnowski \\
         Nearby Young Moving Groups & \autoref{anc:nymg} & {\tt APOGEE2\_TARGET3} = 14 & J.~Downes \\
         Multiple Populations in NGC\,6791 & \autoref{anc:multipops} & {\tt APOGEE2\_TARGET3} = 15 & D.~Geisler \\
         A Library of Reference Stars & \autoref{anc:refstars} & {\tt APOGEE2\_TARGET3} = 16 & M.~Ness \\
         Faint \kepler\ Giants & \autoref{anc:faintkepler} & {\tt APOGEE2\_TARGET3} = 17 & M.~Pinsonneault \\
         The W3/4/5 Star Forming Complexes & \autoref{anc:w345} & {\tt APOGEE2\_TARGET3} = 18 & A.~Roman Lopes \\
         The Galaxy's Evolved Massive Stars & \autoref{anc:evolvedstars} & {\tt APOGEE2\_TARGET3} = 19 & G.~Stringfellow \\
         The APOGEE Reddening Survey & \autoref{anc:redsurvey} & {\tt APOGEE2\_TARGET3} = 20 & E.~Schlafly  \\
         M dwarf \kepler\ Objects of Interest & \autoref{anc:mdwarfs_koi} & {\tt APOGEE2\_TARGET3} = 21 & V.~Smith\\
         AGB Stars and post-AGB Stars & \autoref{anc:agbstars} & {\tt APOGEE2\_TARGET3} = 22 & O.~Zamora, A.~Manchado \\
         \hline
         M\,33 Globular Clusters & \autoref{anc:m33} & {\tt APOGEE2\_TARGET3} = 23 & B.~Anguiano \\
         Cepheid Calibrators & \autoref{anc:cephcalib} & NMSU 1-meter Program$^{a}$ & R.~Beaton \\
         Brown Dwarfs & \autoref{anc:browndwarf} & {\tt APOGEE2\_TARGET3} = 24 & A.~Burgasser \\
         Distant Halo Giants & \autoref{anc:distanthalo} & {\tt APOGEE2\_TARGET3} = 25 & P.~Harding \\ 
         The Young Galaxy & \autoref{anc:younggal} & {\tt APOGEE2\_TARGET3} = 26 & L.~Inno \\ 
         Kapteyn's Selected Areas & \autoref{anc:sa57} & {\tt APOGEE2\_TARGET3} = 27 & S.~Majewski \\
         Tidally Synchronized Binaries & \autoref{anc:tlockbinary} & {\tt APOGEE2\_TARGET2} = 12 & G.~Simonian \\
         M dwarfs in \ktwo & \autoref{anc:mdwarfk2} & {\tt APOGEE2\_TARGET3} = 27 & V.~Smith \\
         Substellar Companions & \autoref{anc:substellar} & {\tt APOGEE2\_TARGET3} = 30 & N.~Troup \\
%         MaStar& \autoref{anc:mastar} &  & R.~Yan \\
         Stellar Populations in Integrated Light & \autoref{anc:m31} & {\tt APOGEE2\_TARGET3} = 30 & G.~Zasowski, D.~Bizyaev \\
         \hline \hline
         \multicolumn{4}{l}{(a) This can be identified with {\tt TELESCOPE} of ``apo1m'' and {\tt FIELD} of ``cepheid.''} 
    \end{tabular}
    \end{rotatetable}
    \end{mdframed}
\end{table*} 
%%%%%%%%%%%%%%%%%%%%%%%%%%%%%%

%%%%%%%%%%%%%%%%%%%%%%%%%%%%%%%%%%%%%%%%%%%%%
\subsection{The Bright Time Extension} \label{ssec:btx_overview}
%%%%%%%%%%%%%%%%%%%%%%%%%%%%%%%%%%%%%%%%%%%%%
The \revise{Bright Time Extension} (BTX) was an expansion of APOGEE-2N programs to fill an excess of bright time anticipated toward the end of the SDSS-IV survey as a result of improved observational efficiencies and better than average weather at APO during the first few years of the survey. 
Planning for the BTX began in mid-2017 and fiber-hours were awarded through an open call across SDSS-IV that anticipated approximately $\sim$1200 hours would be available (corresponding to the equivalent of $\sim$1.5 years, \revise{or 20\%}, of APOGEE-2 bright time observations). 
This time was divided between a mix of programs initiated by the APOGEE-2 team and jointly with the ``After SDSS-IV (AS4)'' scientific collaboration. 
The latter has since been formally established as SDSS-V\footnote{\url{https://www.sdss5.org/}}, but the implementation of these programs used the acronym ``as4'' and throughout this paper we will use both AS4 and SDSS-V when referring to these programs.

The general APOGEE-2N BTX strategy had three primary components:
\begin{enumerate}
    \item expanding ``core'' programs and modifying their target selection to better meet the strategic science objectives of the APOGEE-2 survey (\autoref{sec:strategy}), 
    \item the construction of new programs as a reaction to developments in the scientific community, to secure a more comprehensive legacy of the APOGEE survey (\autoref{sec:new_programs}), or to build synergy with the After Sloan-IV collaboration, and
    \item the expansion of some programs to better meet their overall scientific goals (\autoref{sec:expansion}).
\end{enumerate}
All observations for the BTX have `\_btx' appended to their \texttt{FIELD} and \texttt{PROGRAMNAME} tag; one exception to this policy was the ``odisk'' program.\footnote{We note that the Outer Disk program was a completely new program and did not need `\_btx' appended to differentiate it and its targeting strategy from a similar non-BTX program.}

%%%%%%%%%%%%%%%%%%%%%%%%%%%%%%%%%%%%%%%%%%%%%
\subsection{The Final Field Plan} \label{ssec:field_plan}
%%%%%%%%%%%%%%%%%%%%%%%%%%%%%%%%%%%%%%%%%%%%%

\begin{figure*} %%%%%%%%%%%%%%%%%%%%%%%%%%%%%%%%%%%%%%%%%%%%%%%%%%%%%%%%%%%%%%%%%%%%%%%%%%%%%%%%%%%%%%%%%%%%%
    \begin{mdframed}
    \centering
    \includegraphics[width=\textwidth]{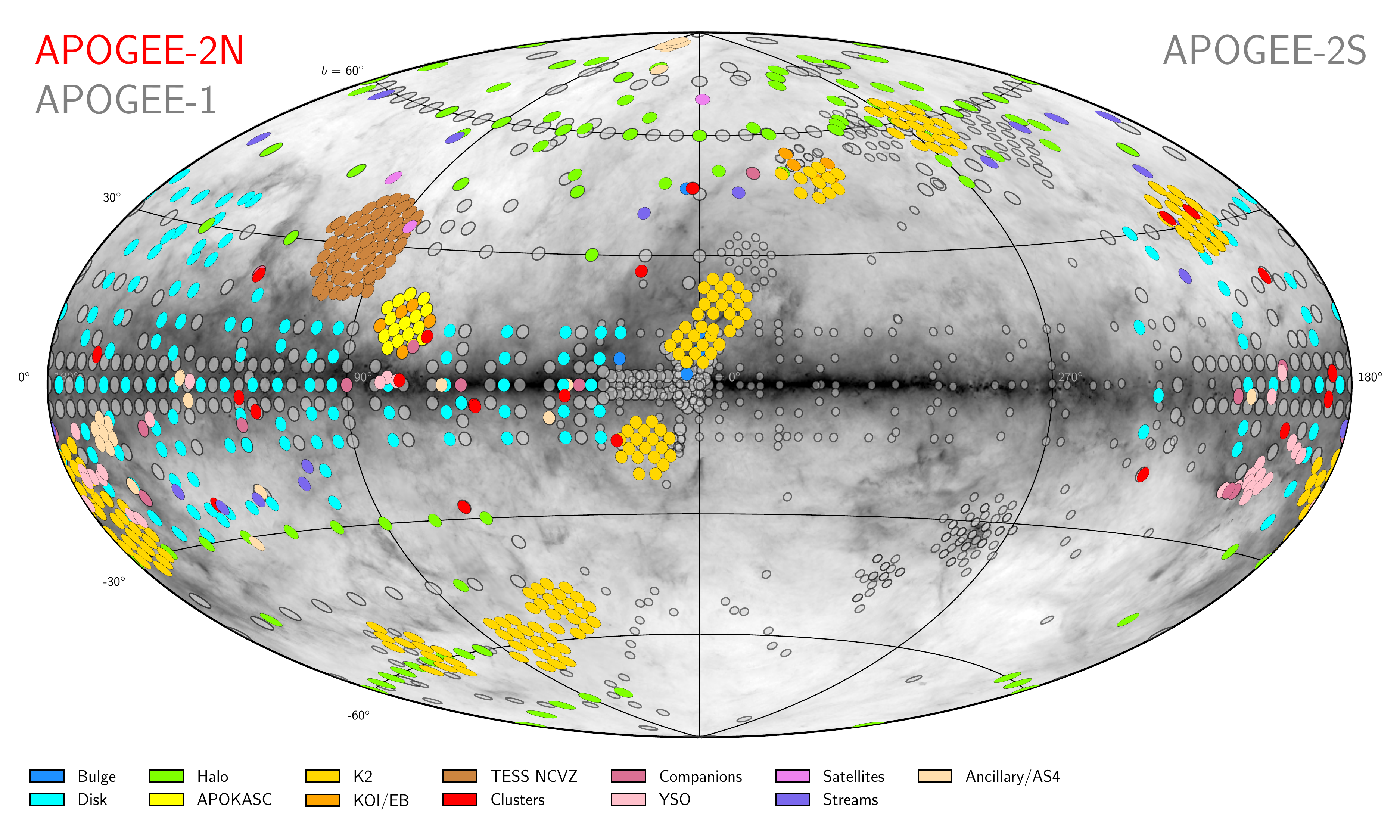}
    \caption{The complete APOGEE-1 and APOGEE-2 field plan overlaid on the \citet{sfd98} all sky infrared dust map and shown in Galactic coordinates. 
    Fields designed and observed for APOGEE-2N are color-coded by program and fields designed and those observed in APOGEE-1 or APOGEE-2S are shown in gray.
    The FOV for APOGEE-S is smaller than APOGEE-N, such that APOGEE-S pointings can be distinguished from APOGEE-N pointings by the point size. 
    The general motivations for this field plan are given in \citetalias{zasowski_2013} and \citetalias{zasowski_2017}, whereas this paper describes the Ancillary Science Programs and the BTX. 
    }
    \label{fig:fieldmap}
    \end{mdframed}
\end{figure*}

\autoref{fig:fieldmap} shows the final field plan for the APOGEE-2 survey overlaid on the \citet{sfd98} all sky infrared dust map in Galactic coordinates, with colored circles representing APOGEE-2N (this paper and \citetalias{zasowski_2017}) and grey circles representing APOGEE-1 \citepalias{zasowski_2013} and APOGEE-2S \santanatext. 
The color-coding in \autoref{fig:fieldmap} is used to indicate the primary scientific program for a given field, which, in some cases, combine programs core to APOGEE-2 (e.g., disk, halo) and BTX expansions. 
The labels roughly correspond to terms used in the \texttt{programname} tag in the summary files produced for the SDSS-IV data releases \citep[e.g.,][]{Jonsson_2020}\footnote{See also the DR16 documentation: \url{https://www.sdss.org/dr16/irspec/dr_synopsis/}}; the tag(s) specific to a program will be given with its description. 

 %%%%%%%%%%%%%%%%%%%%%%%%%%%%%%%%%%%%%%%%%%%%%%%%%%%%%%%%%%%%%%%%%%%%%%%%%%%%%%%%%%%%%%%%%%%%%

%%%%%%%%%%%%%%%%%%%%%%%%%%%%%%%%%%%%%%%%%%%%%%%%%%%%%%%%%%%%%%%%%%%%%%%%%%%%%%%%%%%
\subsection{APOGEE Datasets Used in this Paper} \label{ssec:datasets}
%%%%%%%%%%%%%%%%%%%%%%%%%%%%%%%%%%%%%%%%%%%%%%%%%%%%%%%%%%%%%%%%%%%%%%%%%%%%%%%%%%%

\revise{Throughout this paper, we will show results from APOGEE-1 and APOGEE-2 using the final sample and pipeline to be released as Data Release 17 (DR17) planned for December 2021 \holtzmanprep. 
Each of the image reduction, radial velocity measurements, spectral combination, and the derivation of stellar parameters and chemical abundances have been modified from that of Data Release 16 described in \citet{Jonsson_2020}. 
For the purposes of this paper, we typically show targets using targeting-related planes (typically \emph{Gaia} color-magnitude diagrams), but for the evaluation of some targeting strategies, we will use the stellar parameters (\logg\ and \teff).
From the perspective of what is used in this paper, however, the impacts are overall small and the reader may use intuition from \citet{Jonsson_2020} to understand these limited ASPCAP-results when presented.
}

We will often refer to ``tags'' or ``fields'' that occur in various APOGEE-2 data products using their official names.
Such ``tags'' are found in multiple APOGEE data products, as in the headers affiliated with APOGEE spectra, as well as in the more commonly used summary files, \texttt{allStar} and \texttt{allVisit}. 
Generally the ``tags'' described here will refer to those in the summary files, unless otherwise noted and tags will be referred to in true-type fonts, e.g., {\tt APOGEE\_ID}. 
A full description of the data products and their affiliated data models are given in the online documentation for the Data Release.\footnote{For DR16: \url{https://www.sdss.org/dr16/irspec/spectro_data/}}

To estimate how successful our targeting methods were, throughout the paper we will classify stars using their DR17 ASPCAP \logg\ into dwarfs, sub-giants, and giants. 
To determine the numbers of stars observed in each of these stellar classifications, we will only compare the numbers of those stars with calibrated measurements \citep[see][\holtzmanprep]{Jonsson_2020}, e.g., those with the \texttt{LOGG} tag populated. 
We will use the following definitions:
    a dwarf is a star with \logg \textgreater\ 4.1,
    a giant has -1 \textless \logg \textless 3.5, and 
    subgiants have  3.5 \textless \logg \textless 4.1. 

\revise{Because spectro-photometric can be determined for stars that are well beyond the current reach of trigonometric parallaxes from \gaia, will also use spectro-photometric distances} following the methods of \citet{Rojas-Arriagada_2017,Rojas-Arriagada_2019,Rojas-Arriagada_2020}, but applied to the \sout{internal, incremental data release mentioned above that uses the DR16 pipeline} %\citep{Jonsson_2020} 
\revise{final DR17 dataset described above (e.g. not the specific datasets described in those works, but using the same spectro-photometric distance method)}. 
\revise{As described in \citet[][their Section 2.2]{Rojas-Arriagada_2020}, the APOGEE spectroscopic parameters (\teff, \logg, and [M/H]) are used to match an observed star to potential absolute magnitudes on PARSEC stellar evolution tracks \citep{Bressan_2012,Marigo_2017}.
Using the observed 2MASS photometry ($JHK_s$), the distance and extinction can be determined.}
A detailed comparison of these distances to those derived by other studies \revise{including \gaia~trigonometric parallaxes} is given in \citet[][their appendix A]{Rojas-Arriagada_2020}.

%%%%%%%%%%%%%%%%%%%%%%%%%%%%%%%%%%%%%%%%%%%%%%%%%%%%%%%%%%%%%%%%%%%%%%%%%%%%%%%%%%%%%%%%%%
\section{Targeting Strategy Changes in the Bright Time Extension} \label{sec:strategy}
%%%%%%%%%%%%%%%%%%%%%%%%%%%%%%%%%%%%%%%%%%%%%%%%%%%%%%%%%%%%%%%%%%%%%%%%%%%%%%%%%%%%%%%%%%

This section discusses changes to the general targeting strategy established in \citetalias{zasowski_2017} as they apply to observations planned in the BTX. Fields that are subject to these changes in strategy have `\_btx' appended to their field name (and given in the \texttt{FIELD} tag) and also have `btx' appearing in the \texttt{PROGRAMNAME} tag.\footnote{There is one exception to this rule in the case of the new Outer Disk BTX program which can be identified according to it's `odisk' \texttt{PROGRAMNAME} tag (\autoref{sec:odisk}).} 
\revise{Two strategy modifications are described: (1) a change in priority star selection (\autoref{sec:mastar}) and (2) the remaining sections discuss a major targeting change aimed at bolstering the sample of distant halo stars.}

%%%%%%%%%%%%%%%%%%%%%%%%%%%%%%%%%%%%%%%%%%%%%
\subsection{Telluric Selection and MaStar Co-Targeting} \label{sec:mastar}
%%%%%%%%%%%%%%%%%%%%%%%%%%%%%%%%%%%%%%%%%%%%%

The SDSS-IV MaNGA Stellar program \citep[MaStar;][]{yan_2019} uses the MaNGA fiber bundles \revise{\citep{bundy_2015}} to collect stellar spectra; this is distinct from APOGEE-2N co-targeting with MaNGA galaxy observations that is described in \citetalias{zasowski_2017}. 
\revise{The aim of these observations is to build a \emph{fully empirical}  stellar library with a broad span in stellar type and abundance for use in the MaNGA project \citep[][]{yan_2016,yan_2019}.}
Because the observations for MaStar occur in tandem with the APOGEE-2N observations using the same plates, the plate design process takes into account the locations of MaStar targets, and this can influence APOGEE-2N targeting.

Prior to the BTX, the MaStar targets were included on the plate at the lowest priority, with the intent of having the smallest impact on the APOGEE-2 targeting. 
However, this resulted in the under-sampling of some of some of MaStar's target classes due to targeting collisions or conflicts with APOGEE-2N targets. 
One example deficiency in the MaStar sampling occurred for very luminous B and A type stars with low foreground extinction; this is a natural consequence of APOGEE's reliance on these stars as telluric calibrators, which are the highest priority targets.

To remedy the lack of such stars in the MaStar sample, the priority for plate design within the BTX fields was altered so that the highest priority MaStar targets were selected first. 
The revised default priority scheme for the BTX is as follows: 
\begin{enumerate} \itemsep -2pt
    \item MaStar high priority targets (but see description of reconciliation process below),
    \item APOGEE telluric calibration stars,
    \item APOGEE targets (but see description of reconciliation process below),
    \item MaStar low priority targets,
    \item MaStar standard stars,
    \item MaStar sky fibers,
    \item APOGEE sky fibers.
\end{enumerate}
Though this scheme solves the MaStar under-sampling issue, it poses a potential problem for APOGEE-2N science goals.

There are two impacts: the selection of the same target (a conflict) and the selection of a neighboring target the precludes another target (a collision).
For the latter, the MaStar fiber-bundle collision radius (102\arcsec) is larger than that for the APOGEE-N fibers (72\arcsec) and, as a result, MaStar targets are not a one-to-one replacement of an APOGEE-N target in a given plate design; this is particularly challenging for plates with spatially clustered targets. 
Some of the more complicated cases include targeting for star clusters, photometric objects of interest (e.g., \kepler, \ktwo, and \tess), and confirmed distant halo stars; in most of these cases, the scientific motivation to collect data for a specific target is similar for MaStar and APOGEE-2N. 
This competition for targets created a logistical and managerial challenge.

Thus, a target reconciliation process was constructed that compared the highest priority MaStar and APOGEE-2N targets for a given plate to determine conflicts (same target desired by both surveys) and collisions (where the MaStar fiber bundle precludes an APOGEE-2N target).
A team of MaStar and APOGEE-2N scientists carefully evaluated these issues (at a typical rate of only a 1-2 incidents per plate, but potentially dozens for the ensemble of plates being designed at a given time) with the aim that both programs were ensured success of their science goals.
Solutions to these conflicts and collisions included: 
    (i) one survey ceding the target, 
    (ii) for a multi-visit field, splitting the visits between MaStar and APOGEE-2N data collection, 
    (iii) inclusion of an additional design or visit to satisfy the needs of both surveys. 
This process was time-consuming, but assured mutual success of the respective goals for each survey. 

While we anticipate the net impact of these priority changes to be small, those science investigations requiring a detailed selection function analysis may want to exclude all plates designed with MaStar modifications --- i.e.,  all plates that have ``\_btx'' appended to their field name.

%%%%%%%%%%%%%%%%%%%%%%%%%%%%%%%%%%%%%%%%%%%%%%%%%%%%%%%%%%%%%%%%%%%%%%%%%%%%%%%%%%%%%%%%%%
\begin{figure*}[h]
    \begin{mdframed}
    \centering
    \includegraphics[width=1.0\textwidth]{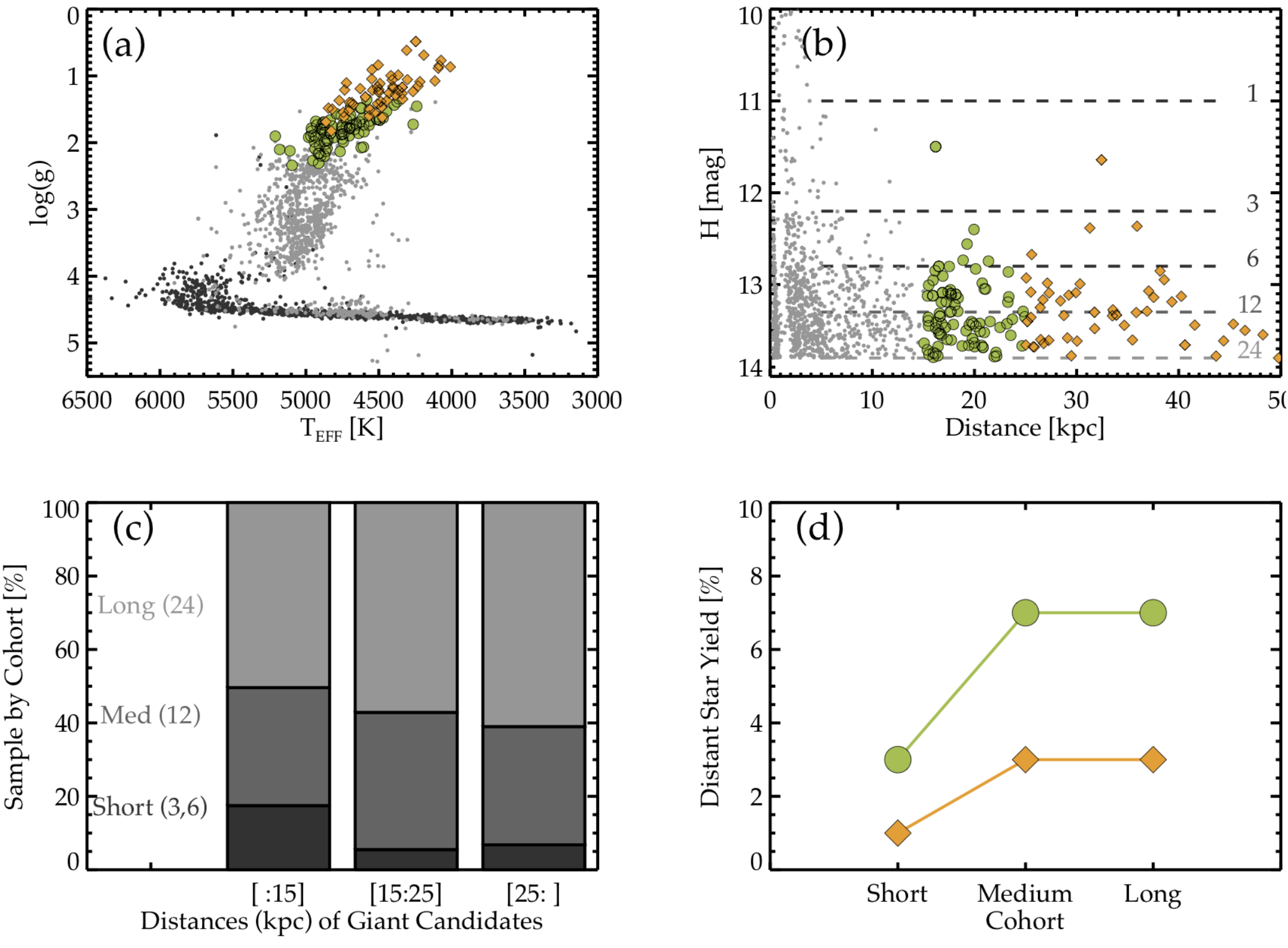}
    \caption{Evaluation of the Washington+$DDO51$ giant pre-selection in the APOGEE-2N Main Survey halo program for distant halo stars (e.g., \texttt{PROGRAMNAME}=``halo'' and \texttt{TELESCOPE}=``apo25m''). 
    (a) Spectroscopic \logg\ versus \teff\ for APOGEE-observed giant candidates (grey) and dwarf candidates (black) demonstrating the overall effectiveness of the Washginton$+DDO51$ pre-selection technique. 
    Stars with distances between 15 and 25 kpc are shown as green circles and those with distances greater than 25 kpc as yellow diamonds.
    (b) Spectro-photometric distance against apparent $H$ magnitude for giant candidates (grey) and the distant star samples (green and yellow). 
    Horizontal dashed lines indicate the magnitude limits for targeting at the specified number of visits (labeled to the right).
    (c) After separating the giant candidate sample into distance bins ($d$~\textless~15~kpc, 15 \textless~$d$~\textless~25~kpc, 25~\textless~$d$), the fraction of the sample identified in the long (24-visit), medium (12-visit) and short (3- or 6-visit) ``cohorts'' on the plate designs. 
    (d) The yield of distant stars within each ``cohort'' defined as $N_{\rm distant}/N_{\rm total}$, with $N_{\rm total}$ being the number of giant candidates.
    }
    \label{fig:halo_bad}
    \end{mdframed}
\end{figure*}
%%%%%%%%%%%%%%%%%%%%%%%%%%%%%%%%%%%%%%%%%%%%%%%%%%%%%%%%%%%%%%%%%%%%%%%%%%%%%%%%%%%%%%%%%%

%%%%%%%%%%%%%%%%%%%%%%%%%%%%%%%%%%%%%%%%%%%%%%%%%%%%%%%%%%%%%%%%%%%%%%%%%%%%%%%%%%%%%%%%%%
\subsection{Sampling Stars in the Distant Halo} \label{sec:halo}
%%%%%%%%%%%%%%%%%%%%%%%%%%%%%%%%%%%%%%%%%%%%%%%%%%%%%%%%%%%%%%%%%%%%%%%%%%%%%%%%%%%%%%%%%%

A high-level scientific goal of APOGEE is to define the chemo-dynamical fingerprint for stars in all of the structural components of the Milky Way. 
One particularly difficult component to sample is the distant halo, both because it is sparsely populated and because the stars are fainter due to their distance. 
To sample this distant and diffuse component of the Milky Way, the APOGEE-2 Science Requirements Document (SRD) set a benchmark APOGEE-2N goal of 1000 stars at distances beyond 15 kpc, such that at least 100  were ``distant'' stars beyond 25~kpc and the remaining $\sim$900 were at ``intermediate'' distances between 15 and 25 kpc.
The SRD was specific that these stars were to be identified from ``halo'' targeting (e.g., not counting stars in dwarf satellite galaxies, but including deliberately targeted streams or serendipitous targets in the background or foreground of dwarf satellite galaxies).  
  %%
  % From Zasowski et al. 2013, Sect 7.1 
  %%
  % Washington+DDO51giants(Section3.3) with(J−Ks)0  >0.3,
  % “faint” Washington+DDO51 giants with (J − Ks )0   0.3,
  % red ([J − Ks ]0   0.5) targets without Washington+DDO51 giant/dwarf classification,
  % blue (0.3   [J − Ks]0 < 0.5) targets without Washington+DDO51 giant/dwarf classification, and
  %  Washington+DDO51 dwarfs with (J − Ks )0 > 0.3.
  %%
  
Because only giants can be seen at these distances (given APOGEE's magnitude limits), giants are the targets of interest to probe the Milky Way halo, however, local dwarf stars in the disk of the Milky Way provide significant levels of contamination, providing a challenge to identify giants of interest.  
To mitigate this foreground contamination, APOGEE-1 used the Washington+$DDO51$ dwarf-giant separation technique \citep[as defined by][]{majewski_2000} to pre-select likely giant stars.
This technique combines the $DDO51$ intermediate-band filter \citep{McClure_1973}, which is centered on the surface gravity sensitive Mgb triplet at 5051~\AA, with the Washington $M$ and $T_2$ filters \citep{Canterna_1976}, which are optimized for temperature separation in late-spectral types.
As demonstrated in \citetalias{zasowski_2013} for the APOGEE-specific use of  Washington+$DDO51$, a star is classified as a likely dwarf or giant based on its location in the $M-DDO51$ versus $M-T_2$ color-color diagram.

As discussed in detail in \citetalias{zasowski_2017}, the APOGEE-1 Washington+$DDO51$ dwarf-giant separation methodology was adopted for APOGEE-2N. 
Those stars meeting our NIR color-magnitude criteria {\it and} classified as giants using the  Washington$+DDO51$ criterion were targeted at the highest priority for halo fields, with the likely dwarf stars targeted at lowest priority \citepalias[see Section~7.1 of][]{zasowski_2013}.
The APOGEE-1 strategy was continued in APOGEE-2 \citepalias{zasowski_2017} in new 3-visit ``short'' fields and additional visits to APOGEE-1 fields (so-called ``deep-drill fields'') for a total of 24 visits per field (to reach $H\sim13.8$).
The ``short''  fields had a single short cohort of 3-visits, while the ``deep-drill fields'' had four short cohorts observed for 6 visits each, two medium cohorts of 12 visits, and a single long cohort of 24 visits.

As discussed in \autoref{sec:badhalo}, at early targeting reviews, it was clear that the halo program was deficient in its number of observed giants in both the intermediate and distant samples defined in the SRD. 
The subsections that follow first explain the halo-targeting problem in more detail (\autoref{sec:badhalo}), then describe our BTX targeting scheme where we attempted to remedy this sampling problem (\autoref{sec:known_giants} and \autoref{sec:new_halo_targeting}), evaluate the new targeting scheme (\autoref{sec:how_did_we_do}), and summarize our investigation into halo targeting (\autoref{sec:halo_sum}). 

%%%%%%%%%%%%%%%%%%%%%%%%%%%%%%%%%%%%%%%%%%%%%%%%%%%%
\subsection{Detailed Evaluation of Halo Targeting} \label{sec:badhalo}
%%%%%%%%%%%%%%%%%%%%%%%%%%%%%%%%%%%%%%%%%%%%%%%%%%%%

The spectroscopic \teff--\logg\ diagram for the stars in the APOGEE-2N halo program (selected using the \texttt{PROGRAMNAME} `halo' and \texttt{TELESCOPE} `apo25m'), for which target selection relied on Washington$+DDO51$ photometry, is shown in \autoref{fig:halo_bad}a.
The data points in \autoref{fig:halo_bad}a are color-coded by their Washington$+DDO51$ photometric classification as a dwarf (black filled circle; \texttt{APOGEE2\_TARGET1} bit 8) or a giant (grey filled circle; \texttt{APOGEE2\_TARGET1} bit 7).
%%
% RLB 23 Nov 2020:
% -----------------
% 1471 w_gi with params; 
%1183 with -2 < logg < -4.1, 1288 with -2 < logg < -3.5
% now, 80% are giants; and 88% are sub-giants.
%%

For our initial, targeting review evaluations of the efficacy of our halo targeting and whether we were meeting our SRD goal of probing the more distant halo, we computed spectro-photometric distances of our observed halo stars.  
Here we reproduce that initial assessment from our targeting reviews, but using updated distances \citep[using the methods of][]{Rojas-Arriagada_2020} and sub-divide the stars into the three SRD-relevant distance bins as shown in \autoref{fig:halo_bad}a:
   (i) stars with $d$~\textless~15 kpc (black for dwarfs and grey for giants), 
  (ii) stars with 15~\textless~$d$~\textless~25 kpc (green circles), and 
 (iii) stars with 25~\textless~$d$ (orange diamonds). 
As visible in \autoref{fig:halo_bad}a, the distant stars (green and orange) are among the most intrinsically luminous in this sample.

Overall, the Washington$+DDO51$ pre-selection has proven very effective at identifying giant stars, since, of the stars with spectroscopically measured parameters, only 12\% of those pre-selected to be giant candidates turned out to be dwarf stars. 
Many of the pre-selected giants do not fall into these intermediate and distant halo bins, and instead appear to be at distances associated with the thin or thick disk. 
\autoref{fig:halo_bad}b compares spectro-photometric distances, $d$, of stars to their apparent $H$ magnitude (note that extinction is negligible in these fields) for the giant candidate sample having reliable ASPCAP results and distances \revise{(1214 stars)} with the same color-coding as \autoref{fig:halo_bad}a.
The dashed horizontal lines in \autoref{fig:halo_bad}b indicate the magnitude limits from \citetalias{zasowski_2013} and \citetalias{zasowski_2017} to obtain $S/N\sim$100 in the labeled number of visits. 
The majority of stars with $d > $1~kpc are fainter than $H$=12.2 (the 3-visit depth), but the $d$\textgreater~25~kpc stars are largely fainter than $H$=12.8 (the 6-visit depth). 

%% THIS IS FOR WASHINGTON SELECTION ONLY
%% DOES NOT INCLUDE red-star sample mag-color only selection
%% Final Numbers: DR17, Alvaro Distances (4/21/2021)
% Counts in the cohorts Alvaro Distances
%Short (3,6)      11         195
%Med (12)         12         395
%Long (24)        13         624
%% 
% TOTAL: Sht+Med = 195+395 = 590
%%
%OutPut [Sht, Med, Lng] 
% Nearest Stars      186     342     536 (1064 total)
% Middle Stars         5      34      52   (91 total)
% Farthest Stars       4      19      36   (59 total)
%%% 

\autoref{fig:halo_bad}c dissects the sample into the three distance bins by the targeting cohort; specifically, there were \revise{195, 395, and 624 stars} in the short (3- or 6-visit), medium (12-visit) and long (24-visit) cohorts, respectively.
\revise{The fractions of the stars in the nearest distance-bin were targeted using the magnitude selection for the the short, medium, and long cohorts were 17\%, 32\%, 50\%, respectively}. 
For the intermediate and distant distance-bins, \revise{just over} half of the sample was targeted using the long cohort strategy \revise{(57\% and 61\% )} with the other half coming from the sum of the medium and short cohorts \revise{(43\% and 39\%)}.
Despite the greater 0.7 mag greater $H$ limit and, especially, the commitment of $\sim$12 hours more observing time, \autoref{fig:halo_bad}c demonstrates that the long, 24-visit cohort is not more effective at yielding distant halo stars than the combination of medium and short cohorts.

\autoref{fig:halo_bad}d, which shows the intermediate and distant star targeting efficiency for each targeting cohort, further clarifies the optimal observing strategy to yield such stars. 
While the efficiency in the medium cohort is double (triple) that of the short for intermediate (distant) stars, there is no commensurate gain in efficiency from the long cohort. 
Thus, despite requiring two to four times more fiber hours per target, the long cohort did not provide a commensurate gain in the harvest of intermediate and distant star samples.  
Furthermore, while the medium cohorts yield a higher percentage of distant stars than the short cohorts, these also required more visits to reach their requisite $S/N$, meaning that for the same amount of time, the short cohorts were nearly as efficient at accumulating distant halo stars.
The ultimate reason for this is difficult to diagnose, but it could be reflective of a number of operational reasons (specific fields targeted, the depth of the Washington$+DDO51$ photometry, etc.) or astrophysical reasons (halo giant luminosity function,  global density law, the presence of halo substructure, variations of chemistry and luminosity functions in accreted systems, etc.).

Of the $\sim$900 intermediate distance stars and the 100$+$ distant stars required by the SRD, there are only \revise{91 stars in the intermediate bin} and \revise{59 in the distant bin} for the data shown in \autoref{fig:halo_bad} (\revise{this is only for the Washington-selected sample and} most of this data was in hand by the time of BTX planning). 
That means that around the time of the BTX planning, only \revise{$\sim$15\% of the goal} was met for stars beyond 15~kpc and \revise{59\% of the goal} for beyond 25~kpc.
Fortunately, the BTX program offered us an opportunity not only to add to, but also to course correct, our halo targeting to help meet our SRD goal. 

Because of the diminishing returns of the medium and long cohorts (corresponding to the deep-drill fields in the original APOGEE-2N halo plan), we opted to take a ``broad and shallow'' approach to the BTX halo targeting program (corresponding roughly 1/3 of the APOGEE-2N BTX allocation). 
We used 6-visit, single cohort halo fields rather than deeper 12- or 24-visit fields, which, despite their comparable effective yields, are more challenging to schedule and complete. 
Combined with this alteration in how observing hours were distributed by field, we also modified our target selection strategy, with the goal of a higher yield of more distant stars collected with a higher efficiency.
This modified target selection strategy employed a two-tiered prioritization scheme, with one tier focusing on fields containing confirmed distant halo stars previously identified by the SEGUE survey (\autoref{sec:known_giants}), and the second tier relying on new giant star candidate selection criteria that exploited \revise{the (then) newly available {\it Gaia}~DR1 astrometry} (\autoref{sec:new_halo_targeting}). 

%%%%%%%%%%%%%%%%%%%%%%%%%%%%%%%%%%%%%%%%%%%%%%%%%%%%%%%
\subsection{BTX Halo Targeting Method 1: Known Halo Giants in SEGUE} \label{sec:known_giants}
%%%%%%%%%%%%%%%%%%%%%%%%%%%%%%%%%%%%%%%%%%%%%%%%%%%%%%%
Coincident with the BTX planning was the submission of an Ancillary Science Program on the distant halo, described in \autoref{anc:distanthalo}, that intended to use spectro-photometric distances derived from SEGUE and SEGUE-II observations in SDSS-II and SDSS-III, respectively, to expand upon the APOGEE-2 halo targeting at large heliocentric distances \citep[][\inprep{C.~Rockosi et al.~in prep}]{yanny_2009, eisenstein_2011}. 
We opted to fold the Ancillary Science Program into the BTX halo program, making the former much larger in scope and implemented at a high priority.

\citet{xue_2014} identified over 6000 giants in the SEGUE sample, many of which were both metal poor and distant; exploiting this dataset is an ideal way to ensure that APOGEE-2 adequately samples the chemical fingerprint of the distant halo.
APOGEE-2N fields were selected to contain 2-3 of the \citeauthor{xue_2014} K-giants. 
We did not place constraints on the distance or metallicity of the \citeauthor{xue_2014} stars and, instead, opted to target any stars from the \citeauthor{xue_2014}, as any overlap with SEGUE expands upon our inter-survey cross-targeting (\autoref{sec:cross}). 

In the figures developed in the discussion to follow, the targets from \citeauthor{xue_2014} are always shown as filled black symbols.
Because we consider these targets part of our ``BTX Halo'' strategy, they will be included in efficiency metrics, in part because all of the \citet{xue_2014} targets would have been selected from our algorithmic targeting strategy (\autoref{sec:new_halo_targeting}) and, in effect, we have just prioritized their selection by including them as special targets. 

Stars targeted from \citet{xue_2014} have \texttt{APOGEE2\_TARGET2} bit 20 set and, because a different scheme was used for the APOGEE-2S Halo Program, will have a \texttt{TELESCOPE} tag of ``apo25m''. 

%%%%%%%%%%%%%%%%%%%%%%%%%%%%%%%%%%%%%%%%%%%%%%%%%%%%%%%%%%%%%%%%%%%%%%%%%%%%%%%%%%%%%%%%%%%%%%%%%%%%%%%%%%
\begin{figure*}[h]
    \begin{mdframed}
    \centering
    \includegraphics[width=1.0\textwidth]{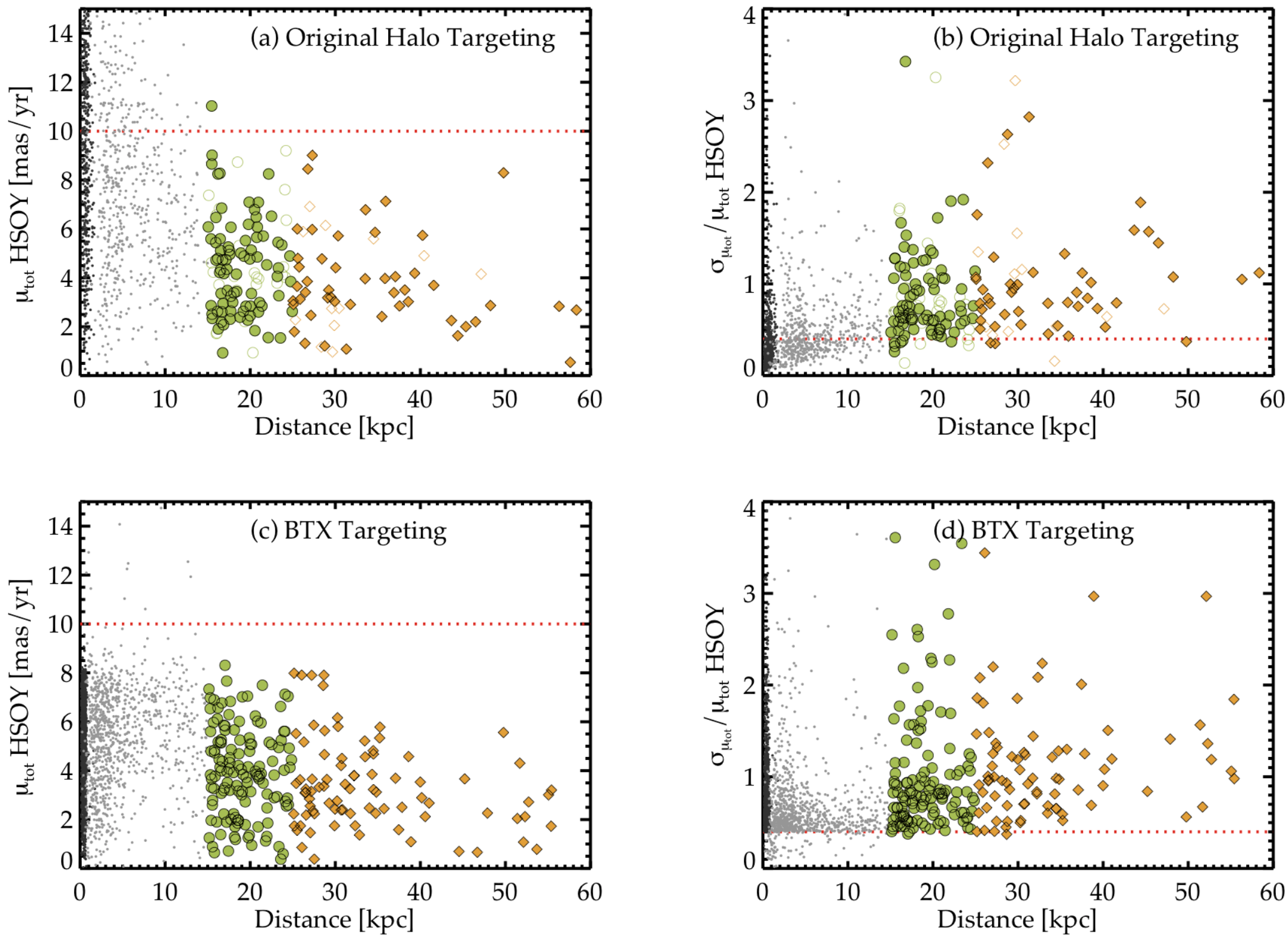}
    \caption{Giant star pre-selection using the total proper motion ($\mu_{\rm tot}$) and fractional proper motion uncertainty ($\sigma_{\mu_{tot}}/ \mu_{tot}$) from HSOY \citep{altmann_2017,altmann_2017_catalog}. 
    The original halo targeting, (a) and (b), was evaluated in all ``deep'' fields to demonstrate that distant giants almost always have $\mu_{\rm tot}$ \textless~5~mas yr$^{-1}$ and $\sigma_{\mu_{tot}}/ \mu_{tot}$ \textgreater~0.4 (indicated by dotted lines in these panels). 
    Similar to \autoref{fig:halo_bad}, confirmed giant candidates are in grey, confirmed dwarf candidates in black, and the distant samples are in green for 15~\textless~$d$~\textless~25~kpc and yellow for 25~\textless~$d$~kpc. 
    Open symbols are colored by distance bin and represent stars from the ``main red star sample'' that were targeted without Washington$+DDO51$ classification. 
    The result of using this proper motion-based strategy in the BTX is shown in panels (c) and (d), with spectroscopically-confirmed giants in grey, dwarfs in black, and the distant samples as in (a) and (b).
     }
    \label{fig:halo_pm}
    \end{mdframed}
\end{figure*}
%%%%%%%%%%%%%%%%%%%%%%%%%%%%%%%%%%%%%%%%%%%%%%%%%%%%%%%%%%%%%%%%%%%%%%%%%%%%%%%%%%%%%%%%%%%%%%%%%%%%%%%%%%

%%%%%%%%%%%%%%%%%%%%%%%%%%%%%%%%%%%%%%%%%%%%%%%%%%%%%
\subsection{BTX Halo Targeting Method 2: Exploiting Astrometry} \label{sec:new_halo_targeting}
%%%%%%%%%%%%%%%%%%%%%%%%%%%%%%%%%%%%%%%%%%%%%%%%%%%%%
Due to the short time between the BTX planning and the beginning of its observations, it was impossible to acquire and process Washington$+DDO51$ pre-imaging for plate design. 
Thus, we used the existing APOGEE observations, spectro-photometric distances, and, at that time, the recently released Hot Stuff for One Year \citep[HSOY;][]{altmann_2017,altmann_2017_catalog} catalog of proper motions, combining URAT1 \citep{urat1} and \gaia~DR1 \citep{gaia_dr1} astrometry, to see if an alternative strategy for increasing our yield of distant halo stars could be constructed.
Our original investigations of how to identify distant giants effectively explored  any fields in APOGEE reaching sufficient depth (e.g., 6, 12, or 24 visits) and included tests with a number of different spectro-photometric distance codes; but for simplicity, we will demonstrate some of the test results using the original APOGEE-2 halo sample and the spectro-photometric distances from \autoref{fig:halo_bad}. 

For these stars, \autoref{fig:halo_pm}a compares the spectro-photometric distance to the total proper motion from HSOY, where $\mu_{\rm tot}$ is taken to be the quadrature sum of $\mu_{\alpha}$ and $\mu_{\delta}$.
\autoref{fig:halo_pm}a demonstrates that the vast majority of the Washington$+DDO51$-selected distant stars (green circles and orange diamonds as in \autoref{fig:halo_bad}) have $\mu_{\rm tot}$ \textless 10 mas year$^{-1}$. 
\autoref{fig:halo_pm}b compares the derived distance to the fractional proper motion uncertainty from the HSOY measurements. 
We found that all of our distant stars not only had small $\mu_{\rm tot}$ but also had large uncertainties relative to those motions.  
These studies on data in-hand led to an algorithmic strategy that had three criteria to select halo candidates:
\begin{enumerate} \itemsep -2pt
    \item $J-K_{\rm s}$ \textgreater\ 0.5 
    \item $\mu_{\rm tot}$ \textless\ 10 mas year$^{-1}$ 
    \item $\sigma_{\mu_{tot}}/ \mu_{tot}$ \textgreater\ 0.4
\end{enumerate}
The latter two, proper motion criteria are illustrated in \autoref{fig:halo_pm}c and \ref{fig:halo_pm}d, along with the targets observed in BTX halo fields, again color-coding distant halo stars (green circles and orange diamonds as in \autoref{fig:halo_bad}).
    
To further increase our yield and acknowledging that ASPCAP produces reliable stellar parameters and chemical abundances for spectra with $S/N$ as low as 70 \citep[][]{holtzman_2015,holtzman_2018,Jonsson_2020}, we also increased the $H$-magnitude limits to $H$=13.5 in 6-visit fields (i.e., that corresponding to $S/N\sim$70 per pixel for the faintest stars; \autoref{tab:maglims}) for any halo star candidates; stars drawn from the main red star sample were still restricted to $H$=12.8 for a 6-visit field (\autoref{tab:maglims}).
Targets selected following this scheme have \texttt{APOGEE2\_TARGET2} bit 21 set and, because a different method was used for APOGEE-2S, will have \texttt{TELESCOPE} of ``apo25m''.

\begin{figure*}[ht] %%%%%%%%%%%%%%%%%%%%%%%%%%%%%%%%%%%%%%%%%%%%%
    \begin{mdframed}
    \centering
    \includegraphics[width=1.0\textwidth]{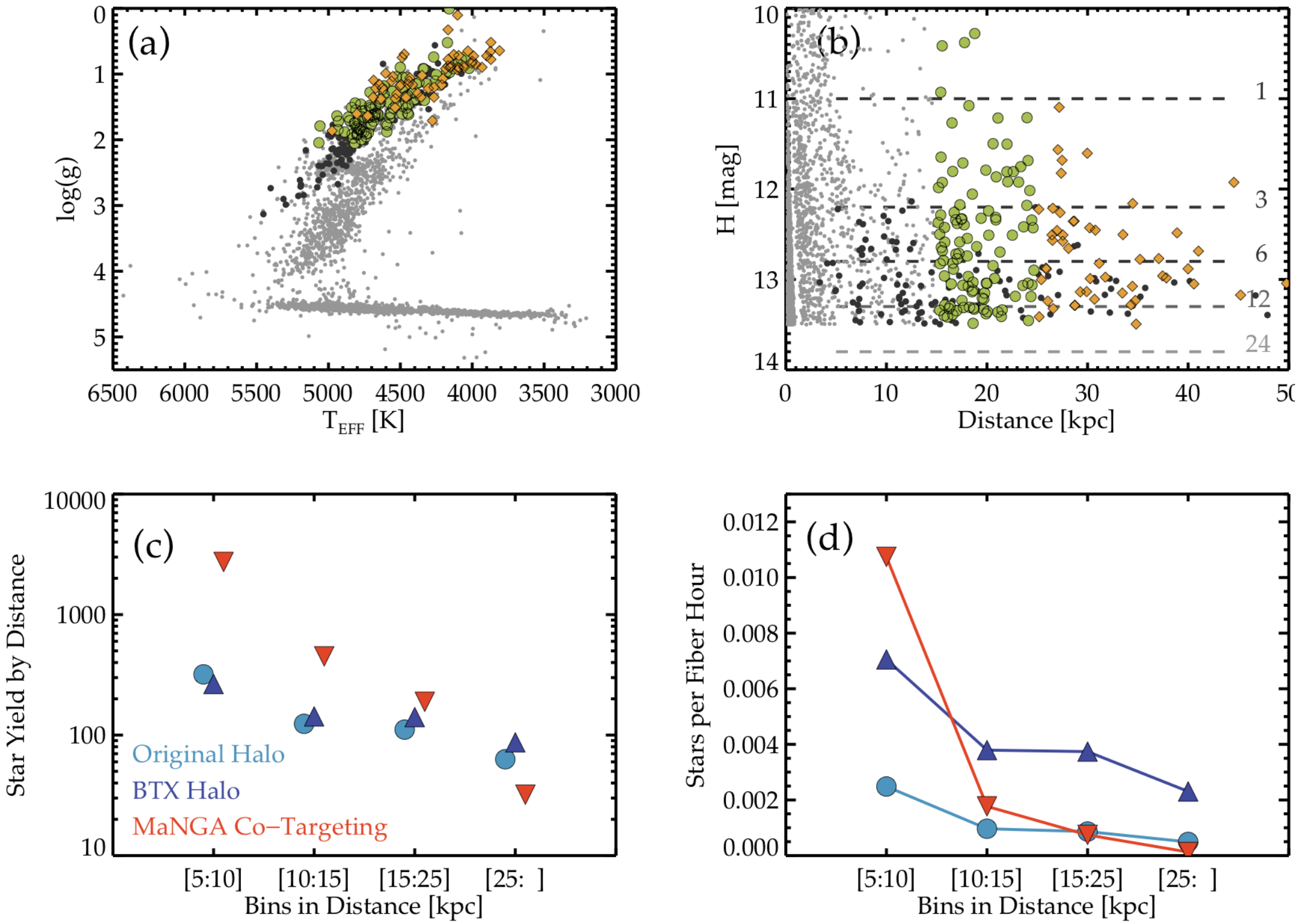} 
    \caption{
    Evaluation of the new halo targeting using HSOY proper motions described in \autoref{sec:new_halo_targeting} \citep{altmann_2017}.
    (a) Kiel diagram for BTX halo targeting. 
        Giant candidates are shown in grey, SEGUE targets are shown in black, and the distant samples are in green for the intermediate bin and yellow for the distant bin.
    (b) Spectro-photometric distance versus apparent $H$ magnitude with color-coding as in (a). 
        The depth of various targeting strategies is labeled.
        These panels can be compared with their counterparts for the original halo targeting \autoref{fig:halo_bad}a and \ref{fig:halo_bad}b.
    (c) Total number of stars in distance bins for three targeting methods: the original halo selection scheme in blue circles, the BTX scheme in purple triangles, and stars collected via MaNGA co-targeting in red down-triangles. 
        As discussed in the text, the strategies are quite different, but produce similar numbers of distant stars.
    (d) The number of stars in a given bin normalized by the total fiber hours applied to a given targeting scheme, following the color-coding in (c). 
        Here the increased yield from the BTX halo targeting is evident; indeed, the BTX targeting improved over the original halo targeting \revise{by a factor of $\sim$3$\times$ to $\sim$5$\times$ for all distance bins and by 5$\times$ to 18$\times$ over the MaNGA Co-Targeting strategy in the more distant bins.}
    Broadly we interpret the efficiency panels to indicate that dampening the dwarf-star-foreground has led to our gains.
    The data used for panels c and d is given in \autoref{tab:halo_distant_stars}.
    }
    \label{fig:halo_newtarg}
    \end{mdframed}
\end{figure*} %%%%%%%%%%%%%%%%%%%%%%%%%%%%%%%%%%%%%%%%%%%%%

%%%%%%%%%%%%%%%%%%%%%%%%%%%%%%%%%%%%%%%%%%%%%
%% RLB: Additional Notes: 
%==> UPDATE for DR17 (4/19/2021)
%  52 fields in Original Survey
%  32 fields in BTX
% 606 fields in MaNGA. 
%%
% 
%  MaNGA fields are 3 visits, so targets/fiber hours => approach 1/3 = 33%
%  Original Halo are a mix of 3 and 24 with lots of cohorts, not obvious what that should be. 
%  BTX fields are 6 visits, so targets/fiber hours => approach 1/6 = 17%
%  
%%
\begin{table}[h]
    \begin{mdframed}
    \centering
    \caption{Data used to Compare Targeting Strategies in APOGEE-2 Halo-focused Programs}
    \begin{tabular}{r c c c | c c c}
    \hline \hline
         & \textbf{MaNGA}    & \textbf{Original} & \textbf{BTX}  & \textbf{MaNGA}    & \textbf{Original} & \textbf{BTX} \\
         & \textbf{co-targ.} & \textbf{Halo}     & \textbf{Halo} & \textbf{co-targ.} & \textbf{Halo}     & \textbf{Halo} \\
    \hline \hline
    \multicolumn{2}{l}{{\it Star Counts}} & & & \multicolumn{3}{l}{{\it Fiber Hours$^{a}$}} \\
    All Targets$^{b}$   &  95740 & 12180 & 7068 &  254353 & 128461 & 37702   \\
         Giants$^{c}$   &  44189 &  3029 & 2618 &  119773 &  14581 & 24788   \\
    \hline 
    $ 5 < D < 10$   &   2732 &   319 &  266 &    8589 &   4081 &   1559   \\
    $10 < D < 15$   &    450 &   124 &  143 &    1303 &   2094 &    831   \\
    $15 < D < 25$   &    189 &   111 &  141 &     598 &   1710 &    844   \\
    $25 < D~~~~~~~$ &     32 &    63 &   87 &      98 &    985 &    513   \\
    \hline \hline
    \multicolumn{3}{l}{{\it Stars Normalized by Total Fiber Hours}} & & \multicolumn{3}{l}{{}} \\
    All Targets$^{b}$   &  0.38 &  0.09 & 0.19 &   &  &    \\
         Giants$^{c}$   &  0.37 &  0.20 & 0.11 &   &  &    \\
    \hline 
    $ 5 < D < 10$   &   0.0107 &   0.0025 &  0.0071 &   &  &     \\
    $10 < D < 15$   &   0.0018 &   0.0010 &  0.0038 &   &  &     \\
    $15 < D < 25$   &   0.0007 &   0.0009 &  0.0037 &   &  &     \\
    $25 < D~~~~~~~$ &   0.0001 &   0.0005 &  0.0023 &   &  &     \\
    \hline
    \multicolumn{3}{l}{{\it BTX Halo Improvement}} & & \multicolumn{3}{l}{{}} \\
    $ 5 < D < 10$   &   0.7 &   2.8 &  -- &   &  &     \\
    $10 < D < 15$   &   2.1 &   3.9 &  -- &   &  &     \\
    $15 < D < 25$   &   5.0 &   4.3 &  -- &   &  &     \\
    $25 < D~~~~~~~$ &  18.3 &   4.7 &  -- &   &  &     \\
    \hline
    \multicolumn{7}{l}{\it Note: All distances, $D$, are in kpc.}\\
    \multicolumn{7}{l}{\it (a) Fiber hours are the number of visits contributing to the final spectrum.}\\
    \multicolumn{7}{l}{\it (b) All targets are all stars with stellar parameters, excluding duplicates and tellurics.}\\
    \multicolumn{7}{l}{\it (c) Giants are defined to have \logg~\textless~4.1.}\\
    \end{tabular}
    \label{tab:halo_distant_stars}
    \end{mdframed}
\end{table}

%%%%%%%%%%%%%%%%%%%%%%%%%%%%%%%%%%%%%%%%%%%%%

%%%%%%%%%%%%%%%%%%%%%%%%%%%%%%%%%%%%%%%%%%%%%%%%%%%%%%%%%%%%%%%%%%%%%%%%%%%%%%
\subsection{Effectiveness of the BTX Halo Targeting Strategies} \label{sec:how_did_we_do}
%%%%%%%%%%%%%%%%%%%%%%%%%%%%%%%%%%%%%%%%%%%%%%%%%%%%%%%%%%%%%%%%%%%%%%%%%%%%%%

After several years of observing, it is now possible to evaluate the effectiveness of the strategies employed for the BTX observing.
\autoref{fig:halo_newtarg} illustrates the current status of the BTX program by way of the metrics used to understand the original halo targeting (e.g., \autoref{fig:halo_bad}a and \autoref{fig:halo_bad}b).
\autoref{fig:halo_newtarg}a shows the Kiel diagram for all targets in halo fields (grey), SEGUE distant halo stars (black), and distant stars identified based on our altered BTX targeting scheme described in Section 3.5 (green and orange). 
As may be seen in \autoref{fig:halo_newtarg}a and \autoref{fig:halo_bad}a, the distant stars in the BTX targeting span a larger range in \logg, meaning that they provide a more broadly representative sample of giant stars, not just the most intrinsically luminous ones, as in the original scheme. 
\autoref{fig:halo_newtarg}b and \autoref{fig:halo_bad}b also show that the BTX targeting has identified stars over a wide range of apparent magnitudes than in the original halo targeting scheme.
Based on \autoref{fig:halo_newtarg}, we believe our targeting criteria are actually building an overall less biased sample of stars at large distances.

The remaining panels of \autoref{fig:halo_newtarg} show metrics to quantify the relative success of our BTX halo targeting strategy against those of the other targeting scenarios used in APOGEE-2. 
Here we focus on three in particular:
\begin{itemize} \itemsep -2pt
    \item {\it MaNGA Co-Targeting:} As described in \citetalias{zasowski_2017}, APOGEE-2 observations were obtained in tandem with MaNGA observations of their galaxy sample. 
    These fields are in the North Galactic Cap, which is in our halo ($\ell$,$b$) range. 
    The APOGEE-2N targeting strategy for these ``free fibers'' adopted the high-latitude halo color-cut (\autoref{tab:colorcuts}), but a faint limit of $H\sim$11.5.\footnote{The pointings for MaNGA co-targeting are not shown in \autoref{fig:fieldmap} because their positions are not determined by APOGEE-2; these fields and their data are shown in the data release papers, see e.g., \citet[][see their Figure 1]{Jonsson_2020} and \inprep{S.~Majewski et al. (in prep.)}.} These pointings can be identified using the \texttt{PROGRAMNAME} of ``manga'' and the program is comprised of 3-visit fields. 
    While not formally part of the APOGEE-2N Halo Program, the MaNGA Co-Targeting sample is a good comparison set as it is a true \revise{near-infrared-based} magnitude-color selection without any other interventions or modifications aiming to isolate giants. \revise{There were 606 individual, but sometimes overlapping, fields in the MaNGA Co-Targeting program all to equal depth and using identical selection.}
    \item {\it Original Halo:} As described both in \citetalias{zasowski_2017} and in this paper, the original halo strategy is a combination of the Washington+$DDO51$ strategy and the color-magnitude criteria (\autoref{tab:colorcuts}). The results of this strategy are summarized in \autoref{fig:halo_bad}. This program consists of 52 pointings with a range of depths. 
    \item {\it BTX Halo:} Halo giant candidates selected according to the criteria given in \autoref{sec:known_giants} and \autoref{sec:new_halo_targeting} or as filler according to typical color-magnitude criteria (\autoref{tab:colorcuts}). This program is comprised of 32 pointings of 6-visits each.
\end{itemize}
\autoref{tab:halo_distant_stars} summarizes statistics for these three strategies, including the total number of targets and total number of fiber hours. 
The row ``All Targets'' is the number of targets with calibrated stellar parameters (e.g., {\tt LOGG} and {\tt TEFF} tags in the summary file), but excluding duplicates ({\tt EXTRATARG} bit 4) and telluric observations ({\tt EXTRATARG} bit 2) \citep{Jonsson_2020}. 
We use the same spectro-photometric distances to separate targets into four distance bins: 
    (1) 5 \textless~$d$~\textless 10 kpc, 
    (2) 10 \textless~$d$~\textless 15 kpc,
    (3) 15 \textless~$d$~\textless 25 kpc, and 
    (4) $d$ \textgreater 25kpc. 
In \autoref{fig:halo_newtarg}c, the total number of targets in each of these distance bins is plotted for each of the three strategies; here we see that the MaNGA Co-Targeting strategy actually produces a large sample of distant stars in all distance bins, larger even than the focused halo plates, with the exception of the most distant bin.
However, the raw counts of stars is not necessarily the best metric of success and we should take into account the ``resource cost'' for different means of targeting.

In \autoref{fig:halo_newtarg}d, the number of targets in a given distance bin is normalized by the total fiber hours in the program; this normalization takes into account the different spatial areas (e.g., number of distinct fields) and magnitude-depths of the strategies. 
To compute fiber hours, we sum the number of visits contributing to the final spectrum ({\tt NVISITS}) for all targets meeting the criteria of the sample; the total fiber hours is the total {\tt NVISITS} for all targets (not just those in a given distance bin). 
We note that the MaNGA Co-targeting and the Original Halo programs used approximately the same number of total fiber hours (see \autoref{tab:halo_distant_stars}), but used them differently, with MaNGA Co-targeting accumulating \revise{606 unique, but sometimes overlapping, fields} to a depth of $H\sim$11.5 and the Original Halo only targeting 52 fields, but employing the wedding-cake targeting strategy of deeper cohorts going as faint as $H\sim$13.8 (see \autoref{fig:halo_bad}b, and the star counts enumerated in \autoref{tab:halo_distant_stars}).
Here, the overall inefficiency of the MaNGA Co-Targeting for distances beyond 10~kpc is evident, however, these fields were not designed specifically for reaching distant halo stars, and, remarkably, the Original Halo strategy does not perform distinctly better. 
In the end, these performance metrics indicate that distant stars are sufficiently sparse on the sky that that a shallow-depth, but wide-area sampling strategy has benefits in terms of overall sample size.

In contrast, the BTX Halo program used \revise{about $\sim$30\% of the fiber hours} as employed in the original halo targeting (\revise{in contrast, MaNGa Co-Targing used 2x the fiber hours}), and employed them to target only 32 fields (61\% of fields in original halo \revise{and 5\% of MaNGA Co-Targeting}). 
The efficiency of the BTX halo program, as shown in the \autoref{fig:halo_newtarg}d, is \revise{$\sim$3$\times$ to $\sim$5$\times$} more efficient than the original halo targeting and \revise{from $\sim$2$\times$ to $\sim$18$\times$} more efficient than the MaNGA co-targeting strategy (the exact numbers are given in \autoref{tab:halo_distant_stars}). 
The greatest BTX gains are at the largest distances and, using \autoref{fig:halo_newtarg}b, we can see that the bulk of the distant stars have apparent magnitudes fainter than the depth probed by the MaNGA co-targeting.

%%%%%%%%%%%%%%%%%%%%%%%%%%%%%%%%%%%%%%%%%%%%%%%%%%%%%%%%%%%%%%%%%%%%%%%%%%%%%%
\subsection{Summary of the Efficacy of APOGEE-2N Halo Targeting} \label{sec:halo_sum}
%%%%%%%%%%%%%%%%%%%%%%%%%%%%%%%%%%%%%%%%%%%%%%%%%%%%%%%%%%%%%%%%%%%%%%%%%%%%%%

Thanks to routine evaluations of its targeting strategies, the APOGEE-2 targeting team was able to identify the shortcomings of the original halo targeting strategy, and then determine a set of new strategies to  reduce the effect of the dominant foreground contamination of disk stars, and increase both the total number of distant halo stars in the sample as well as the efficiency at which they were accrued. 
As a result, the BTX program netted a comparable number of distant stars as the original strategy, \revise{but with only $\sim$30\% of the total fiber hours.}
It is worth noting that the strategy was successful not so much through more efficient, deliberate targeting of distant halo stars, but rather by more effectively limiting the amount of foreground disk contamination.
We also compromised between depth and area, choosing to increase to an $H\sim$13.5 limit using 6-visits, rather than using past strategies that prioritized depth with $H\sim$13.8 limit in 24-visit fields.
In the end, the BTX Strategy of dampening the foreground improved our yield per fiber hour \revise{by factors of $\sim$3 to $\sim$5} over that of the original halo targeting strategy for stars with distances greater than 5~kpc.

APOGEE-2 also sampled the halo using a simple magnitude-color criterion with the MaNGA co-targeting. 
This sample represents a wide-area, but shall depth ($H\sim$11.5) targeting strategy that differs from the original APOGEE-2 strategy and that adopted in the BTX using proper motions. 
We find that the MaNGA co-targeting produced comparable, or even substantially larger, samples of stars at heliocentric distances greater than 5~kpc.  
However, this was due to the sheer number of fibers allocated in this way, and the halo stars accumulated were collected very inefficiently; \revise{the BTX strategy is more efficient by factors of $\sim$2 to $\sim$18} over the MaNGA co-targeting program in terms of yield per fiber hour (\autoref{tab:halo_distant_stars}). 

The counts of the MaNGA sample suggest that even relatively shallow but wide-area strategies can accumulate large numbers of stars at halo-relevant distances; however, the stars are accumulated highly inefficiently and adding in a proper-motion based criteria would improve the efficiency at reaching distant halo stars.
We note that in many cases the highest-latitude fields are so sparse in stars of any type that cohorted targeting strategies like the Original Halo will ``run-out'' of suitable bright targets.
This too motivates the sample building potential of a wide-and-shallow approach, like those taken in the MaNGA co-targeting or BTX halo.

Future surveys aiming to target the halo will need to weigh their observing strategies carefully, as we did here, but also take into account the impacts on the selection function.
For APOGEE-2, we deliberately altered our halo targeting strategy during the course of the survey to explicitly build a sufficiently large sample (e.g., hundreds of stars) at large distances so that we might better probe the chemical distributions of stars in the outer halo.
We made this choice while fully recognizing the impact it would have in complicating our selection function and other scientific investigations.
For example, the proper motion-prior imposed in the BTX halo targeting could impact dynamics-focused studies of the halo using APOGEE-2N data.  On the other hand, the halo star sample that resulted is far less-biased in \logg\ and \teff, and therefore better samples the luminosity function. 
And for all of the subsamples of halo stars within the APOGEE dataset, there are complex distance-luminosity biases imposed by the varying magnitude limits employed.
Sampling the halo is difficult and compromises have to be made that are driven by the scientific aims.

%%%%%%%%%%%%%%%%%%%%%%%%%%%%%%%%%%%%%%%%%%%%%%%%%%%%%%%%%%%%%%%%%%%%%%%%%%%%%%%%%%%%%%%%%%
\section{New Programs in the Bright Time Extension} \label{sec:new_programs}
%%%%%%%%%%%%%%%%%%%%%%%%%%%%%%%%%%%%%%%%%%%%%%%%%%%%%%%%%%%%%%%%%%%%%%%%%%%%%%%%%%%%%%%%%%
This section discusses programs undertaken in the BTX that add new scientific objectives to the APOGEE-2N program.
For each, a scientific motivation is provided to place the targeting needs and constraints in context.
\revise{There are four new programs:
 (1) mapping of the California Giant Molecular Cloud (\autoref{sec:calicloud}), 
 (2) coverage in the TESS Continous Viewing Zone (\autoref{fig:tess}),
 (3) probing the outer Galacitic disk (\autoref{sec:odisk}), 
 (4) main sequence calibrations (\autoref{sec:mainsequence}).
 }

%%%%%%%%%%%%%%%%%%%%%%%%%%%%%%%%%%%%%%%%%%%%%
\subsection{Mapping the Interstellar Medium in the California Giant Molecular Cloud}\label{sec:calicloud}
%%%%%%%%%%%%%%%%%%%%%%%%%%%%%%%%%%%%%%%%%%%%%

This program is a pathfinder to understand the observational limits for a larger program to map Giant Molecular Clouds (GMCs) planned for SDSS-V \citep{sdssv}.
To understand GMC evolution, one needs to understand better the relative importance of colliding flows, gravitational contraction, magnetic support, and turbulence over the range of physical scales and internal conditions spanned by GMCs \citep[e.g.,][]{VazquezSemadeni_2007_GMCformation,Clark_2012_GMCformation,Heitsch_2013_GMCformation,Fujimoto_2014_GMCformation}.
The velocity field of the GMC's environment may hold the key: for a GMC to contract and form stars, material must flow together.
A promising way to probe these flows around the GMC formation regime is to use the  1.5272 \micron\ Diffuse Interstellar Band (DIB) present in the APOGEE spectral range \citep{Zasowski_2015a}, in combination with line-of-sight dust column measurements \citep{Zasowski_2019_highDdust}.

The California GMC\footnote{As noted in \citet{Harvey_2013}, this cloud is called ``Auriga'' in the \spitzer\ Legacy survey. \citet{Harvey_2013} use the name ``Auriga-California Cloud.'' We stay consistent with the naming convention within APOGEE-2N of ``California Cloud'' following \citet{lada_2009}.} is a nearby ($d_{\rm los} = 400$~pc), massive (10$^{5}$ $M_{\odot}$), isolated, quiescent GMC with a large reservoir of surrounding dust and gas \citep{lada_2009,Harvey_2013}. 
\gaia~DR2 \citep{gaia_dr2} provides distances accurate to 10~pc for a large number of stars within about 50~pc of the California GMC. 
Thus, APOGEE spectra can be used to measure the mean line-of-sight velocity of the 1.5272~$\mu$m DIB, map the flow of material onto or away from the molecular cloud, and thus probe the dynamical environment of the California GMC. 
Many of the stars in the vicinity of the California GMC are either F-type stars or red giants with $H$ \textless\ 11.5 that can be targeted with single, one-hour visits to provide $S/N\sim$100 spectra suitable for measuring the DIB in $A_{\rm V}\sim1$ material \citep[a discussion of observational requirements for DIB measurements can be found in][]{Zasowski_2015a}. 
A target density of at least 1 star per 10 pc$^{3}$ is estimated to meet these scientific aims, which requires $\sim$1200 stars within 50 pc of the California GMC.
Later catalogs from \gaia\ will be released during the operation of SDSS-V and are likely to provide comparable stellar distance accuracy to GMCs that are $\sim$1 kpc away, enabling an expansion of this program to other spiral arms and star formation environments \citep[SDSS-V Science Drivers are explained in][]{sdssv}.
This pilot program was designed to determine what quantities can be recovered from different observing strategies and thus optimize the upcoming SDSS-V observations. 

This program was implemented with seven dedicated 1-visit fields with field names of the form `CA\_$lll$-$bb$\_btx', where $lll$-$bb$ are the Galactic coordinates of the field center. 
The \texttt{PROGRAMNAME} is `as4\_btx,' where `as4' stands for After Sloan-IV, the project now established as SDSS-V. 
Individual stars observed as part of this program have \texttt{APOGEE2\_TARGET2} bit 1 set. 

% The seven fields are:
%CA_160-09_btx; CA_161-06_btx; CA_162-11_btx; CA_163-05_btx; CA_163-08_btx; CA_165-07_btx; CA_165-10_btx; 	

%%%%%%%%%%%%%%%%%%%%%%%%%%%%%%%%%%%%%%%%%%%%%
\subsection{\tess\ Northern Continuous Viewing Zone} \label{sec:tessncvz}
%%%%%%%%%%%%%%%%%%%%%%%%%%%%%%%%%%%%%%%%%%%%%

\begin{figure*} %%%%%%%%%%%%%%%%%%%%%%%%%%%%%%%%%%%%%%%%%%%%%
    \begin{mdframed}
    \centering
    \includegraphics[width=0.95\textwidth]{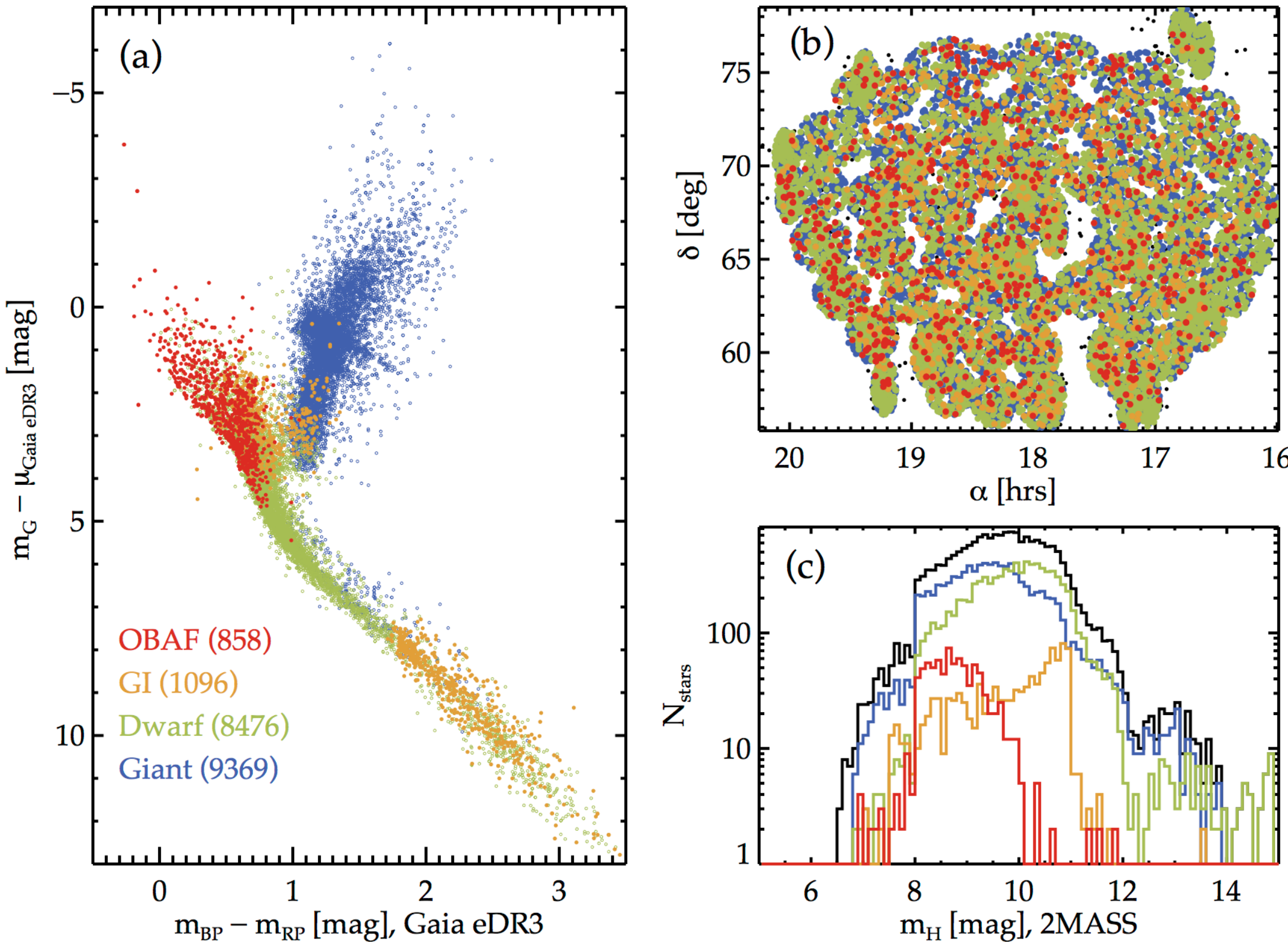}
    \caption{Overview of the \tess~N-CVZ program in APOGEE-2N. 
    (a) CMD from \gaia~DR2 photometry and using \gaia~DR2 parallaxes to compute a distance modulus ($\mu_{\rm \gaia~DR2}$). 
    The color-coding indicates the four targeting categories of 
        (1) OBAF stars (red), 
        (2) stars from the \tess~Guest Investigator programs (orange), 
        (3) stars identified as dwarf-type from the TESS Candidate Target List (CTL; green),and 
        (4) stars identified as giants in the CTL (blue). 
    (b) Sky distribution of targets. 
    (c) Total $m_{\rm H}$ distribution of targets (black), as well as histograms for each of the the four targeting categories separately (colored lines).
    }
    \label{fig:tess}
    \end{mdframed}
\end{figure*} %%%%%%%%%%%%%%%%%%%%%%%%%%%%%%%%%%%%%%%%%%%%%

As described in \citet{ricker2015}, the \tess\ mission was approved to enter Phase B implementation in 2013 with a launch no earlier than March 2018.\footnote{ \url{https://tess.mit.edu/science/}}.
This timeline precluded large-scale consideration of the \tess\ mission for APOGEE-2 planning given that APOGEE-2N operations began in 2014 (APOGEE-2S began in 2017), and with the APOGEE-2 science requirements, field plan, and targeting schema largely in place prior to even the \tess\ Phase B approval.
Thus, no specific effort to coordinate with \tess\ observations was included in the original APOGEE-2 targeting plan \citepalias{zasowski_2017}.

The BTX thus provided a key opportunity to capitalize on the scientific opportunities feasible from a joint-analysis of \tess\ and APOGEE data products.
Such opportunities include, but are not limited to: 
  characterization of planet-hosting stars \citep[e.g.,][]{canas_2019}, 
  radial velocity monitoring \citep[e.g.,][]{troup_2016}, 
  binary star identification \citep[e.g.,][]{El-Badry_2018}, 
  and stellar astrophysics \citep[e.g.,][]{apokasc,apokasc2}. 
Many of these science goals are fundamental rationales for SDSS-V \citep{sdssv}, and indeed SDSS-V planned to synergize with the \tess\ mission to help address them. 
The BTX represented a timely opportunity to test several strategies for SDSS-V, and this motivated a joint effort from APOGEE-2 and the AS4 teams.  
Because SDSS-V will operate with robotic fiber positioners that lend it a greater ability to survey sparsely-spaced \tess\ targets across the full sky, the APOGEE-2 observations of \tess\ targets have focused on the \tess\ Continuous Viewing Zones (CVZ, hereafter) in the Northern and Southern Hemispheres, which correspond to a circular area 15$\degs$ in diameter around the ecliptic poles.  
The latter regions of sky have received 365 day coverage at 30-minute cadence by \tess\ in the original two year mission, with additional observations occurring in the on-going \tess\ extended mission.
While similar in concept and yielding complementary data, the Southern \santanapar\ and Northern \tess\ CVZ programs differ in their logistical implementation.

A \tess-focused APOGEE-2N program is challenging because, despite spanning a large area on the sky, the CVZ is only accessible for observations over a limited range of local sidereal time (LST) for ground-based observations (with a bulk of the N-CVZ accessible to the APOGEE-N spectrograph from roughly at LSTs of 17 to 19 hours).
Moreover, this LST range is already over-subscribed in APOGEE-2N owing to the location of the \kepler\ field (R.A.,Dec = 19:22:40,+44:30:00) and the cadence requirements for observing programs in the \kepler\ field (see \autoref{sec:koi} and discussion in \citetalias{zasowski_2017}). 
Thus, only a limited number (75) of hours were available and allocated to this program over the 1.5 years of the BTX, and these observations had to be spread out over several years because of these LST constraints.

Planning how to implement the 75 1-visit fields was a joint effort between the scientific teams within SDSS-IV and SDSS-V, who identified four classes of targets to observe, and which are, in priority order (see Fig. \ref{fig:tess}a): 
\begin{enumerate} \itemsep -2pt
   \item hot stars of OBAF spectral type (\texttt{APOGEE2\_TARGET2} bit 27), 
   \item stars on \tess\ 2-min cadence, largely those either from the \tess\ Guest Investigator programs or candidates for such programs (\texttt{APOGEE2\_TARGET2} bit 28), 
   \item dwarf stars in the  Asterosiesmic Target List \citep[ATL;][]{tess_atl} or Candidate Target List \citep[CTL;][]{stassun_2019} (\texttt{APOGEE2\_TARGET2} bit 29),
   \item giant type stars generally meeting the specifications of the APOGEE `main red star sample' \revise{and drawn from the TESS Input Catalog (TIC;} \texttt{APOGEE2\_TARGET2} bit 30). 
\end{enumerate}
An effort was made to have roughly 50\% dwarfs and 50\% giants on a given plate to ensure broad coverage in the Hertzsprung-Russell diagram.
These four classes of science targets are briefly described below, before discussion of the detailed implementation of the \tess\ N-CVZ program (\autoref{sec:tessncvz_implement}). 

%%%%%%%%%%%%%%%%%%%%%%%%%%%%%%%%%%%%%%%%%%%%%%%%%%%%%%%
\subsubsection{OBAF Stars} \label{sec:tessncvz_obaf}
%%%%%%%%%%%%%%%%%%%%%%%%%%%%%%%%%%%%%%%%%%%%%%%%%%%%%%%
High-resolution spectra for oscillating OBAF stars are the foundation for an goal of ``dynamical asteroseismology'' for stars with convective cores; this program was designed in synergy with the program planned for SDSS-V. 
These stars contribute to the dynamical and chemical evolution of galaxies, but the models of their stellar structure and evolution are known to be inadequate. 
By comparing seismically determined parameters with spectroscopic parameters and dynamical masses from modelling multi-epoch radial velocities, we will infer precise constraints, for example on the size of the convective core.
Such constraints are necessary for the calibration and improvement of the present-day models of stellar structure and evolution for these stellar types \citep[see e.g.,][]{Pedersen_2018}. 
The SDSS-IV observations for the OBAF stars provide a first epoch radial velocity measurement to aid in the orbital determinations as well as allow for a first pass on their stellar parameters.
These stars have \texttt{APOGEE2\_TARGET2} bit 27 set.  

%%%%%%%%%%%%%%%%%%%%%%%%%%%%%%%%%%%%%%%%%%%%%%%%%%%%%%%
\subsubsection{Stars with 2-min Cadence Observations} \label{sec:tessncvz_gi}
%%%%%%%%%%%%%%%%%%%%%%%%%%%%%%%%%%%%%%%%%%%%%%%%%%%%%%%

\tess\ Cycle 1 and Cycle 2 observations produced two data products: full-frame images (FFIs) at 30 minute cadence and ``postage stamp'' images at 2 minute cadence. 
The former category ensures that every bright star has some data, whereas the latter category was reserved for Guaranteed Time Observations (GTO) from the \tess\ team and Guest Investigator (GI) observations from the broader astronomical community, with the latter awarded through competitive proposal cycles. 
Thus, a portion of our program was reserved for stars with 2-minute cadence observations from Guest Investigator samples and given high priority to ensure these rare-target classes would be sufficiently sampled. 
Such targets include known planet hosts, cool dwarfs, and subgiants. 
These targets are all identified by \texttt{APOGEE2\_TARGET2} bit 28. 

%%%%%%%%%%%%%%%%%%%%%%%%%%%%%%%%%%%%%%%%%%%%%%%%%%%%%%%
\subsubsection{Dwarf Stars} \label{sec:tessncvz_ctl}
%%%%%%%%%%%%%%%%%%%%%%%%%%%%%%%%%%%%%%%%%%%%%%%%%%%%%%%

After the rare target classes were selected, roughly half of the fibers ($\sim$125) were reserved for dwarf-type stars drawn from two \tess\ scientific target lists: 
    (i) first, stars from the \revise{Asterosiesmic Target List \citep[ATL;][]{tess_atl}} and then 
    (ii) stars from the \revise{Candidate Target List \citep[CTL;][]{stassun_2019}}.

Stars with solar-like oscillations were selected from the \tess\ ATL produced by the \tess\ Asteroseismic Science Consortium (TASC)\footnote{\url{http://tasoc.dk}} as of Version 4.\footnote{This is Version 4 of the ATL produced in $\sim$October 2017 (see TASOC website \url{https://tasoc.dk/wg1/Targetselection}) that implemented discussion from the TASC3/KASC10 workshop; \url{https://www.tasc3kasc10.com/}} 
The final ATL sample and its detailed derivation is described in \cite{tess_atl}, but we provide a brief summary of the intermediate catalog and priority scheme that was available for our plate design.
Stars were selected based on their assigned priority in the ATL. 
The ATL priorities that we used were assessed based on a term known as $P_{\rm mix}$ that is a linear combination of the likelihood of the detection of seismic modes and the likelihood that such modes are falling-off beyond detectability based on the stellar parameters relative to the known space in the Hertzsprung-Russell diagram with seismic modes.\footnote{$P_{\rm mix} = (1-\alpha)~P_{vary} + \alpha~P_{\rm fix}$, where $\alpha$=0.5 in the version of the catalog used for APOGEE-2 targeting.}
These parameters were selected to balance competing scientific objectives in the ATL.
This is similar to what is described in \citet[][their Subsubsection 4.3.3]{tess_atl} and the APOGEE-2 sample should mimic the overall distribution of the larger ATL. 
At the time of plate design, $\sim$900 ATL stars with solar-like oscillations were in the \tess~N-CVZ footprint. 
The entire list was folded into the priority schema and the ATL stars will have \texttt{APOGEE2\_TARGET2} bit 28 and \texttt{APOGEE2\_TARGET2} bit 29 set because they were targeted as part of a GI program {\it and} counted toward the dwarf-stars quota for each plate.

After selecting candidates from the ATL, the remaining fibers for dwarf-type targets were assigned from the CTL version 8 (CTL8) \citep[][]{stassun_2019}.
As described in \citet{stassun_2018, stassun_2019}, the CTL is a set of stars selected from the \revise{\tess~Input Catalog (TIC)} that are ideal candidates for the \tess\ planet-finding mission;
CTL stars are high-likelihood main-sequence or sub-giant type stars with their likelihood determined from the broad range of photometric and astrometric measurements collated into the TIC.

A major component of the classification of stars from the TIC and into the CTL was the likelihood that a star was a dwarf.  
Accordingly, a key component of this determination was the use of the reduced proper motion diagram (RPM), more specifically the NIR version of RPM developed by \citet{colliercameron_2007} known as RPMJ. 
This means that astrometric information (proper motions) are important for the CTL and the classification of sources may have some dependence on the astrometric catalog being adopted \citep[for evaluations pre- and post- adoption of \gaia\ astrometry see][]{stassun_2019}. 
We will return to this in \autoref{sec:tessncvz_implement} for the \tess~N-CVZ program and, because the RPMJ technique is used in the Outer Disk Program, more discussion is included in \autoref{sec:outerdist_assess}.
APOGEE-2N targets selected from the CTL will have \texttt{APOGEE2\_TARGET2} bit 29 set. 

Two sets of fiber assignment were performed from the CTL: 
    a ``bright'' set limited to targets with $H<$ 12 and 
    a ``faint'' set of targets than with $12 < H < 14$; 
the latter ``faint'' selection was performed after selecting ``bright'' giants but before selecting ``faint'' giants as described in the next subsection.
In either selection round, the stars in the CTL were ranked by their CTL-priority \citep[see][their Section 3.4]{stassun_2018}, which, briefly, is the probability of detecting a transit signal from a small, rocky planet from a typical \tess\ 2-min postage stamp observation.

%%%%%%%%%%%%%%%%%%%%%%%%%%%%%%%%%%%%%%%%%%%%%%%%%%%%%%%
\subsubsection{Giant Stars} \label{sec:tessncvz_giants}
%%%%%%%%%%%%%%%%%%%%%%%%%%%%%%%%%%%%%%%%%%%%%%%%%%%%%%%

After the rare target classes were selected, roughly half of the fibers ($\sim$125) were reserved for candidate giant-type stars drawn from the TIC  \citep{stassun_2018,stassun_2019}. 
Two rounds of fiber assignment were performed, a ``bright'' with $H<$13 and a ``faint'' set with $13 < H < 14$. 
These two rounds occurred subsequent to the``bright'' and ``faint'' dwarf samples, respectively, to both prioritize dwarfs over giants while prioritizing bright stars over faint stars. 
In both cases, the giant-candidates were selected following the dwarf-giant classification criteria given in \citet{stassun_2018} using reduced-proper motion diagrams \citep[the criterion of][]{colliercameron_2007}. 
Giants selected from the TIC have \texttt{APOGEE2\_TARGET2} bit 30 set. 

After all of the above \tess\ target selections were completed, any remaining fibers were assigned following the standard ``main red star sample'' criteria in APOGEE-2 \citepalias[][]{zasowski_2013,zasowski_2017} and are flagged accordingly.

%%%%%%%%%%%%%%%%%%%%%%%%%%%%%%%%%%%%%%%%%%%%%%%%%%%%%%%
\subsubsection{Implementation of the N-CVZ Program} \label{sec:tessncvz_implement}
%%%%%%%%%%%%%%%%%%%%%%%%%%%%%%%%%%%%%%%%%%%%%%%%%%%%%%%
The first year of APOGEE-2N observations of the \tess~N-CVZ was planned in advance of the \tess~Guest Investigator cycles and the release of \gaia~DR2 \citep{gaia_dr2}, both of which were deemed to likely have a significant impact on how to optimize our limited number of fiber hours to apply to the \tess\ science cases.
Thus, the first set of APOGEE-2N observations of the \tess~N-CVZ were drilled around the locations of rare, O and B type stars (priority 1), with the remaining fibers assigned following our general schema but using the Tycho-\gaia\ Astrometric Solution from \gaia~DR1 \citep[TGAS][]{gaia_dr1} for the RPMJ computations;
21 such plates were designed around these OB stars (\texttt{FIELD} names of CVZ\_OB\#\#\_btx) plus three additional plates (\texttt{FIELD} names of CVZ\_FILL\#\#\_btx).

A second set of plates were designed to attain more-or-less uniform spatial coverage of the N-CVZ; 
51 plates were required and these have \texttt{FIELD} names CVZTILE\_$lll\pm bb$\_btx, with $lll\pm bb$ representing the Galactic coordinates of the field center. 
These plates were designed simultaneously from a consistent list of input targets and relied on \gaia~DR2 \citep{gaia_dr2} for the RPMJ computation; the plates, themselves, were drilled over time to optimize our use of this LST range (e.g., selecting Hour Angles of observation that optimized our observing schedule).
All plates from this program have \texttt{PROGRAMNAME} `cvz\_btx' with individual targets flagged as described above.

\autoref{fig:tess} gives a summary of the targeting for the \tess~N-CVZ program.
\autoref{fig:tess}a shows a \gaia~DR2 color-absolute magnitude diagram with the distinct targeting classes identified using colors (OBAF stars are red, GI are orange, dwarfs are green, and giants are blue). 
\autoref{fig:tess}b provides the sky distribution of the targets.
During the first year of observations, an error in target lists led to spatial distributions for science targets that did not fill the full plate footprint. These plates are visible as oblong footprints in \autoref{fig:tess}b, surrounded by telluric standards that did utilize the full circular footprint of the plate. 
\autoref{fig:tess}c provides histograms of the magnitude range spanned by each of the four targeting classes; these plates were only intended to have single visits, but span a larger range of magnitudes than typical 1-visit plates in APOGEE-2, as a test of strategies planned for SDSS-V.

%%%%%%%%%%%%%%%%%%%%%%%%%%%%%%%%%%%%%%%%%%%%%
\subsection{Probing the Outer Disk} \label{sec:odisk}
%%%%%%%%%%%%%%%%%%%%%%%%%%%%%%%%%%%%%%%%%%%%%

The advent of photometric and spectroscopic surveys over large areas of the sky have led to a revolution in our understanding of the structure of the outer disk.
Far from there being a slow ramp down of the inner disk properties, the outer disk is abundant with star clusters \citep[e.g.,][]{Zasowski_2013clusters}, apparent substructure \citep[e.g.,][]{Slater_2014}, and perhaps additional, yet undiscovered features.  
With the advent of large-area imaging surveys, such as SDSS and 2MASS, star-count maps revealed apparent stellar overdensities in the outer disk; these include the 
    Monoceros Ring \citep[sometimes included into the ``Galactic Anticenter Stellar Structure,'' or GASS;][]{Newberg_2002,crane_2003,yanny2003}, 
    Canis Major \citep{martin_2004}, 
    Triangulum-Andromeda \citep[TriAnd;][]{majewski_2004,rocha-pinto_2004}, 
    and A13 \citep{Sharma_2010,li_2017}, 
    among other, smaller features \citep[for a more exhaustive list see the overview from][]{grillmair_carlin_2016}.  

Due to their discovery at a time when clearly identifiable stellar streams from dwarf galaxies were also being uncovered, such as the Sagittarius stream \citep{ivezic_2000,Newberg_2002,majewski_2003}, these overdensities were broadly interpreted to be debris from dwarf satellite mergers \citep{rocha-pinto_2003,penarrubia_2005,chou_2010,chou_2011,sollima_2011,sheffield_2014}. 
Spectroscopic follow-up of such features largely seemed to justify these interpretations \citep{crane_2003,chou_2011,sheffield_2014, deason_2014}, but typically limited themselves to only sampling the most likely member stars and not broadly examining the larger scale behavior of the stars in these areas of the Milky Way.

While the dwarf galaxy debris origin was originally favored, evidence grew that these Galactic Anticenter overdensities could also be related to the Milky Way disk, in particular as perturbations to the disk excited by orbiting dwarf galaxies \citep{kazantzidis_2008,purcell_2011,gomez_2013,gomez_2016,price-whelan_2015,xu_2015,li_2017,newberg_2017,laporte_2018}.  
In this picture, density waves in the Milky Way disk would be expressed as vertical oscillations of the disk midplane and and the crests and troughs of these waves would appear as apparent overdensities \citep[although in reality, because of the low density in the outer disk, these overdensities may instead appear to be more feathery, spiral arm-like features rather than a strictly rippled disk; for more realistic examples see][]{laporte_2018}.  

Various studies argued for this revised picture of the Galactic Anticenter overdensities, given that
    (1) the ratio of RR Lyrae to M-giant type stars implied a lack of old stars, which are normally seen in dwarf spheroidal galaxies \citep{price-whelan_2015}, 
    (2) star counts suggest a connection between the apparent substructures that could be interpreted as one continuous feature originating from the Galactic disk \citep{xu_2015}, and 
    (3) we may expect to see these kinds of corrugations across the outer disk excited by known dwarf galaxy satellites like the Sagittarius dSph \citep[e.g.,]{laporte_2018}.  
However, the few existing spectroscopic analyses of the Galactic Anticenter overdensities largely focused on small numbers of stars, were of moderate spectral resolution, and lacked strong sampling of the chemo-dynamics of the outer disk, where these overdensities are located, so that chemistry could not decisively be brought to bear on the origin of these overdensities.

The original targeting plan for APOGEE-2N contained five pointings on one of these overdensities, Triangulum-Andromeda \citepalias[referred to as TriAnd in][]{zasowski_2017}, each of which contains only a handful of confirmed members of the TriAnd feature \citep[drawn from][]{chou_2011,sheffield_2014}, but designed with the hope that the strategy for the main star red sample would naturally identify additional members \revise{over the extent of TriAnd on the sky (from 100$\degs$\textless$\ell$\textless150$\degs$ and -50$\degs$\textless$b$\textless-15$\degs$).}
Despite the relatively small sample of TriAnd stars as well as the limited sample of APOGEE targets in the outer disk available at that time, \citet{hayes_2018} used samples of TriAnd and available outer disk stars to show convincingly that the TriAnd chemistry in [$\alpha$/Fe]-[Fe/H] space is  a natural extension of disk chemical patterns to lower metallicity, a finding in agreement with the concurrent study by \citet{bergemann_2018}.
These high resolution spectroscopic studies revealed the potential for chemistry to be used to help clarify the origin of these other Galactic Anticenter overdensities and better understand their evolution, and motivated a more thorough probe of the Milky Way's outer disk with available time in the BTX.

\begin{figure*} %%%%%%%%%%%%%%%%%%%%%%%%%%%%%%%%%%%%%%%%%%%%%%%%%%%%%%%%%%%%%%%%%%%%%%%%%%%%%%%%%%%%%%%%%%
    \begin{mdframed}
    \centering
    \includegraphics[width=0.95\textwidth]{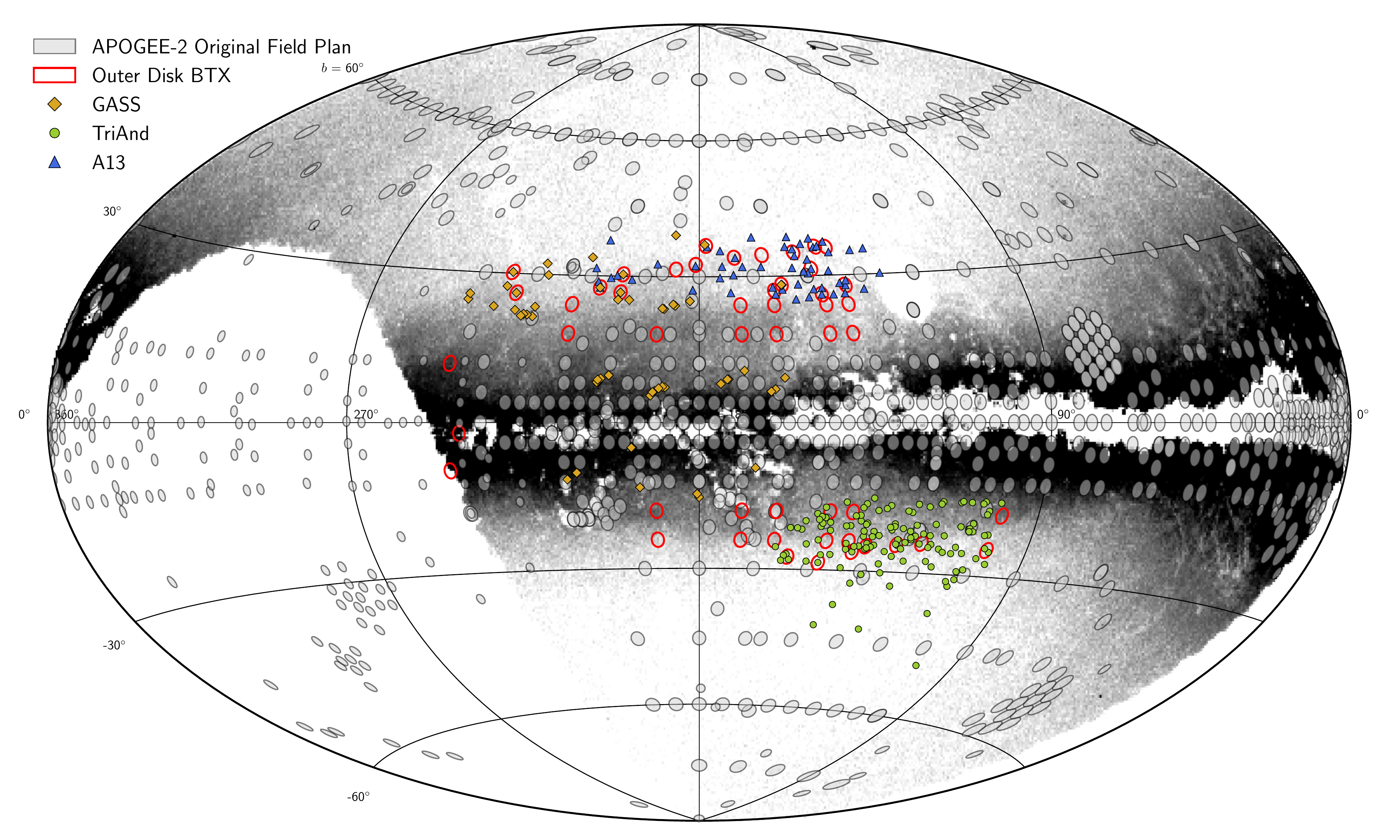}
    \caption{
    PanStarrs DR1 star-count map \citep[for stars at distances between 7.6 $-$ 11.0 kpc, adapted from][]{Slater_2014} showing also the main survey APOGEE-2 fields (grey) and the outer disk fields for the BTX (red). 
    Previously identified member stars in the outer disk substructures are indicated as well; more specifically, 
        GASS \citep{crane_2003} in yellow diamonds, 
        A13 \citep{li_2017} as blue triangles, and
        TriAnd \citep{rocha-pinto_2004,sheffield_2014} as green circles.
    }
    \label{fig:outerdisk_map}
    \end{mdframed}
\end{figure*}  %%%%%%%%%%%%%%%%%%%%%%%%%%%%%%%%%%%%%%%%%%%%%%%%%%%%%%%%%%%%%%%%%%%%%%%%%%%%%%%%%%%%%%%%%%

%%%%%%%%%%%%%%%%%%%%%%%%%%%%%%%%%%%%%%%%%%%%%%%%%%%%%%%%%%%%%%%%%%%%%%%%%%%%%%%%%%%%%%%%%%
\subsubsection{Outer Disk Program Implementation} \label{sec:outerdist_implement}
%%%%%%%%%%%%%%%%%%%%%%%%%%%%%%%%%%%%%%%%%%%%%%%%%%%%%%%%%%%%%%%%%%%%%%%%%%%%%%%%%%%%%%%%%%

The \citet{hayes_2018} study motivated a larger-scale effort to use the APOGEE-2N BTX program to trace the outer disk using a more deliberate targeting strategy; this was essentially accomplished by systematically extending the APOGEE disk field grid out to larger Galactic latitudes ($|b|$~\textgreater~20$\degs$), while also pursuing a focused targeting of previously studied stars in GASS, TriAnd, and A13 to define robustly the multi-abundance ``chemical fingerprint'' of these systems relative to the disk \citep[with targets drawn from][]{crane_2003,chou_2011,sheffield_2014,li_2017}.
Stars targeted from these prior surveys have bit 7 set in \texttt{APOGEE2\_TARGET2}.

The BTX Galactic Anticenter plan encompasses 50 fields, with 26 having $|b|$~\textless~$24\degs$ but that expand the disk grid out of the Galactic plane to probe lower latitude features like the Monoceros Ring \citep{Newberg_2002,yanny2003}, another 19 fields with $|b|$ $\gtrsim 24\degs$ that target known members of these features (described below), and two fields, 162$+$34\_btx and 186$+$31\_btx, designed to fall on regions of the sky along the AntiCenter Stream \citep[ACS;][]{grillmair_2006}.
The three remaining fields, 108$-$31\_btx, 114$-$25\_btx, and 129$-$21\_btx, were intended to target known TriAnd members but due to an error in the design of these fields, instead probe the lower latitude disk closer to the disk midplane. 

Each of these pointings is comprised of two short (3-visit) cohorts (7 \textless\ $H$ \textless\ 12.2) and 1 medium (6-visit) cohort (12.2 \textless\ $H$ \textless\ 13.3). 
Known members of anticenter structures were always given the highest priority.
Next, likely dwarf stars were identified and removed using a reduced proper motion (RPM) diagram in the NIR; we followed the the example of \citet{colliercameron_2007}, who adapted a \emph{Tycho}-based algorithm established by \citet[][]{gould_2003} to the 2MASS $J$ filter \citep{Skrutskie_06_2mass}. 
We used the best fit from \citet{colliercameron_2007} to separate dwarfs from giants as follows.
First, the RPMJ for a star is defined as follows, 
    \begin{equation} %%%%%%%%%%%%%%%%%%%%%%%%%%%%%%%%%%%%%%%%%%%%%%%%%%%%%%%%%%%%%%%%%%%%%%%%%%%%%%%%%%%%%%%%%%
        \label{eq:rpmj_def}
        RPMJ = m_{\rm J} + 5\log(\mu) \\
    \end{equation}
\noindent and the empirical division between giant- and dwarf-like RPMJ, is 
    \begin{equation}
        \label{eq:rpmj_division}
        RPMJ \ {\rm Division} = -141.25 (J-H)^{4} + 473.18 (J-H)^{3} - 583.6 (J-H)^{2} + 313.42 (J-H) - 58.
    \end{equation} %%%%%%%%%%%%%%%%%%%%%%%%%%%%%%%%%%%%%%%%%%%%%%%%%%%%%%%%%%%%%%%%%%%%%%%%%%%%%%%%%%%%%%%%%%
Thus, for source $i$, if RPMJ$_{i}$ is brighter than \autoref{eq:rpmj_division}, then the star is a RPMJ giant candidate and, conversely, if its RPMJ$_{i}$ is fainter, then the star is a RPMJ dwarf candidate; we will refer to \autoref{eq:rpmj_division} this as the ``RPMJ Division'' in the text that follows.
Any source defined as an RPMJ dwarf candidate using the URAT1 proper motions \citep{urat1} was removed from initial consideration and only the remaining, giant candidates were available for the first phase of targeting; however, the dwarf candidates would still be eligible for selection in``main red star sample.'' 
Stars targeted as RPMJ giant candidates have bit 8 set in \texttt{APOGEE2\_TARGET2}. 

For any remaining fibers, targets were selected largely following the normal red star sample color-cuts (without consideration its RPMJ; \autoref{eq:rpmj_def}). 
Fields with $|b|$\textless $24\degs$ followed a disk-like cut with \jk \textgreater\ 0.5 and a requirement that 50\% of the stars were assigned to each the short and medium samples per design.
Fields with $|b|$\textgreater $24\degs$ followed a halo-like cut with \jk \textgreater\ 0.3, but also had a 50:50  short/medium cohort requirement.
The targeting flags for these selections are as given in \autoref{tab:colorcuts}. 

%%
% We don't think that this is relevant, but in case it is asked, here is the info:
%%
%The timing of the \gaia~DR2 release was unfortunately too late to enable a modification to our targeting. At that time approximately 37 of the 52 fields had been observed, which effectively ``locked in'' a medium-cohort where the contamination from foreground dwarfs was the largest (and by proxy where the RPMJ cut less reliable due to uncertainties in the PMs). 

%%%%%%%%%%%%%%%%%%%%%%%%%%%%%%%%%%%%%%%%%%%
\subsubsection{Outer Disk Program Assessment} \label{sec:outerdist_assess}
%%%%%%%%%%%%%%%%%%%%%%%%%%%%%%%%%%%%%%%%%%%

The implementation of the foreground removal using the RPMJ division (\autoref{eq:rpmj_division}) represents a different targeting strategy than that typically employed in APOGEE \citepalias{zasowski_2013,zasowski_2017}. 
Thus, we provide evaluation of the targeting strategy in three ways:
   (i) by the fraction of spectroscopic giants and dwarfs obtained using the RPMJ strategy compared to that obtained in the Main Survey,
  (ii) by the reliability of the RPMJ giant selection using newer, and more precise, proper motions, and  
 (iii) by the fraction of targets at the distances of the overdensities of interest.
\autoref{fig:pm_comp} is used to illustrate these evaluations, with the columns in this figure representing the BTX targeting (\autoref{fig:pm_comp}a and \ref{fig:pm_comp}c) and a selection of stars from the ``main red star sample'' at a similar location in the sky, i.e., in original APOGEE-1 and -2 fields, which we refer to as the ``Main Survey'' sample below (\autoref{fig:pm_comp}b and \ref{fig:pm_comp}d). 

The BTX Sample is selected using \texttt{PROGRAMNAME} of ``odisk''.
The ``Main Survey'' sample covers the same area on the sky (see \autoref{fig:outerdisk_map}; $90\degs < \ell < 220\degs$ and $5\degs<|b|<40\degs$), but we have eliminated fields dominated by special targeting (e.g., using \texttt{PROGRAMNAME} to remove stars in the young stellar clusters, radial velocity monitoring, and contributed programs, among others) and removed the mid-plane to avoid extremely dusty sightlines ($|b| > 5^{\circ}$). 
The Main Survey and BTX Samples used to compare the Outer Disk targeting each have $\sim$18,000 stars.
Because the BTX targeting (\autoref{fig:outerdisk_map}) was designed to target high-latitude fields, the samples are not perfectly matched in terms of the underlying stellar density, but the samples are comparable enough for present purposes.

The rows in \autoref{fig:pm_comp} show the differences in the RPMJ (\autoref{eq:rpmj_def}) distribution obtained using the URAT1 proper motions adopted for the BTX design \citep{urat1} (a and b) and what would be obtained using the \revise{\gaia~eDR3 proper motions \citep{gaia_edr3}} (c and d).
The \citet{colliercameron_2007} RPMJ division between dwarf- and giant- candidates (\autoref{eq:rpmj_division}) is shown in each panel as the dashed line.
The color-coding is based on spectroscopic luminosity class using the calibrated \logg, such that dwarfs are blue, giants are red, and subgiants are green (see \autoref{ssec:datasets} for the \logg\ limits). 
Giant-type stars in the spectro-photometric distance catalog with heliocentric distances larger than 12~kpc, placing them at the distance of the TriAnd overdensity (or beyond), are shown as the larger, yellow symbols. 

 %%%%%%%%%%%%%%%%%%%%%%%%%%%%%%%%%%%%%%%%%%%%%%%%%%%%%%%%%%%%%%%%%%%%%%%%%%%%%%%%%%%%%%%%%%
\subsubsection{RPMJ Efficacy for Identifying of Giants}
\begin{figure}[h] %%%%%%%%%%%%%%%%%%%%%%%%%%%%%%%%%%%%%%%%%%%%%%%%%%%%%%%%%%%%%%%%%%%%%%%%%%%%%%%%%%%%%%%%%%
    \begin{mdframed}
    \centering
    \includegraphics[width=0.8\textwidth]{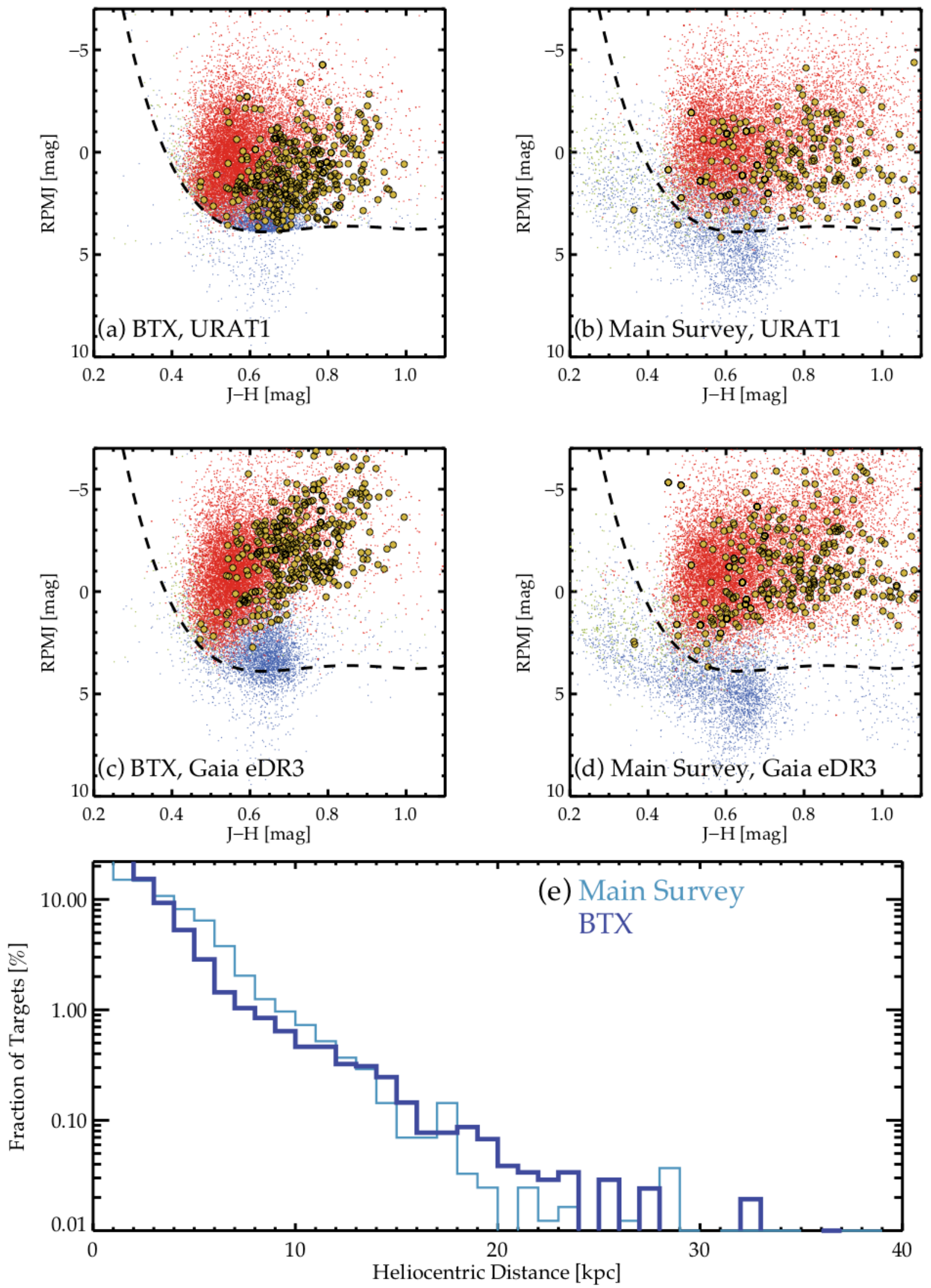}
    \caption{ 
    RPMJ selection criteria applied in the outer disk (left column) and the same visualized for the main survey sample in the same region of the sky (right column) for different proper motion catalogs, at top URAT1 and middle \revise{\gaia~eDR3 \citep{gaia_edr3}}. 
    The dashed line shows the RPMJ division given in \autoref{eq:rpmj_division}.
    The points are color-coded by spectroscopic giants (red), spectroscopic sub-giants (green), and spectroscopic dwarfs (blue) with the larger symbols indicating spectroscopic giants at heliocentric distances greater than 12 kpc. 
    The BTX and Main Survey samples both have $\sim$18,000 stars. 
    The BTX sample has systematically bluer colors than the main survey, though the main survey sample does have higher-extinction sight-lines. 
    (e) Fraction of targets at a given heliocentric distance for the main survey (thin blue) and the BTX (thick purple). 
    }
    \label{fig:pm_comp}
    \end{mdframed}
\end{figure}

 %%%%%%%%%%%%%%%%%%%%%%%%%%%%%%%%%%%%%%%%%%%%%%%%%%%%%%%%%%%%%%%%%%%%%%%%%%%%%%%%%%%%%%%%%%
\autoref{fig:pm_comp}a shows the BTX Sample with the RPMJ determined from the URAT1 proper motions; as designed, 98\% of the targets are giant-candidates using the RPMJ division (above the dashed line; \autoref{eq:rpmj_division}) and the 2\% of targets that are classified as RPMJ dwarf candidates are all from the ``main red star sample'' (using the targeting flags).
In \autoref{fig:pm_comp}b, the Main Survey sample (that only followed the ``main red star sample'' color-magnitude criteria) RPMJ-color distribution using URAT1 is shown. 
For the Main Survey sample, only \revise{81\% of the targets} would have been classified as giants and \revise{19\% being classified} as dwarfs following the RPMJ division (\autoref{eq:rpmj_division}).
Thus, \revise{19\%} of the targets in the Main Survey sample could have been excluded were the RPMJ criterion employed, and opened up fibers for more giant candidates to be observed.

In the end, the efficacy of the RPMJ criterion can be evaluated using the spectroscopic parameters. 
Of the BTX Sample, \revise{69\% of the stars} are spectroscopic giants comprise whereas 72\%  of the targets from the Main Survey sample were giants; thus, the end yield of giant stars were similar from either program.
However, it is important to note that the bulk of the BTX Sample were targeting regions of dramatically lower stellar density than the Main Survey. 
The initial targeting simulations of APOGEE-1 illustrated that the main red star sample would achieve a $\sim 50-75$\% giant fraction in traditional disk fields, but at higher latitudes such as the BTX outer disk fields, the respective giant fraction was significantly lower, around $\sim 25-50$\% \citep[see Appendix D. of][]{majewski_2017}.  
Therefore, achieving the same fraction of giants in the BTX and Main Survey samples is impressive.
Perhaps most importantly, 75\% of the spectroscopic dwarfs from the Main Survey sample would have been identified as dwarf-candidates from using the RPMJ division.
%%
% In case the RED specks that aren't RPMJ giants come up in relation to the figure:
%
%Of the targets in the BTX Sample selected following the ``main red star sample'' color-magnitude limits and not classified as RPMJ giant candidates, 22 spectroscopic giants were identified (0.1\% of all targets and 0.2\% of the spectroscopic giants); notably, these giants are not at distances of interest for this program. 
%%

  %%%%%%%%%%%%%%%%%%%%%%%%%%%%%%%%%%%%%%%%%%%%%%%%%%%%%%%%%%%%%%%%%%%%%%%%%%%%%%%%%%%%%%%%%%

%%%%%%%%%%%%%%%%%%%%%%%%%%%%%%%%%%%%%%%%%%%%%%%%%%%%%%%%%%%%%%%%%%%%%%%%%%%%%%%%%%%%%%%%%%
\subsubsection{Reliability of URAT1 RPMJ}
%%%%%%%%%%%%%%%%%%%%%%%%%%%%%%%%%%%%%%%%%%%%%%%%%%%%%%%%%%%%%%%%%%%%%%%%%%%%%%%%%%%%%%%%%%
While scientifically interesting, the BTX Outer Disk program was designed to fill an imminent need for observations in specific regions of the sky and it was one of the first programs planned in the BTX.
At that time, the URAT1 proper motions were the best available and, for the purposes of screening out nearby stars, should have been more than sufficient given our goals.
Since that time, \gaia~eDR3 proper motions were released that attain significantly higher precision \citep{gaia_edr3}.
\autoref{fig:pm_comp}c and \ref{fig:pm_comp}d investigate \revise{if the more precise, \gaia~eDR3 proper motions could have impacted our selection}.\footnote{\revise{We also performed this exercise with \gaia~DR2 \citep{gaia_dr2} and drew identical conclusions.}}

In the BTX Sample, the number of targets classified as RPMJ giant candidates changes from 98\% using URAT1 to 92\% using \gaia~eDR3 (\revise{note $\sim$1\%} of the targets did not have proper motions in \gaia~eDR3, so if we only consider the sample of stars with measured proper motions, the  number of giant candidates \revise{increases to 93\%}).
Thus, using the same RPMJ division and the more precise \gaia~DR2 proper motions would reclassify $\sim$6\% of the targets.  
However, or the BTX Sample, 99.8\% of the spectroscopic giants were RPMJ giants \revise{in \gaia~eDR3} (with a similar fraction from the Main Survey), so therefore, this $\sim$6\% reclassification is predominantly changing spectroscopic dwarfs from giant-candidates to dwarf-candidates and all coming from the ``main red star sample''. 

Because the majority of the fields in the BTX Sample still had open fibers after selecting all of the available RPMJ giant candidates and the \gaia~eDR3 proper motions tended to only reclassify ``main red star sample'' dwarfs, it is unclear that \gaia~eDR3 proper motions would have had a strong impact on the BTX targeting.
Thus, we see no significant impact to our targeting by having used the less precise URAT1 motions over those of \gaia~eDR3, largely because we use these not to select the giants, but to suppress the dwarf foreground and the dwarf foreground has sufficiently large proper motions that the precision of the underlying astrometric catalog has less of an impact.

We further note that the RPMJ division defined by \citet{colliercameron_2007} was designed to construct a ``pure'' sample of dwarf stars and our comparisons reinforce the reliability of this tool for that purpose. 
However, inspection of \autoref{fig:pm_comp}c and \ref{fig:pm_comp}d suggests that the \gaia~eDR3  proper motions are sufficiently precise that the RPMJ division could be refined.
The refinements could act to build a more complete dwarf sample by including the ``cloud'' of spectroscopic dwarfs just above the RPMJ division and, in turn, to make a more pure giant sample with their exclusion. 

 %%%%%%%%%%%%%%%%%%%%%%%%%%%%%%%%%%%%%%%%%%%%%%%%%%%%%%%%%%%%%%%%%%%%%%%%%%%%%%%%%%%%%%%%%%
\subsubsection{Using RPMJ to Access Distant Stars}
 %%%%%%%%%%%%%%%%%%%%%%%%%%%%%%%%%%%%%%%%%%%%%%%%%%%%%%%%%%%%%%%%%%%%%%%%%%%%%%%%%%%%%%%%%%
Lastly, \autoref{fig:pm_comp}e is a histogram of fraction of targets at a given heliocentric distance for the Main Survey (thin, light blue) and BTX outer disk sampling (thick, purple). 
Overall, the BTX targeting scheme has a higher fraction of stars at larger distances ($d>15$~kpc) and a similar fraction of stars from $10<d<15$~kpc. 
For stars $d<10$~kpc, the Main Survey has a larger fraction of stars, which likely is due to the BTX focusing on fields with larger $|b|$ that skim the disk distribution until larger distances rather than looking ``through'' the disk at lower $|b|$.  

In the Main Survey Sample \revise{236} stars were identified beyond 12~kpc with \revise{91\% being giant candidates} in RPMJ based on URAT1 (\revise{98\% in \gaia eDR3}).
From the BTX sample, \revise{315 stars were found beyond 12~kpc}, all of which were giant candidates in RMPJ based on URAT1.
Taken together, we see considerable benefit to using the RPMJ division to screen out nearby dwarf stars and permit allocation of fibers to stars more likely to meet the science goals of the program.

%%%%%%%%%%%%%%%%%%%%%%%%%%%%%%%%%%%%%%%%%%%%%%%%%%%%%%%%%%%%%%%%%%%%%%%%%%%%%%%%%%%%%%%%%%
\subsection{Calibrations Clusters for Main Sequence Stars} \label{sec:mainsequence}
%%%%%%%%%%%%%%%%%%%%%%%%%%%%%%%%%%%%%%%%%%%%%%%%%%%%%%%%%%%%%%%%%%%%%%%%%%%%%%%%%%%%%%%%%%

M~dwarf stars are among the most numerous in the Milky Way and among the best observational targets to reveal the physics of planet formation. 
While not explicitly targeted in APOGEE-1, M dwarfs are the dominant contaminant population in efforts to target the main red star sample, and the avoidance of such stars was a major component of the initial APOGEE targeting strategies for a large fraction of the survey \citepalias{zasowski_2013,zasowski_2017}. 
Despite such efforts to avoid M dwarfs, \citet{Birky_2020} identified over 5,000 M dwarfs in the DR14 sample \citep{dr14,holtzman_2018}. 

Early work in APOGEE-1 \citep{deshpande_2013} aimed at testing the ability of APOGEE spectra to characterize M~dwarfs found that additional special techniques were required. 
Recently, however, \citet{souto_2017} was able to use APOGEE spectra to extract detailed chemical abundances that paved the way for adaptations in the ASPCAP methodologies that address the atmospheres of cool dwarfs. 
As a result, the DR16 ASPCAP pipeline included a multi-year effort for extensive expansion of the input linelist to improve the underlying spectral synthesis to expand the range of \teff\ with reliable ASPCAP results (these efforts are described in \citep{Jonsson_2020} and \inprep{V.~Smith et al (in prep.)}). 
At the same time, the work of \citet{Souto_2020} and \citet{Birky_2020} establishes key calibrator datasets for late type dwarfs, although the number of appropriate calibration stars is still small in number compared to the target sample for which they are needed.
Indeed, it is expected that, by the end of survey, a sample of 10's of thousands M~dwarfs will exist in the APOGEE dataset.

Anticipating these gains in APOGEE's ability to utilize APOGEE spectra for such stars, a number of programs within APOGEE-2 have operated to target cool stars; these efforts can be found in Section 4.10 of \citetalias{zasowski_2017} as well as \autoref{anc:mdwarfs_koi} and \autoref{anc:mdwarfk2} here. 
Many of these programs piggyback on detailed characterization from the literature and, in particular, the on-going field star targeting in \kepler\ and \ktwo\ motivated to characterize planet-hosting stars.
However, at the time that the BTX was planned we had no significant observations of main sequence stars in star clusters, which are ideal calibrators, because they have a known metallicity and age (which can be measured from other, better understood cluster members).  Therefore, this lack of late-type cluster calibrator stars limited our ability to fine tune the ASPCAP software to work on late-type dwarf field stars. 

Designing a program that will provide calibrations for main sequence stars requires an understanding of how the ASPCAP pipeline calibrates its results. 
The calibration strategy is multi-pronged: 
    (i) use of \textgreater\ 10 member stars in well understood stellar clusters that provide a sense of the internal scatter for stars with similar patterns  \citep[D14 and prior; see discussion in][]{holtzman_2015,holtzman_2018}, 
    (ii) comparison of ``duplicate'' observations of the same star that are processed independently through the pipeline \citep[DR16 and after;][]{Poovelil_2020,Jonsson_2020},
    (iii) and an absolute correction to the chemical trends in the solar neighborhood \citep{Jonsson_2020}. 
While the latter two aspects of calibration use natural occurrences in the APOGEE targeting, the acquisition of cluster member-stars, in particular member stars on the main sequence, requires a coordinated and large-scale effort. 

\begin{figure*} %%%%%%%%%%%%%%%%%%%%%%%%%%%%%%%%%%%%%%%%%%%%%%%%%%%%%%%%%%%%%%%%%%%%%%%%%%%%%%%%%%%%%%
    \begin{mdframed}
    \centering
    \includegraphics[width=1.0\textwidth]{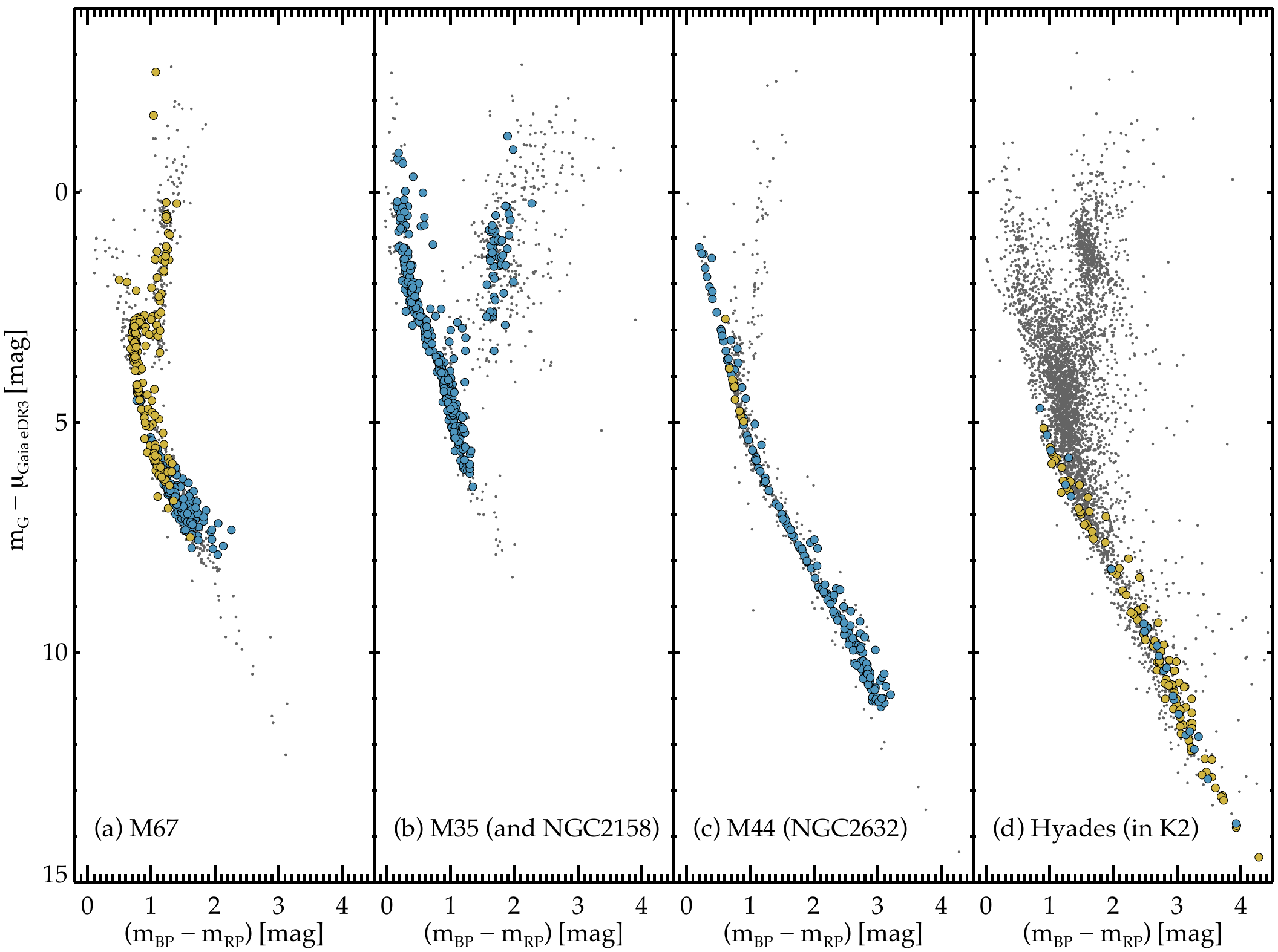}
    \caption{Color-absolute magnitude diagrams in the \gaia~Photometric Systems for the four key clusters \revise{designed} to calibrate ASPCAP results for main sequence. 
    The term $\mu_{\rm \gaia~DR2}$ is the distance modulus computed through inversion of the \revise{\gaia~eDR3} parallax.
    The clusters are: 
        (a) M\,67, 
        (b) M\,35, which also contains the red giant branch of NGC\,2158 in the background, 
        (c) M\,44 (also known as NGC\,2632), and 
        (d) the Hyades (targeted within the \ktwo\ program for C4 and C13). 
    Each panel reflects the APOGEE dataset in the vicinity of these clusters (through APOGEE-1, the original APOGEE-2 program, and the BTX), 
    with all targets in the fields
    shown as small grey points. 
    Special targets are shown as larger, colored symbols with confirmed cluster members (\texttt{APOGEE2\_TARGET2} bit 10) in yellow and candidate cluster members (\texttt{APOGEE2\_TARGET1} bit 9) in blue. 
    Combining these clusters, it is possible to compare pipeline results against external data for these clusters to understand better the ASPCAP pipeline systematics.}
    \label{fig:m_dwarf}
    \end{mdframed}
\end{figure*}

An APOGEE-2N program to target cluster mains sequence stars is necessarily limited to only the nearest clusters, which, unfortunately, will impart an unavoidable age- and metallicity-bias to the calibration sample.
The following clusters were targeted in the BTX specifically to reach main sequence stars: 
  M\,44 (Praesepe, Beehive, NGC\,2632), 
  Ruprecht\,147\footnote{Although, unfortunately, due to COVID-19 closures, no observations were obtained of this field.}, 
  the Hyades (which falls in two  \ktwo\ campaigns), 
  M\,67, 
  and M\,35 (NGC\,2158 is in the background of this cluster and was co-targeted); 
  their key information is summarized in \autoref{tab:m_dwarf}. 
We waited for the release of \gaia~DR2 to implement this program so that its precise proper-motions could help isolate the stellar sequences at these faint magnitudes.
Color-absolute magnitude diagrams in the \gaia\ filter system are given in \autoref{fig:m_dwarf} and the $M_{\rm G}$ targeting limit is given in \autoref{tab:m_dwarf} to provide a sense the mass limits for the APOGEE-2N main sequence calibration program.

The targeting in these cluster fields is generally performed following the prescriptions given in \citet[][DR13]{frinchaboy_2013}, \citet[][DR14]{donor_2018}, and \citet[][DR16]{donor_2020}; more specifically, setting known members from the literature to the highest priority (with stars having high quality spectroscopic information marked by \texttt{APOGEE2\_TARGET2} bit 2, and stars with any spectroscopic information marked by \texttt{APOGEE2\_TARGET2} bit 10), and then using a suite of auxiliary data to select candidates (\texttt{APOGEE2\_TARGET1} bit 9).
Unused fibers are then back-filled following the criteria for the main survey red star sample, which may also yield serendipitous members.
For the BTX targeting, the highest priority was given to the lowest mass stars, with other higher-mass cluster members being used to boost our sample of members across \teff-\logg\ space. 
We aim to reach $S/N=$ 100 per pixel for as many stars as possible, even though for M\,67 this would require 36 visits.  
Unfortunately, because of the observing demands by competing fields at similar right ascensions, many of the clusters were incompletely observed, some particularly impacted by the spring 2020 COVID19 closure at APO.
In the case of M\,67, to recover this particularly vital cluster (which contains stars spanning both the red giant branch as well as an extensive part of the main sequence down to the M dwarfs), we redesigned a new  36-visit plate to be observed by APOGEE-2S, for which operations for SDSS-IV have been extended into early 2021.
As discussed in \santanatext, the differing plate scale and radius for APOGEE-2S imposed some changes to the final targets in this field, but it otherwise followed the same target selection process as described here.

In the end, while our attempts to improve our calibration of dwarf stars through use of observations of cluster main sequences may not be fully realized due to unanticipated circumstances, we have at least established a firm start along this path that can be completed in SDSS-V and beyond.

  %%%%%%%%%%%%%%%%%%%%%%%%%%%%%%%%%%%%%%%%%%%%%%%%%%%%%%%%%%%%%%%%%%%%%%%%%%%%%%%%%%%%%%

%%%%%%%%%%%%%%%%%%%%%%%%%%%%%%%%%%%%%%%%%%
%%%%%%%%%%%%%%%%%%%%%%%%%%%%%%%%%%%%%%%%%%%%%%%%%%%%%%%%%%%%%%%%%%%%%%%%%%%%%%%%%%%%%%%
\begin{table*}[h]
    \begin{mdframed}
    \centering
    \caption{Calibration Clusters for Main Sequence Stars}
    \label{tab:m_dwarf}
    \begin{tabular}{l cccc l}
        \hline \hline
        Cluster Name &  Age$^{a}$  & Distance$^{a}$ & Metallicity & M$_{G}$ Limit  & APOGEE-2N BTX \\
                     & [Gyr]       & [kpc]          &  [dex]      & [mag]          &  Field Names\\
        \hline \hline
        M\,44$^{b}$                          & 0.68 & 0.183 &  0.14 & 11.5 & NGC2632\_btx \\
        M\,67                                & 4.27 & 0.889 &  0.01 &  9.8 & M67\_btx     \\
        Hyades                               & 0.79 & 0.047 &  0.15$^{c}$  & 14.5 & K2\_C4\_$lll$$\pm$$bb$\_btx or \\ 
                                             &      &       &              &      & K2\_C13\_$lll$$\pm$$bb$\_btx \\
        Ruprecht 147                         & 3.02 & 0.323 &  0.12 &  ... & Rup\_147\_btx \\
        M\,35 (NGC\,2168)                    & 0.15 & 0.906 & -0.12 &  6.3 & M35N2158\_btx \\
        \hline \hline
        \multicolumn{6}{l}{(a) Ages and Distances are taken from \citet{cantat_gaudin_2020}.} \\ 
        \multicolumn{6}{l}{(b) Alternate names include: Beehive, Praesepe, and NGC\,2632. } \\
        \multicolumn{6}{l}{(c) From \citet{Dutra-Ferreira_2016}. }
    \end{tabular}
    \end{mdframed}
\end{table*}
%%%%%%%%%%%%%%%%%%%%%%%%%%%%%%%%%%%%%%%%%%%%%%%%%%%%%%%%%%%%%%%%%%%%%%%%%%%%%%%%%%%%%%%
%%%%%%%%%%%%%%%%%%%%%%%%%%%%%%%%%%%%%%%%%%

%%%%%%%%%%%%%%%%%%%%%%%%%%%%%%%%%%%%%%%%%%%%%%%%%%%%%%%%%%%%%%%%%%%%%%%%%%%%%%%%%%%%%%%%%%
%%%%%%%%%%%%%%%%%%%%%%%%%%%%%%%%%%%%%%%%%%%%%%%%%%%%%%%%%%%%%%%%%%%%%%%%%%%%%%%%%%%%%%%%%%
\section{Expansion of Existing Programs} \label{sec:expansion}
%%%%%%%%%%%%%%%%%%%%%%%%%%%%%%%%%%%%%%%%%%%%%%%%%%%%%%%%%%%%%%%%%%%%%%%%%%%%%%%%%%%%%%%%%%
%%%%%%%%%%%%%%%%%%%%%%%%%%%%%%%%%%%%%%%%%%%%%%%%%%%%%%%%%%%%%%%%%%%%%%%%%%%%%%%%%%%%%%%%%%

This section discusses programs that were previously described in \citetalias{zasowski_2017} but that were expanded upon in the BTX.
\revise{These program expansions largely applied to ``Goal Programs'' that were outside of the ``main red star sample'' that formed the core of the targeting strategy.
The following programs or objectives were expanded: 
 (1) Open Clusters (\autoref{sec:clusters}), 
 (2) APOGEE-K2 Survey (\autoref{sec:k2}), 
 (3) \kepler\ Objects of Interest (\autoref{sec:koi}),
 (4) Eclipsing Binaries (\autoref{sec:eb}),
 (5) Young Stellar Clusters (\autoref{sec:youngclusters}), 
 (6) SubStellar Companions (\autoref{sec:substellar}), 
 (7) Dwarf Spheroidal Galaxies (\autoref{sec:dsph}), and
 (8) cross-survey calibration (\autoref{sec:cross}). 
}
\revise{For each program the scientific motivation is summarized; this is similar to that presented in \citetalias{zasowski_2017} but updated with more recent results, publications from the program, or discussion of other modifications to its strategic objectives. 
The degree of augmentation varies from the addition of only a few additional objects to complete a key sample (e.g., Ecclipsing Binaries in \autoref{sec:eb}) to the inclusion of thousand of additional stars (e.g., \ktwo\ in \autoref{sec:k2}).}

%%%%%%%%%%%%%%%%%%%%%%%%%%%%%%%%%%%%%%%%%%%%%
\subsection{Open Clusters} \label{sec:clusters}
%%%%%%%%%%%%%%%%%%%%%%%%%%%%%%%%%%%%%%%%%%%%%

APOGEE has long placed special attention on the open cluster population in the Milky Way. 
Each open cluster is a set of stars that formed from the same star formation event and have evolved together over time.
As a result, star clusters represent sources for which both ages and chemistry may be determined accurately, and the large scale study of open clusters with homogeneously derived spectroscopic quantities presents significant scientific opportunities to trace chemical abundance and age gradients in the Galactic disk. 
\citet{frinchaboy_2013,Cunha2016,donor_2018,donor_2020} have each presented the status of the Open Cluster Chemical Abundance and Mapping (OCCAM) Survey (within APOGEE using successive data releases DR10, DR12, DR14, and DR16, respectively). 
As of DR16, 128 open clusters have had some APOGEE data collected in their spatial footprints, with 71 clusters having sufficient member stars sampled to be used as reliable data points for measuring gradients in Galactic properties \citep{donor_2020}.

Open cluster observations for the BTX were designed around the study of \citet{donor_2018} using the DR14 census of open clusters. 
Using their work, specific aspects of the sample, such as Galactocentric distance, age, and metallicity, were evaluated and clusters were selected to better span the full range physical parameters.  
The BTX open cluster targeting followed prior targeting procedures \citep{frinchaboy_2013,zasowski_2013,zasowski_2017}, while also incorporating the newly available proper motions from \gaia~data releases, which provided a heightened ability to select high-likelihood cluster members. 

With the goal of increasing APOGEE's Galactic radius coverage, newly targeted distant outer disk clusters include Berkeley 2, Berkeley 18, Berkeley 20, Berkeley 21, and Berkeley 22, while Berkeley 81 was targeted toward the inner Galaxy.
Additional visits were obtained for N6819 as part of the SubStellar Companions program (\autoref{sec:substellar}), with NGC\,188, NGC\,2158, NGC\,2420, NGC\,6791, NGC\,7789, and M\,71 obtaining additional visits for both additional radial velocity monitoring and to increase sampling of both the red giant branch and the upper main sequence calibrators. 
The cluster NGC\,752 was targeted with the particular aim of exploring the effects of atomic diffusion on surface chemistry for stars along the red giant branch, at the main sequence turn-off and down the main sequence. 
Aligned with the goals discussed in \autoref{sec:mainsequence}, M\,35, NGC\,2632, Ruprecht\,147, and M\,67 were targeted with attention to stars as far down the mass function as possible as constrained by the observing time available, while the Hyades were targeted within the context of the \ktwo\ Targeting for Campaigns 4 and 13. 
Stars selected as potential open cluster members have the targeting bit 9 in \texttt{APOGEE2\_TARGET1} set.

\begin{figure}[h] %%%%%%%%%%%%%%%%%%%%%%%%%%%%%%%%%%%%%%%%%%%%%%%%%%%%%%%%%%%%%%%%%%%%%%%%%%%%%%%%%%%%%%
    \begin{mdframed}
    \centering
    \includegraphics[width=\textwidth]{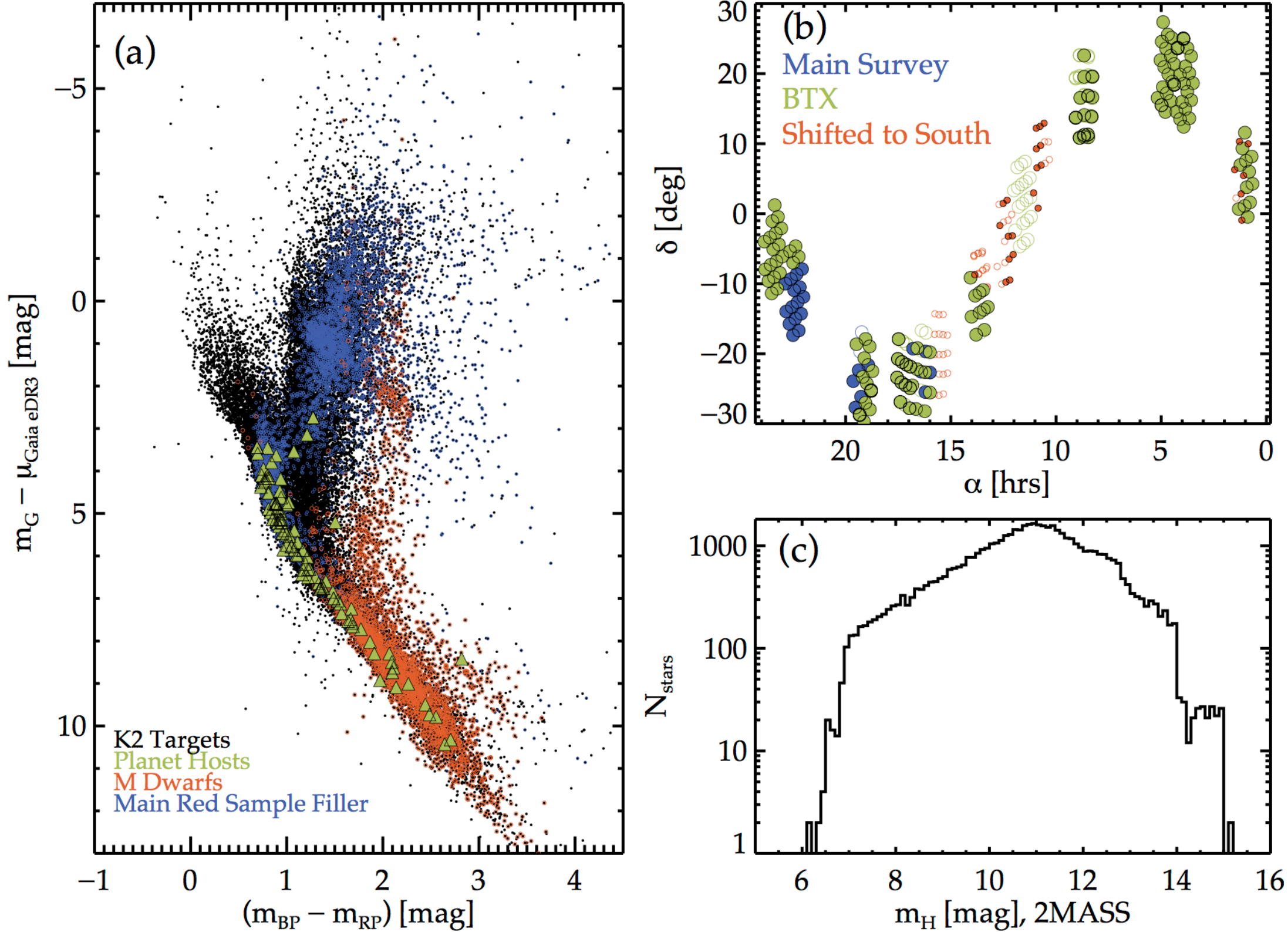}
    \caption{Overview of the \ktwo\ program. 
    (a) \revise{\gaia~eDR3} color-absolute magnitude diagram for all observed \ktwo\ targets. 
    All GAP targets are in black, with main red star sample giants in blue, M~dwarfs in orange, and planet hosts in green. 
    (b) Sky distribution of \ktwo\ fields that were designed for the main survey (blue), the BTX (green), and \ktwo\ fields shifted to the Southern survey due to limited observing time available in the North. 
    Filled symbols indicate fields with observations and open circles have no observations to date. 
    The symbol size is not to scale, but the relative sizes of the symbols are. 
    (c) Distributions of apparent $H$ magnitude for the targets. in each of the main targeting classes. 
    }
    \label{fig:k2}
    \end{mdframed}
\end{figure} %%%%%%%%%%%%%%%%%%%%%%%%%%%%%%%%%%%%%%%%%%%%%%%%%%%%%%%%%%%%%%%%%%%%%%%%%%%%%%%%%%%%%%
\begin{figure}[h] %%%%%%%%%%%%%%%%%%%%%%%%%%%%%%%%%%%%%%%%%%%%%%%%%%%%%%%%%%%%%%%%%%%%%%%%%%%%%%%%%%%%%%
`\begin{mdframed}
    \centering
        \includegraphics[width=\textwidth]{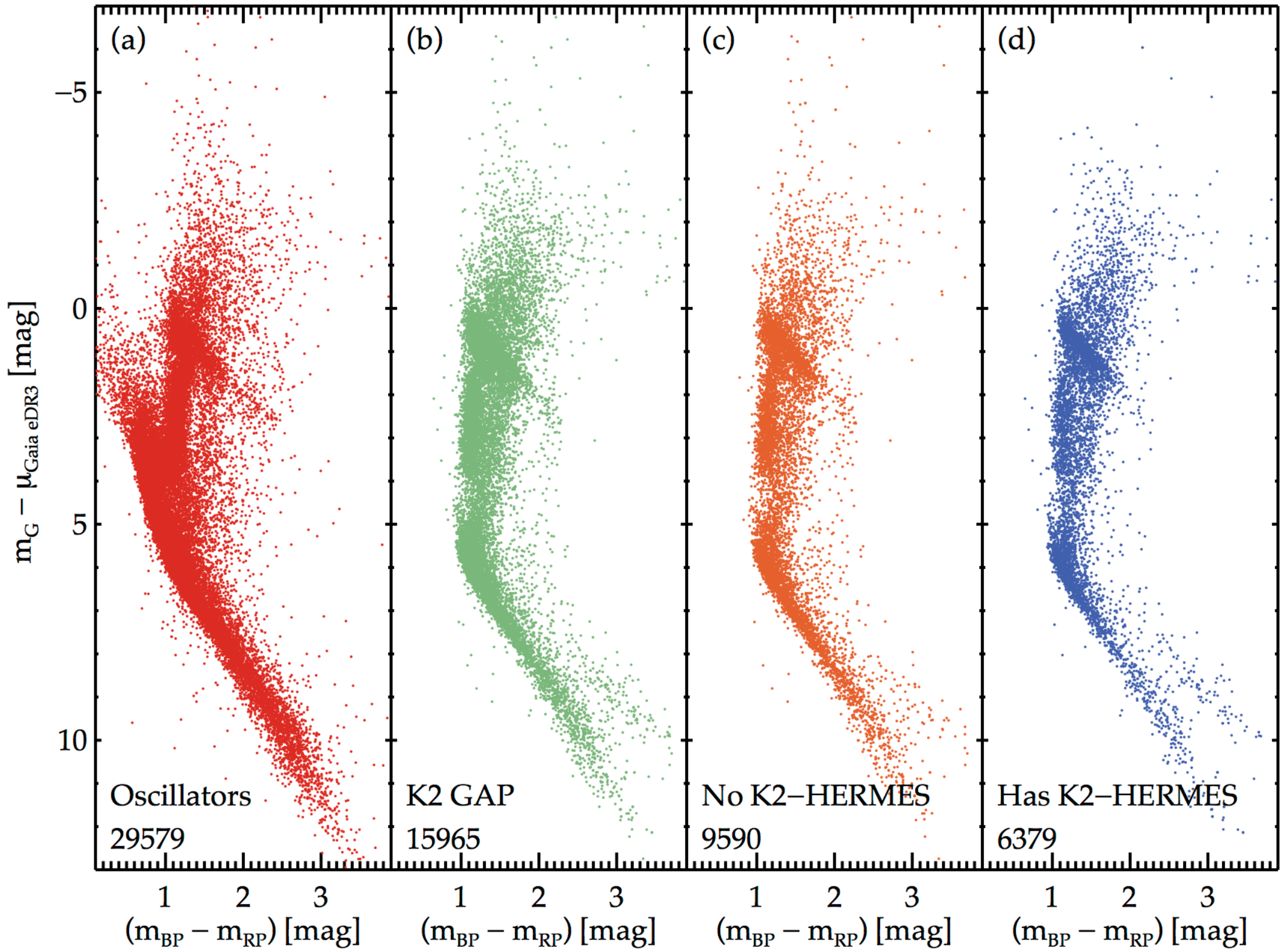}
        \caption{Overview of stars in the APOGEE-2 \ktwo\ program that were targeted to coordinate with observations in other surveys. 
        Note that a given star may be displayed in multiple panels based on its multiple classifications. 
        (a) Stars determined to be oscillators in \ktwo\ light curves \citep[following][]{Hon_2019} (\texttt{APOGEE2\_TARGET3} bit 6 and \texttt{APOGEE2\_TARGET1} bit 30).
        (b) Stars targeted as part of the \ktwo~GAP program \citep{Stello_2017} (\texttt{APOGEE2\_TARGET3} bit 6 and \texttt{APOGEE2\_TARGET2} bit 0). 
        (c) Stars without prior \ktwo-HERMES observations (\texttt{APOGEE2\_TARGET3} bit 6 and \texttt{APOGEE2\_TARGET2} bit 0 and bit 17).
        (d) Stars with prior \ktwo-HERMES observations \citep[see][]{Wittenmyer_2018,Sharma_2019} (\texttt{APOGEE2\_TARGET3} bit 6 and \texttt{APOGEE2\_TARGET2} bit 0). 
    }
    \label{fig:k2_2}
    \end{mdframed}
\end{figure} %%%%%%%%%%%%%%%%%%%%%%%%%%%%%%%%%%%%%%%%%%%%%%%%%%%%%%%%%%%%%%%%%%%%%%%%%%%%%%%%%%%%%%

%%%%%%%%%%%%%%%%%%%%%%%%%%%%%%%%%%%%%%%%%%%%%%%%%%%%%%%%%%%%%%%%%%%%%%%%%%%%%%%%%%%%%%%%%%
\subsection{APOGEE-\ktwo\ Survey} \label{sec:k2}
%%%%%%%%%%%%%%%%%%%%%%%%%%%%%%%%%%%%%%%%%%%%%%%%%%%%%%%%%%%%%%%%%%%%%%%%%%%%%%%%%%%%%%%%%%

% \textcolor{red}{NEED TO ADD REFERENCE TO \citet{Zinn_2020}}

\kepler's primary mission, the continuous multi-year monitoring of a single field of stars, concluded due to a spacecraft hardware failure; the \ktwo\ project that followed using the two reaction wheel spacecraft \citep[][]{k2_overview} observed a set of fields along the ecliptic plane for a shorter period of time than \kepler's primary field ($\sim$80~days each instead of $\sim$4~years for the original \kepler\ field). 
The stars observed by \ktwo\ add considerably to the programs in the \kepler\ field, in particular by spanning a large range of Galactic sub-components and their underlying stellar populations, as well as several notable star clusters. 

There are two primary differences between the \ktwo\ targeting from \citetalias{zasowski_2017} and the BTX. 
First, we know which stars were targeted with the optical HERMES spectrograph \citep{Sheinis_2015,Sheinis_2016} using the GALAH survey setup \citep{Wittenmyer_2018,Sharma_2019}.
Second, we were able to identify \revise{stars exhibiting oscillation signals} {\it before} targeting in APOGEE-2N using the \revise{methods described in} \citet{Hon_2019}. 
\citeauthor{Hon_2019} searched the $\sim$197,000 targets from the \ktwo\ mission using custom apertures to detect solar-like oscillations for 21,914 stars, with another 600 serendipitous oscillating giants that contaminated the postage stamp image targeting a different \ktwo\ target.
These results enable a holistic and quantifiable target selection employed across APOGEE-2's \ktwo\ program. 

The following prioritization scheme is used for the targeting in each field of the \ktwo\ program. 
First, we eliminate all stars that were already observed by APOGEE-1 (serendipitously) or APOGEE-2N (intentionally). 
The primary target categories were prioritized in the following order: 
\begin{enumerate} \itemsep -2pt
    \item known planet hosts (\texttt{APOGEE2\_TARGET2} bit 11),
    \item stars with a confirmed oscillation or granulation signal (\texttt{APOGEE2\_TARGET1} bit 30),
    \item red giants that were targeted by the \ktwo\ Galactic Archaeology Program \citep[GAP;][]{Stello_2017} and \textit{not} observed with HERMES, 
    \item GAP targets observed by HERMES \citep[see][]{Wittenmyer_2018,Sharma_2019,Zinn_2020}, 
    \item M-dwarfs in the unbiased sample from the Ancillary Science Program described in \autoref{anc:mdwarfk2} (\texttt{APOGEE2\_TARGET3} bit 28).
    \item stars meeting the criteria for the ``main red star sample'' (\texttt{APOGEE2\_TARGET1} bit 14). 
\end{enumerate}
Though targets were prioritised in this order, individual targets could exist in multiple target categories.
Given that the \kepler\ spacecraft has no full frame images (FFIs), all targets were proposed for and are available via the \ktwo\ Guest Observer programs for each campaign.\footnote{Available:\url{https://keplerscience.arc.nasa.gov/k2-approved-programs.html}} 

Field centers were chosen to maximize the number of ``primary'' targets, which tended to drive the field centers to the center of the {\it Kepler}-modules, but this was not always the case (we note that \revise{the APOGEE-N spectrograph} has a very similar FOV to a Kepler module).
The \ktwo\ program was allocated 18 1-visit fields per \ktwo\ campaign or roughly one APOGEE-N pointing per \kepler\ module, but due to dysfunctional modules and local target density, all 18 fields were not always required to attain the scientific goals.

All fields in the \ktwo\ program have the \texttt{PROGRAMNAME} `k2\_btx.'
Individual fields have \texttt{FIELD} names of the form `K2\_C\#\_$lll$$\pm$$bb$\_btx', where C\# indicates the \ktwo\ campaign and $lll$$\pm$$bb$ indicate the Galactic coordinates for the field center. 
The following campaigns were targeted as part of the BTX: C1, C2, C3, C4, C5, C6, C7, C8, C11, C12, C13, C16, C18.\footnote{For additional information on these campaigns see: \url{https://keplerscience.arc.nasa.gov/k2-fields.html}} 
We note that there were also main survey (e.g., non BTX) observations that were taken and these are described in \citetalias{zasowski_2017}.  
Due to over-subscription of APOGEE-2N at certain LSTs where \ktwo\ campaigns were located and a corresponding under-subscription at the same LSTs in APOGEE-2S, some \ktwo\ fields were ``moved'' and designed to be observed by APOGEE-2S instead as is discussed in \santanatext; this resulted in a shift of partial or whole Campaigns for the following:  C6, C8, C10, C14, C15, and C17.

Each of the \ktwo\ targeting classes can be identified with targeting bits.\footnote{For DR16, those stars that fall in multiple targeting categories in the \ktwo\ program did not have all of their applicable targeting bits set. This has been corrected for DR17.}
All stars targeted in the \ktwo\ program have \texttt{APOGEE2\_TARGET3} bit 6 set.
Known planet hosts have \texttt{APOGEE2\_TARGET2} bit 11 set.
Any star identified as an oscillator has \texttt{APOGEE2\_TARGET1} bit 30 set. 
Stars in the GAP program have \texttt{APOGEE2\_TARGET2} bit 0 set. 
Stars with K2-HERMES observations have \texttt{APOGEE2\_TARGET2} bit 17 set.
M~dwarfs in the \ktwo\ fields targeted as a part of the Ancillary Science Program have \texttt{APOGEE2\_TARGET3} bit 28 set (\autoref{anc:mdwarfk2}). 

%%%%%%%%%%%%%%%%%%%%%%%%%%%%%%%%%%%%%%%%%%%%%%%%%%%%%%%%%%%%%%%%%%%%%%%%%%%%%%%%%%%%%%%%%%
\subsection{\kepler\ Objects of Interest} \label{sec:koi}
%%%%%%%%%%%%%%%%%%%%%%%%%%%%%%%%%%%%%%%%%%%%%%%%%%%%%%%%%%%%%%%%%%%%%%%%%%%%%%%%%%%%%%%%%%

Over its lifetime, the \kepler\ mission produced a rich sample of confirmed extra-solar planets from a wealth of transit signals detected from its high-precision light curves.
This latter category, broadly known as \kepler\ Objects of Interest or KOIs, contained signals not just from planets, but also eclipsing binaries, strongly spotted stars, low-mass stellar companions, and other classes of objects that have transit-like signals in such light curves. 
The process to confirm a planet typically involves complementary ground-based observations to verify the nature of the signal; multi-epoch radial velocity measurements are key to transit-signal verification. 
While the radial velocity precision of APOGEE is not sufficient to detect small planets on its own, multi-epoch APOGEE spectra can identify the bulk of non-planet signal types whose radial velocity variations range from $\sim$100's of m s$^{-1}$ to 10's of \kms \citep[as demonstrated, e.g.,  in][]{Fleming_2015,canas_2018}. 
In addition to radial velocities, the APOGEE spectra also provide key data that characterize the host star and, thereby, the properties of the planets \citep[e.g.,][]{wilson_2018,canas_2019a,canas_2019}. 

The APOGEE-2N KOI program was designed to use multi-epoch RVs on a sample of confirmed or candidate planet hosts in the \kepler\ field that is matched to a control sample of confirmed non-hosts with the same underlying \teff-\logg\ distribution.
Together this forms a statistical sample through which true \emph{false-positive} rates can be estimated for stellar types as a tool to better frame \emph{occurrence} rates.
The program was designed to sample each system with 18 APOGEE-2N radial velocity epochs and was limited to five \kepler\ modules, but was designed to obtain a sample of $\sim$1000 KOIs and $\sim$200 control objects (with another $\sim$200 KOIs having been observed from APOGEE-1). 

For the BTX, the KOI program was expanded by $\sim$40\% via the inclusion of multi-epoch observations for two additional \kepler\ modules (\texttt{FIELD} of K18\_070+14\_btx and K19\_076+07\_btx).
As with the base program, the NExSCI archive\footnote{http://exoplanetarchive.ipac.caltech.edu} was queried for the current state of KOIs (CONFIRMED versus CANDIDATE) during targeting (approximately May 2018). 
These fields were drilled and designed as early as possible in the BTX implementation to ensure that they obtain the maximal possible APOGEE-2N time baseline ($\sim$2 years).

The BTX fields have \texttt{PROGRAMNAME} `koi\_btx' and the \texttt{FIELD} names have `\_btx' appended.
KOI objects are \texttt{APOGEE2\_TARGET3} bit 0 set and control targets have \texttt{APOGEE2\_TARGET3} bit 2 set. 
An Ancillary Science Program for tidally synchronized binaries have \texttt{APOGEE2\_TARGET2} bit 12 (\autoref{anc:tlockbinary}). 

%%%%%%%%%%%%%%%%%%%%%%%%%%%%%%%%%%%%%%%%%%%%%%%%%%%%%%%%%%%%%%%%%%%%%%%%%%%%%%%%%%%%%%%%%%
\subsection{Eclipsing Binaries} \label{sec:eb}
%%%%%%%%%%%%%%%%%%%%%%%%%%%%%%%%%%%%%%%%%%%%%%%%%%%%%%%%%%%%%%%%%%%%%%%%%%%%%%%%%%%%%%%%%%

Eclipsing binaries (EBs) are valuable systems for the determination of fundamental stellar parameters. 
EBs represent a sub-set of stellar binaries systems in which the orbital plane is nearly parallel to the line-of-sight. 
Thus, in addition to the powerful dynamical constraints obtained from binary systems \citep[e.g., relative masses;][]{troup_2016,price-whelan_2018}, the eclipses render the opportunity to break the degeneracy from dynamical measurements by better understanding the properties of the two stars. 
APOGEE-quality radial velocities paired with high precision light curves permit detailed characterization of systems \citep[e.g.,][]{cunningham_2019}.
Operating in the NIR, APOGEE spectroscopy is able to probe systems that are difficult in the optical due to favorable flux-contrast ratios at infrared wavelengths, such as stars with cool, low-mass, M-dwarf secondaries \citep[][\inprep{K.~Hambleton \& A.~Prsa in prep.}]{Mahadevan_2019}.
Thus, the APOGEE-2 program to characterize EBs opens an interesting window to understand better the low-mass stellar types that are common in planet searches \citep[for occurrence rates see][]{Dressing2015}.

For the BTX, a set of EBs in \ktwo\ Campaign 6 were monitored using five unique fields with the goal to obtain a minimum of nine radial velocity epochs. 
The EBs were selected based on the presence of asteroseismic oscillations in one or both components in the \ktwo\ light curves. 
To test asteroseismic scaling relations, radial velocities can be used as an independent measure of the component masses; moreover, in some cases, the spectra can serve a similar role for the effective temperatures. 
These measurements are vital comparisons, but only $\sim$13 systems have been measured in this way at high precision, and in many cases multiple teams analyzing the same systems have derived results that differ at the 2-$\sigma$ level \citep{Gaulme_2016,Brogaard_2018,themessl_2018}. 
The targets proposed by this program will significantly increase the total sample of EBs with asteroseismic components and precise RVs.

The BTX fields have \texttt{PROGRAMNAME} of `eb\_btx' and the \texttt{FIELD} names have `\_btx' appended.
EB objects are \texttt{APOGEE2\_TARGET3} bit 1 set. 
Because the multi-epoch measurements can also be used to build up $S/N$ on faint stars, unused fibers in these fields were filled with stars from the ``main red star sample'' following the modified halo targeting strategy (\autoref{sec:halo}). 
Other new EB systems will naturally fall into the extension of the KOI sample (\autoref{sec:koi}), but were not intentionally targeted by the EB program. 

%% Actual Field Names -- should I specify?
% K2EB_C06_302+52_btx	
% K2EB_C06_314+48_btx	
% K2EB_C06_320+48_btx	
% K2EB_C06_326+52_btx
% K2EB_C06_327+54_btx
%% 

%%%%%%%%%%%%%%%%%%%%%%%%%%%%%%%%%%%%%%%%%%%%%
\subsection{Young Star Clusters} \label{sec:youngclusters}
%%%%%%%%%%%%%%%%%%%%%%%%%%%%%%%%%%%%%%%%%%%%%

\begin{figure}
    \begin{mdframed}
    \centering
    \includegraphics[width=0.95\textwidth]{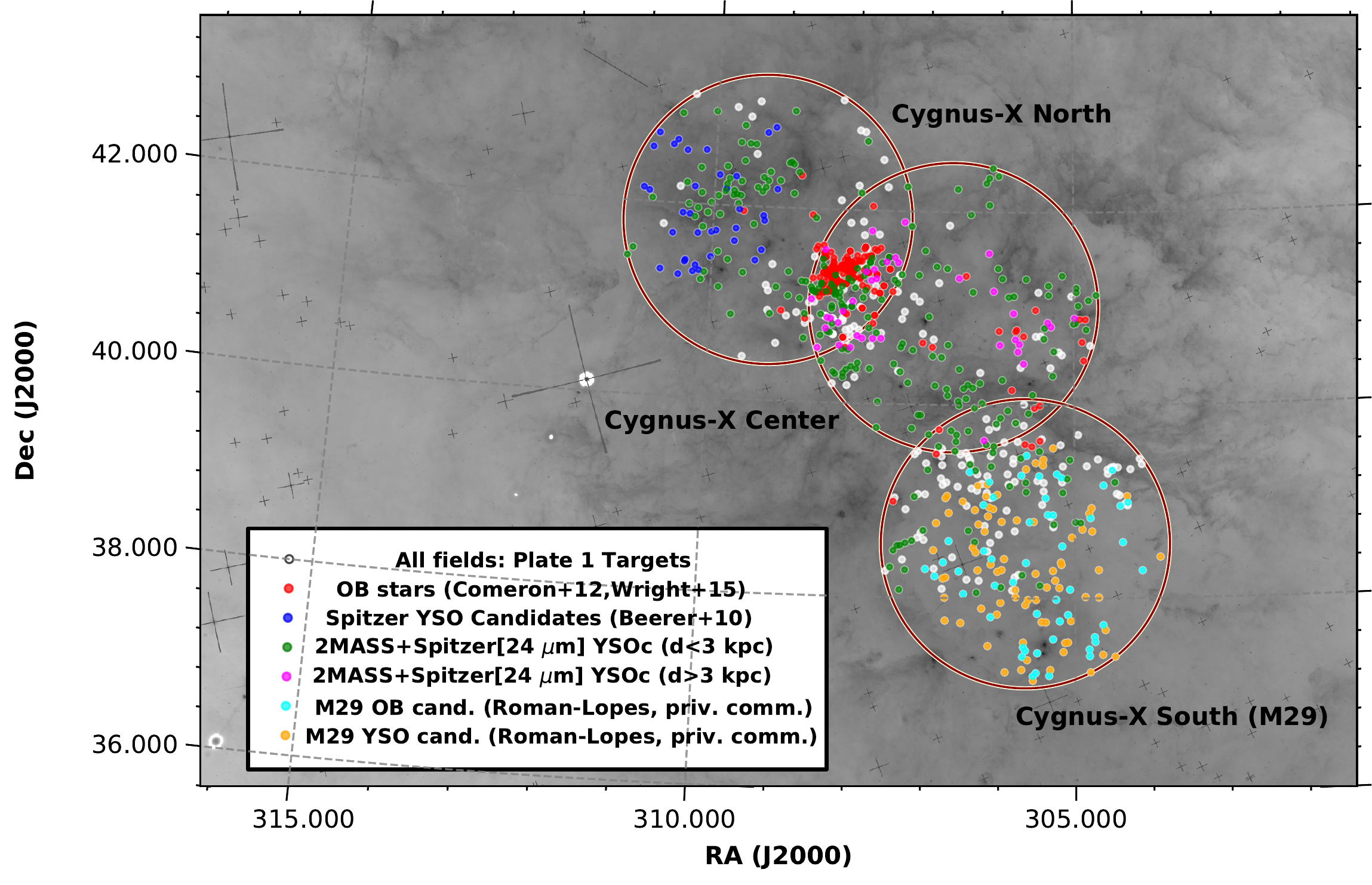}
    \caption{Example of targeting for the three BTX fields in the Cygnus-X star formation complex: Cygnus-X North, Center, and South. 
    The first two comprise the Cygnus OBS and Cygnus-X Complex, while the third one also encloses the M\,29 region. 
    The background grayscale is a WISE 12~\micron\ dust emission from the \citet{Meisner_2014} WSSA Atlas. 
    The large circles show the APOGEE-N target area for each field. 
    All the symbols indicate targets from one plate targeting each of these three fields. 
    Symbol colors indicate examples of targets from different lists used to make the selection, as indicated in the legend.
    Each cluster (of which the Cygnus-X complex, illustrated above is one) was targeted with six plates in total; in crowded regions, sources closer together than the fiber separation limit were observed on separate plates, while in less crowded regions, sources were observed on multiple plates to provide deeper co-added S/N and multi-epoch coverage.
}
    \label{fig:yso_example}
    \end{mdframed}
\end{figure}

Operating in the NIR, the APOGEE spectra are well suited to the study of dusty, obscured regions that are home to young stars. 
APOGEE provides a unique window to determine a systematic census of the stellar characteristics, dynamics, and binary fractions in these clusters. 
APOGEE-2 vividly demonstrated these capabilities in its census of the Orion Complex \citep{cottle_2018}, where stellar properties and radial velocities from APOGEE spectra were combined with \gaia~DR2 astrometry and distances. 
This allowed construction of the first six-dimensional map of the complex that revealed the evolution of the young stellar populations in Orion from its youngest (1-3 Myr) embedded groups that follow closely the kinematics of their parental cloud (located in the outskirts) and are clearly expanding away from the gas \citep{kounkel_2018}.
With the expansion opportunity provided by the BTX, the young clusters program was able to obtain (i) additional pointings in some clusters to both improve overall sample completeness and (ii) radial velocity epochs as necessary. 
A more specific scientific overview of the original APOGEE-2N Orion Complex program is given in \citet{cottle_2018}. 
Deep observations of young star clusters probe main sequence stars at the same stellar masses as the typically older stars in the ``main red star sample.''

Before discussing the specific expansions of the young star cluster program under the BTX, we will briefly note complementary efforts in other programs using APOGEE-2N; \santanatext\ provides an overview of similar efforts with APOGEE-S through the Contributed Programs route. 
We note that the Orion~A and Orion~B fields from the initial survey \citepalias{zasowski_2017} were targeted by the SubStellar companions program (\autoref{sec:substellar}) for radial velocity monitoring.
An Ancillary Science Program (\autoref{anc:w345}) targeted the W3/4/5 star-forming regions for a pilot study that included young stars and also a large population of O and B type sources. The massive star sources for this program were successfully classified by \citet{Roman-Lopes_19} using solely the few Brackett-series lines available in the limited wavelength range of APOGEE: such features define clear semi-empirical spectral type sequences that have been tested against optical counterpart samples and allow the classification of O and B type stars with APOGEE spectra \citep[][]{Roman-Lopes_2018, Ramirez-Preciado_2020}. 
On a related note, it is worth mentioning that APOGEE data have also provided successful identification and classifications (with three new discoveries) of Galactic Wolf-Rayet type stars \citep{Roman-Lopes_20}.

The BTX specifically targeted a set of young stellar clusters that are summarized in \autoref{tab:yso}. These fields were designed in some cases to increase the area coverage and target samples in some regions, as in the Taurus and the W3/W4/W5 Complexes, respectively. 
In other cases, new regions were added to the program to investigate the properties of young stellar clusters in OB association environments, like the Rosette and the Cygnus-X/M29 complexes.
The targeting for the young clusters program is complex, because the target selection is made through a combination of known members and member candidates based on previous photometric and spectroscopic studies, as well as selection based on the expected loci of young clusters in color-magnitude diagrams, variability, infrared excess, and distances. 
The young clusters are typically crowded, and the targeting is typically affected by the collision radii of the APOGEE-N fibers.

For this reason, plate design for this program involves a set of overlapping fields and targets that both sample the full extent of the clusters and progressively build appropriate $S/N$ for individual objects. 
\autoref{fig:yso_example} provides an example of plate design for the Cygnus Complex. 
A first result for that particular survey was the discovery of a new WN4-5 type star (WR 147-1) in the direction of the W75 cluster in the Cygnus-X North region \citep{Roman-Lopes_20}.

The BTX fields have \texttt{PROGRAMNAME} `yso\_btx' and the \texttt{FIELD} names have `\_btx' appended.
Targets selected for the Young Clusters program have \texttt{APOGEE2\_TARGET3} bit 5 set.

%%%%%%%%%%%%%%%%%%%%%%%%%%%%%%%%%%%%%%%%%%%%%%%%%%%%%%%%%%%%%%%%%%%%%%%%%%%%%%%%%%%%%%%
%%%%%%%%%%%%%%%%%%%%%%%%%%%%%%%%%%%%%%%%%%%%%%%%%%%%%%%%%%%%%%%%%%%%%%%%%%%%%%%%%%%%%%%
\begin{table*}[h]
    \centering
    \caption{Young Clusters in the Bright Time Extension}
    \label{tab:yso}
    \begin{tabular}{lcc l}
        \hline \hline
        Cluster Name & Age & Distance & APOGEE-2N Field Names \\
                     & Myr & kpc      & \\
        \hline \hline
%        Eagle Nebula (M\,16-M\,17) & 1-3 & 2.1 & \fixme{????} \\
        Cygnus-X                   & 1-5 & 2.0 & CygnusX\_C\_btx, CygnusX\_S\_btx \\
        Rosette Complex             & 1-3 & 1.6 & Rosette\_btx \\
%        Lagoon Nebula (M\,8)       & 1-4 & 1.2 & \fixme{????} \\
        W3/4/5                     & 1-4 & 2.1 & W34\_btx\\
        Taurus                     & 1-5 & 0.1-0.2 & TAUL1495\_btx, TAUL1517\_btx, TAUL1551\_btx	\\
        \hline \hline 
    \end{tabular}
\end{table*}
%%%%%%%%%%%%%%%%%%%%%%%%%%%%%%%%%%%%%%%%%%%%%%%%%%%%%%%%%%%%%%%%%%%%%%%%%%%%%%%%%%%%%%%
%%%%%%%%%%%%%%%%%%%%%%%%%%%%%%%%%%%%%%%%%%%%%%%%%%%%%%%%%%%%%%%%%%%%%%%%%%%%%%%%%%%%%%%

%%%%%%%%%%%%%%%%%%%%%%%%%%%%%%%%%%%%%%%%%%%%%
\subsection{SubStellar Companions} \label{sec:substellar}
%%%%%%%%%%%%%%%%%%%%%%%%%%%%%%%%%%%%%%%%%%%%%

The radial velocity precision of the APOGEE-N instrument has been particularly useful for the identification of stars with radial velocity variation \citep[e.g.,][]{troup_2016,badenes_2018,price-whelan_2018}. 
A particularly compelling aspect of the APOGEE survey is its ability to detect companions over large range of stellar type, stellar environments, ages, and chemical compositions. 
The quest to characterize the binary fraction as a function of these observable quantities is the underlying goal of the APOGEE-2N SubStellar Companions Program, which is an extension of path-finder work from APOGEE-1 \citep{troup_2016}; the full program will be described in \inprep{N.~Troup et al.~(in prep.)}.

Red giants are particularly challenging targets around which to identify companions due to difficulties in determining the red giant mass (e.g., due to degeneracy in isochrone tracks on the red giant branch) and added noise due to stellar jitter impacting the radial velocity.
Yet, red giants are astrophysically interesting sources due to the evolution of the binary system architecture during stellar evolution (star-planet tidal interactions or even planetary engulfment).
While dwarf-type stars in the solar-neighborhood are the predominant host-stars for planet searches, such stars span a limited range of underlying stellar population properties; yet, there is evidence that planet occurrence may have an underlying age and/or metallicity dependence \citep{wilson_2018}  while also likely having a correlation with the environmental birth-place of the host-star (e.g., distributed star formation in loose associations versus concentrated, cluster formation).
The luminosity of red giants permits exploration of such trends in stars at greater distances, giving access to diverse regions across the Milky Way --- that is if concerns related to system architecture due to evolution can by decoupled statistically. 

For the BTX expansion of the Substellar Companions program, seven fields from APOGEE-1 were included such that radial velocity monitoring for these fields will span nearly a decade. 
These fields are a mix of star clusters (NGC\,6819 in N6819-RV\_btx, NGC\,1333 in N1333-RV\_btx , and NGC\,5634 in N5634SGR2-RV\_btx), pointings in the Galactic disk (090+00\_btx and 203+04-RV\_btx), and young star forming regions (ORIONA-RV\_btx, ORIONB-RV\_btx). 
We also note that some fields beyond those in \citetalias{zasowski_2017} were added via an approved Ancillary Science Program and are listed in \autoref{anc:substellar}. 
The targets were held fixed, modulo MaStar co-targeting (\autoref{sec:mastar}), to those targets selected in APOGEE-1 \citepalias{zasowski_2013}. 
The BTX fields have \texttt{PROGRAMNAME} `sub\_btx' and the \texttt{FIELD} names have `\_btx' appended.
Targets selected for the substellar companions program have \texttt{APOGEE2\_TARGET3} bit 4 set.

% THE FIELDS ARE: 
%	203+04-RV\_btx
%	N5634SGR2-RV\_btx	
%   ORIONA-RV\_btx
%   ORIONB-RV\_btx
%   N6819-RV_btx
%   N1333-RV_btx
%   090+00_btx

%%%%%%%%%%%%%%%%%%%%%%%%%%%%%%%%%%%%%%%%%%%%%
\subsection{Dwarf Spheroidal Galaxies} \label{sec:dsph}
%%%%%%%%%%%%%%%%%%%%%%%%%%%%%%%%%%%%%%%%%%%%%

The program to obtain APOGEE spectroscopy for Milky Way dwarf spheroidal galaxy (dSph) satellites in the Northern Hemisphere was expanded to include at least 12 additional visits during the term of the BTX. 
There were two goals with these allocations: (i) the accumulation of additional radial velocity epochs, and (ii) building $S/N$ for chemical abundance analyses. 
The targeted APOGEE-2N dSphs are Ursa~Major, Draco, and Bo\"{o}tes~I.

Additional radial velocity epochs improves our ability to both detect binary companions on the red giant branch by extending the time baseline for our radial velocity monitoring and to characterize the binary orbits by adding key epochs to improve orbit fitting \citep[see examples in][]{price-whelan_2018, Lewis_2020}. 
Given the small total velocity dispersion of these objects, at the $\sim$10 \kms\ level, the ability to identify sources with radial velocity variability and then discern and use their true systemic velocity greatly improves systematic uncertainties on these measurements, which is important for inferences drawn from the evaluation of internal satellite dynamics.

Additional visits also improve our ability to recover reliable spectra for the faintest stars in these systems. 
Given their distances, the targeting for these objects required probing to very faint magnitudes, even given the depth of a 24-visit field \citepalias[see \autoref{tab:colorcuts}][]{zasowski_2017}.
As such, individual visits have low $S/N$ and additional visits contribute meaningfully to gains in the co-added spectral quality for these faint stars \citep[see discussion in][regarding faint stars]{Jonsson_2020}. 

We note that this BTX program represents re-observations of the initial dSph target selection and that no new plates were designed.   
The plates used were created prior to the release of proper motion measurements from \gaia~DR2, and the original dSph target selection occurred in two tiers: 
 (i) confirmed members flagged with \texttt{APOGEE2\_TARGET1} bit 20 and 
 (ii) candidate members using photometric criteria with \texttt{APOGEE2\_TARGET1} bit 21 set. 
Confirmed members were drawn extensively from the existing literature on these galaxies, whereas candidate members were selected using Washington$+DDO51$ photometry to discriminate between likely foreground Milky Way dwarfs and dSph giants of the same spectral type \citep[for details on the technique, see ][]{majewski_2000}. 
Additional description of the targeting will be reserved for focused science papers describing the results of this project.

%%%%%%%%%%%%%%%%%%%%%%%%%%%%%%%%%%%%%%%%%%%%%
\subsection{Cross-Survey Calibration} \label{sec:cross}
%%%%%%%%%%%%%%%%%%%%%%%%%%%%%%%%%%%%%%%%%%%%%

Another important scientific goal of the BTX was to expand the cross-survey calibration samples between APOGEE and other major existing or planned surveys. 
This occurred naturally across our the various programs previously discussed, but we summarize these key cross-survey calibration samples as follows:
\begin{itemize} \itemsep -2pt
    \item (\autoref{sec:halo}) SEGUE \citep{yanny_2009,eisenstein_2011} overlap was increased by using SEGUE results to target distant halo stars.
    \item (\autoref{sec:k2}) The \ktwo-HERMES program used the GALAH \citep{Wittenmyer_2018,Sharma_2019} configuration of the HERMES instrument \citep{Sheinis_2015,Sheinis_2016} to study large numbers of \ktwo\ targets for the Galactic Astrophysics Program \citep{Stello_2017}. 
    As a result of our complementary \ktwo\ program, APOGEE overlap with GALAH has been increased with \ktwo\ campaigns in common.
    \item (\autoref{anc:refstars}) One of the Ancillary Science Programs is focused on targeting a library of well-known and previously well-measured stars for the purposes of improving ASPCAP calibration.
\end{itemize}

Beyond the spectroscopic surveys, we have also now guaranteed broad spectroscopic coverage for photometric objects of interest coordinated with \kepler, \ktwo, and \tess. 
In particular, our targeting in the BTX relied on the input catalogs or target lists from those photometric surveys such that we could draw from those parent samples using a selection algorithm that enables study of large stellar samples.
Lastly, our datasets have large overlap with stars having \gaia\ astrometry \revise{(either from DR2, provided in DR16, or eDR3, to be provided with DR17)} and we anticipate additional overlap in successive releases from that mission \citep[][\holtzmanprep]{Jonsson_2020}. 
\santanatext\ describes cross-survey calibration efforts for APOGEE-2S.

%%%%%%%%%%%%%%%%%%%%%%%%%%%%%%%%%%%%%%%%%%%%%%%%%%%%%%%%%%%%%%%%%%%%%%%%%%%%%%%%%%%%%%%%%%
\section{Lessons Learned} \label{sec:wrapup}
%%%%%%%%%%%%%%%%%%%%%%%%%%%%%%%%%%%%%%%%%%%%%%%%%%%%%%%%%%%%%%%%%%%%%%%%%%%%%%%%%%%%%%%%%%

Running a large, complex and pioneering survey like APOGEE (APOGEE-1 and APOGEE-2) over more than a decade naturally results in the evolution of a number of operational aspects, but most especially in the targeting strategy, particularly when at least some evolution was built in as a feature of the survey from the start.  
We share here various aspects of our targeting experience --- both planned and learned along the way --- that may prove useful to other survey planners.

\begin{itemize} \itemsep -2pt
    \item \textit{Hold some observing time in reserve:} Hold some observing time in reserve, both for unanticipated projects and to accommodate developments in the field and the turn-over in collaboration members over time. Such projects can only enhance the scientific impact of the survey by pushing into new territory and maintaining a timely edge. 
    Just as important is to build in a mechanism to respond to and integrate in these new directions.  
    We found that our open calls for Ancillary Science Programs was an effective means for both soliciting broad input for new and timely science applications and for testing ideas through pilot studies.  
    After their proven success and impact, several Ancillary Science Programs were later cemented into the main survey.  
    For example, if not for the fiber reserves put toward Ancillary Science Programs early in APOGEE-1, a well-considered, tested, and effective APOKASC \citep{apokasc} program in both APOGEE-1 and APOGEE-2 would not have been feasible.
    \item \textit{Review targeting strategies often:}  
    Even the best laid plans can go awry, especially when pushing out into the unknown.  
    Maintaining a well-understood survey selection function is important and motivates resistance to changing targeting criteria mid-survey; however, just as important is ensuring that the survey will meet its science requirements. 
    While often one of the objectives of a general survey is to reveal unknown distribution functions, in the cases where numbers of a specific type of target are sought and expected, diligent monitoring is needed to ensure that the survey is providing the intended yield.  
    If it is not, then consideration of a mid-course correction in targeting, however complicating for the overall survey selection function, is warranted.
    An example in the case of APOGEE was our goal to characterize chemically the Galactic halo, despite the relative rarity of such targets compared to the overwhelming foreground of disk stars at the same magnitudes.
    On paper, our strategies for targeting halo stars should have been effective, but in practice, through continuous analysis of our data product and yearly targeting reviews, we determined that our implemented strategies were harvesting fewer than expected 
    halo stars, and that modifications in our targeting of such stars were needed.  
    These same reviews revealed that, quite serendipitously, the APOGEE-N fibers piggy-backing on the short, less faint MaNGA visits, were achieving a higher yield of halo stars, and this discovery helped guide mitigation of our targeting plan. 
    \item \textit{Engage your science teams for feedback:} 
    Good data and project planning are best revealed by the exceptional science that it produces. 
    Through the targeting reviews we invited the APOGEE data processing and science teams into the process of targeting and observation planning. 
    Because these teams are largely distinct and work for different operational goals, their concerns are not always fully appreciated across the collaboration. 
    To facilitate communication, the targeting reviews were structured so that 
        (a) in advance, the data team was engaged to produce an interim (generally internal) data product sufficient for high-level review, 
        (b) also in advance, the science teams vested in the relevant science programs were given time to assess the full data in hand to extrapolate and evaluate whether the requisite science requirements for their program would be met, and 
        (c) both the data and science teams played significant roles in the targeting review.  In this way, the targeting reviews fostered collaboration-wide communication in a focused setting and recognized the differing perspectives, skills, interests and knowledge of each team.  
    While a particular science team may be uniquely able to articulate a problem and a solution in scientific terms (e.g., that the existing targeting strategy was not sufficiently probing stars in the outer disk, \citet{hayes_2018}), the targeting team, working with the observing team, have pertinent experience to translate a proposed solution into an effective targeting strategy (e.g., using the RPMJ technique to remove foreground dwarfs) and to integrate it with the rest of the survey.
    Throughout this paper and \santanatext, strategies developed in a specific scientific context were cross-pollinated into others.
    \item \textit{Recognize the laws of institutional entropy:} 
    Even the most organized project grows complex over time and the strands that tie together the original to final plan can be difficult to disentangle. 
    The decision making and implementation plans described in this paper spanned over five years -- some of it predating the in-depth participation of the lead author and surviving numerous changes in personnel. 
    Yet, much of the material presented here was documented in detailed meeting minutes, file repositories, and as part of on-going, online documentation efforts. 
    Moreover, in most cases the documentation was detailed and contained reflections on the process of coming to a decision (how something could be completed more efficiently, what roadblocks occurred).  
    Our availability to depend on such material demonstrated clearly that it is not just the documentation, itself, that is vital, but the time and effort put in to making that documentation useful for others not originally involved in or familiar with the specific process.
    \item \textit{Engage your average users, not just your core collaboration:} 
    The APOGEE data user community extends well beyond the SDSS collaboration.
    By staying current with what this broader user community was doing with our public data, we were able to take into account what that community valued.
    As one example, the wealth of recent work on M-dwarfs observations in APOGEE \citep[e.g., the work published in][but that had started much earlier]{Birky_2020} helped to motivated the targeting team to place greater emphasis on ensuring key calibration data were taken. 
    At the same time, individuals in the collaboration interested in pursuing this work were further encouraged.
\end{itemize}

%%%%%%%%%%%%%%%%%%%%%%%%%%%%%%%%%%%%%%%%%%%%%%%%%%%%%%%%%%%%%%%%%%%%%%%%%%%%%%%%%%%%%%%%%%
\section{Summary} \label{sec:summary}
%%%%%%%%%%%%%%%%%%%%%%%%%%%%%%%%%%%%%%%%%%%%%%%%%%%%%%%%%%%%%%%%%%%%%%%%%%%%%%%%%%%%%%%%%%
In this paper, we have described the targeting for 23 Ancillary Science Programs as well as the 1.5 year expansion of the APOGEE-2N survey through the Bright Time Extension program. 
The former were built into the design of APOGEE-2N in part through the fiber hours reserved for two application cycles for the Ancillary Science Program. 
The latter BTX addition was an unanticipated bonus accrued largely due to better than expected observational efficiency.

Together, the Ancillary Science Programs added new classes of targets (like, photometric variable stars or extragalactic star clusters) to the survey that enable new types of scientific investigations (time resolved spectroscopy or integrated stellar population studies). 
The Ancillary Science Programs broadened the impact of some samples, like the expansion of the APOGEE-\emph{Kepler} and red clump samples, while also testing the ability to employ APOGEE data to synergistic investigations of the physical nature of the interstellar medium using absorption features in the spectra. 
Each new avenue that APOGEE explored brought with it new scientific ideas, research methods, and human enthusiasm. 

% sentences on BTX
%BTX also known as Better Than eXpected :D!
The Bright Time Extension enabled revisions to the general targeting strategies that permitted us to make substantial progress toward difficult goals in the Science Requirements Document for constructing samples in the distant halo. 
Moreover, the APOGEE team was able to act in collaboration with scientists in the MaStar and After Sloan-IV efforts, with the latter undertaking pilot observations for some of its key science objectives. 
New programs were initiated to map the California Molecular Cloud, the \tess~N-CVZ, provide a large scale survey of structure in the Outer Disk, and build challenging calibration samples for main sequence stars. 
Many of the existing programs in APOGEE-2N were also given the opportunity to expand their scope, with significant expansions to the Open Cluster, \ktwo, KOI, Eclipsing Binaries, Young Star Cluster, SubStellar Companions, and dSph programs. 
To enable these projects, the APOGEE-2 targeting team performed detailed assessments of existing targeting strategies and devised new ones, as described in this paper. 
Lastly, many of our programs relied more heavily on complementary datasets from other surveys and, as a result, the cross-survey calibration samples were bolstered significantly.

Targets from these two programs represents over $\sim$25\% of the total number of stars observed in the APOGEE-2N and serve to open new scientific explorations and explore unconventional ideas while also reinforcing our core science goals. 
APOGEE's ability to achieve its scientific goals over its decade of operations stems not just from the established and successful organizational infrastructure 
of SDSS, but also the tight collaborative cooperation between its scientific and technical teams, and its adaptations to developments in the broader scientific field.

%%%%%%%%%%%%%%%%%%%%%%%%%%%%%%%%%%%%%%%%%%%%%%%%%%%%%%%%%%%%%%%%%%%%%%%%%%%%%%%%%%%%%%%%%%
\begin{acknowledgements}
%%%
% Personal Acknowledgements:
%%%
% 
The APOGEE project thanks Jeff Munn (NOFS) for collecting Washington$+DDO51$ imaging for large areas of the sky. 
The authors also recognize contributions from prior members of the targeting team and the members of the science working groups in both APOGEE-1 and APOGEE-2. 
We graciously thank the staff at Apache Point Observatory and the SDSS Observers for both their dedication and their remarkable ability to exceed themselves without needing to. 
We also warmly recognize the APOGEE-2N Operations team for their diligence and the SDSS Plate Shop for their enduring patience.
We thank the SDSS-IV Data Team for always remembering and then carefully considering all of the details and doing so without a complaint, albeit often with a stroopwafel. 
We thank the SDSS-IV Management Committee and all SDSS-IV leadership whose commitment to SDSS-IV is literally is the glue that keeps the project going.
Lastly, we warmly thank Jim Gunn and Jill Knapp for the remarkable manner in which they have and continue to impact the world around them.

%%
% FUNDING
%% 
% Beaton
Support for this work was provided by NASA through Hubble Fellowship grant \#51386.01 awarded to R.L.B. by the Space Telescope Science Institute, which is operated by the Association of  Universities for Research in Astronomy, Inc., for NASA, under contract NAS 5-26555.
% Covey
K.R.C acknowledges support provided by the National Science Foundation (NSF) through grant AST-1449476.
% Majewski
S.R.M. acknowledges support through NSF grants AST-1616636 and AST-1909497.
%%
% Now ALPHABETICAL by last name:
% Aerts + Tkachenko
The research leading to these results has (partially) received funding from the European Research Council (ERC) under the European Union's Horizon 2020 research and innovation programme (grant agreement N$^\circ$670519: MAMSIE), from the KU~Leuven Research Council (grant C16/18/005: PARADISE), as well as from the BELgian federal Science Policy Office (BELSPO) through PRODEX grant PLATO.
% Donor + Frinchaboy
J.D. and P.M.F. acknowledge support for this research from the National Science Foundation AAG and REU programs (AST-1311835, AST-1715662, PHY-1358770, \& PHY-1659444).
% Geisler
D.G. gratefully acknowledges support from the Chilean Centro de Excelencia en Astrof\'isica y Tecnolog\'ias Afines (CATA) BASAL grant AFB-170002.
D.G. also acknowledges financial support from the Direcci\'on de Investigaci\'on y Desarrollo de la Universidad de La Serena through the Programa de Incentivo a la Investigaci\'on de Acad\'emicos (PIA-DIDULS).
% Hasselquist
S.H. is supported by an NSF Astronomy and Astrophysics Postdoctoral Fellowship under award AST-1801940. 
% Roman-Lopes
AR-L acknowledges financial support provided in Chile by Comisi\'on Nacional de Investigaci\'on Cient\'ifica y Tecnol\'ogica (CONICYT) through the FONDECYT project 1170476 and by the QUIMAL project 130001.
% Santana
F.A.S. acknowledges support from CONICYT Project AFB-170002. 
AMS gratefully acknowledges funding support through Fondecyt Regular (project code 1180350) and funding support from Chilean Centro de Excelencia en Astrofísica y Tecnologías Afines (CATA) BASAL grant AFB-170002.
% Zasowski
G.Z. acknowledges support from the NSF through grants AST-1203017 and AST-1911129.

%%%
% Data & Mission Acknowledgements
%%%
%% SDSS-IV
Funding for the Sloan Digital Sky Survey IV has been provided by the Alfred P.~Sloan Foundation, the U.S. Department of Energy Office of Science, and the Participating Institutions. SDSS-IV acknowledges support and resources from the Center for High-Performance Computing at the University of Utah. The SDSS web site is \url{www.sdss.org}.

SDSS-IV is managed by the Astrophysical Research Consortium for the Participating Institutions of the SDSS Collaboration including the Brazilian Participation Group, the Carnegie Institution for Science, Carnegie Mellon University, the Chilean Participation Group, the French Participation Group, Harvard-Smithsonian Center for Astrophysics, Instituto de Astrof\'isica de Canarias, The Johns Hopkins University, Kavli Institute for the Physics and Mathematics of the Universe (IPMU) / University of Tokyo, Lawrence Berkeley National Laboratory, Leibniz Institut f\"ur Astrophysik Potsdam (AIP),  Max-Planck-Institut f\"ur Astronomie (MPIA Heidelberg), Max-Planck-Institut f\"ur Astrophysik (MPA Garching), Max-Planck-Institut f\"ur Extraterrestrische Physik (MPE), National Astronomical Observatories of China, New Mexico State University, New York University, University of Notre Dame, Observat\'ario Nacional / MCTI, The Ohio State University, Pennsylvania State University, Shanghai Astronomical Observatory, United Kingdom Participation Group, Universidad Nacional Aut\'onoma de M\'exico, University of Arizona, University of Colorado Boulder, University of Oxford, University of Portsmouth, University of Utah, University of Virginia, University of Washington, University of Wisconsin, Vanderbilt University, and Yale University.

%% 2MASS
This publication makes use of data products from the Two Micron All Sky Survey, which is a joint project of the University of Massachusetts and the Infrared Processing and Analysis Center/California Institute of Technology, funded by the National Aeronautics and Space Administration and the National Science Foundation.

This work is based, in part, on observations made with the \emph{Spitzer} Space Telescope, which is operated by the Jet Propulsion Laboratory, California Institute of Technology under a contract with NASA.

This publication makes use of data products from the Wide-field Infrared Survey Explorer \citep{WISE}, which is a joint project of the University of California, Los Angeles, and the Jet Propulsion Laboratory/California Institute of Technology, funded by the National Aeronautics and Space Administration, and NEOWISE, which is a project of the Jet Propulsion Laboratory/California Institute of Technology. 
WISE and NEOWISE are funded by the National Aeronautics and Space Administration 

This work has made use of data from the European Space Agency (ESA) mission \gaia\ (\url{https://www.cosmos.esa.int/gaia}), processed by the \gaia\ Data Processing and Analysis Consortium (DPAC, \url{https://www.cosmos.esa.int/web/gaia/dpac/consortium}). Funding for the DPAC has been provided by national institutions, in particular the institutions participating in the \gaia\ Multilateral Agreement.

This research has made use of NASA’s Astrophysics Data System.

Many of the acknowledgements were compiled using the Astronomy Acknowledgement Generator. 

This research has made use of the SIMBAD database, operated at CDS, Strasbourg, France. The original description of the SIMBAD service was published in \citet{simbad_2000}.

This research has made use of the \vizier\ catalogue access tool, CDS, Strasbourg, France (DOI: 10.26093/cds/vizier). The original description of the \vizier\ service was published \citet{vizier2000}.

Some of the data presented in this paper were obtained from the Mikulski Archive for Space Telescopes (MAST). STScI is operated by the Association of Universities for Research in Astronomy, Inc., under NASA contract NAS5-26555. Support for MAST for non-HST data is provided by the NASA Office of Space Science via grant NNX13AC07G and by other grants and contracts. 

This research has made use of the NASA Exoplanet Archive, which is operated by the California Institute of Technology, under contract with the National Aeronautics and Space Administration under the Exoplanet Exploration Program.

The Pan-STARRS1 Surveys (PS1) and the PS1 public science archive have been made possible through contributions by the Institute for Astronomy, the University of Hawaii, the Pan-STARRS Project Office, the Max-Planck Society and its participating institutes, the Max Planck Institute for Astronomy, Heidelberg and the Max Planck Institute for Extraterrestrial Physics, Garching, The Johns Hopkins University, Durham University, the University of Edinburgh, the Queen's University Belfast, the Harvard-Smithsonian Center for Astrophysics, the Las Cumbres Observatory Global Telescope Network Incorporated, the National Central University of Taiwan, the Space Telescope Science Institute, the National Aeronautics and Space Administration under Grant No. NNX08AR22G issued through the Planetary Science Division of the NASA Science Mission Directorate, the National Science Foundation Grant No. AST-1238877, the University of Maryland, Eotvos Lorand University (ELTE), the Los Alamos National Laboratory, and the Gordon and Betty Moore Foundation.
\end{acknowledgements}
%%%%%%%%%%%%%%%%%%%%%%%%%%%%%%%%%%%%%%%%%%%%%%%%%%%%%%%%%%%%%%%%%%%%%%%%%%%%%%%%%%%%%%%%%%%%%%%%%%%%%%%%%%%%%%%%%%%%%%%%%%
%%%%%%%%%%%%%%%%%%%%%%%%%%%%%%%%%%%%%%%%%%%%%%%%%%%%%%%%%%%%%%%%%%%%%%%%%%%%%%%%%%%%%%%%%%%%%%%%%%%%%%%%%%%%%%%%%%%%%%%%%%
\facilities{Du Pont (APOGEE), 
            Sloan (APOGEE), 
            \emph{Spitzer}, 
            WISE, 
            2MASS, 
            Pan-STARRS, 
            \gaia, 
            Exoplanet Archive} 

\software{ 
    \revise{Astropy \citep{Astropy_2013,Astropy_2018}, 
    NumPy \citep{vanderWalt_2011,Harris_2020numpy},
    TopCat \citep{topcat}} 
    }

%%%%%%%%%%%%%%%%%%%%%%%%%%%%%%%%%%%%%%%%%%%%%%%%%%%%%%%%%%%%%%%%%%%%%%%%%%%%%%
%\bibliographystyle{aasjournal}
%\bibliography{03_refs}

%%%%%%%%%%%%%%%%%%%%%%%%%%%%%%%%%%%%%%%%%%%%%%%%%%%%%%%%%%%%%%%%%%%%%%%%%%%%%%

%%%%%%%%%%%%%%%%%%%%%%%%%%%%%%%%%%%%%%%%%%%%%%%%%%%%%%%%%%%%%%%%%%%%%%%%%%%%%%%%%%
\begin{appendix}
Ancillary Science Programs from the 2015 and 2017 calls for proposals are given in \autoref{sec:ancillary2015} and \autoref{sec:ancillary2017}, respectively.
A glossary of SDSS or APOGEE specific terms is provided in \autoref{sec:glossary}.
%%%%%%%%%%%%%%%%%%%%%%%%%%%%%%%%%%%%%%%%%%%%%%%%%%%%%%%%%%%%%%%%%%%%%%%%%%%%%%%%%%%%%%%
\section{2015 Ancillary Programs} \label{sec:ancillary2015}
% https://trac.sdss.org/wiki/ApprovedAncillaryPrograms2015
Ancillary programs were solicited via open proposal calls across the SDSS-IV collaboration (i.e., not limited to just APOGEE) and were evaluated for both technical feasibility and scientific merit.
The 2015 proposals were received in April 2015, reviewed and selected by an ad hoc committee during May-June 2015, and implemented starting in the September-October 2015 time frame. 
Many of the 2015 programs continued to be observed through the end of APOGEE-2N observations.
%%%%%%%%%%%%%%%%%%%%%%%%%%%%%%%%%%%%%%%%%%%%%%%%%%%%%%%%%%%%%%%%%%%%%%%%%%%%%%%%%%%%%%%
%%%%%%%%%%%%%%%%%%%%%%%%%%%%%%%%%%%%%%%%%%%%%%%%%%%%%%%%%%%%%%%%%%%%%%%%%%%%%%%%%%%%%%%
\subsection{Quasar Survey}\label{anc:quasar_survey}
Observations were collected for high redshift quasars (2.0\textless $z$~\textless~2.4) selected from SDSS-III/BOSS-DR12Q with the goal of detecting $H\beta$-[OIII] emission that is redshifted into APOGEE's wavelength range. 
These observations will benefit the investigations on three timely topics in the field of cosmology: 
    (i) the possible variation of the fine-structure constant, 
    (ii) calibration of different methods to estimate the mass of supermassive black holes, and 
    (iii) a comparison study of redshifts based on C{\sc IV}, C{\sc III}] and Mg{\sc II} emission lines in BOSS spectra and those derived from [O{\sc III}] in APOGEE spectra.

The selection of targets proceeded as follows: Starting with 297,301 quasars from the SDSS-III/BOSS-DR12Q catalogue \citep{Paris2017}, 71,587 were selected that have a redshift from spectral at visual redshift in the range 2.06$<$z$<$2.38. 
This sample is further restricted by requiring a Vega-based $H_{\rm Vega}$~\textless~20; however, only 20,796 of these have $H$ measurements from UKIDSS \citep{Lawrence_2007}. 
For the remaining quasars, we studied the correlation between UKIDSS $H$  and SDSS $z'$ using those quasars with UKIDSS $H$~\textless~21  (19,993 in total). Thus, for those quasars without $H$ measurements in UKIDSS, we require z-band SDSS magnitude $<$ 21.27. Alternatively, we demand the $H$-magnitude obtained either from UKIDSS or estimated from z-band SDSS measurements to be $z$ \textless~20. 
In each plate, the quasar targets are prioritized according to $H$ magnitude.
Only 971 are located in the available APOGEE-2N halo fields at the time of the ancillary call, which defines our final sample that is distributed across 24 fields (17 halo, 7 halo stream). 
    %In this way, we obtain a sample of 61,075 quasars. 
    %Only 971 are located in the available APOGEE-2N halo fields at the time of the ancillary call, which defines our final sample that is distributed across 24 fields (17 halo, 7 halo stream). 
%Targets from this program have targeting flag bit apogee2\_target3 = 10 set.
%%%%%%%%%%%%%%%%%%%%%%%%%%%%%%%%%%%%%%%%%%%%%%%%%%%%%%%%%%%%%%%%%%%%%%%%%%%%%%%%%%%%%%%
%%%%%%%%%%%%%%%%%%%%%%%%%%%%%%%%%%%%%%%%%%%%%%%%%%%%%%%%%%%%%%%%%%%%%%%%%%%%%%%%%%%%%%%
\subsection{Cepheid Metallicity} \label{anc:cephmetals}

As intrinsically bright standard candles, Cepheid variables are an invaluable tool used to reveal the three-dimensional structure of our own and other nearby galaxies. 
While their use has a long history \citep[beginning with][]{leavitt_1912}, only recently has the dramatically reduced intrinsic scatter for the near- and mid- infrared period-luminosity relationships been exploited for applications to the extragalactic distance ladder \citep[e.g.,][]{freedman_2011,riess_2011} and for the detailed structure of Local Group constituents \citep{monson_2012,scowcroft_2013,scowcroft_2016,scowcroft_2016b}.
This Ancillary Science Program aims to explore the color metallicity relationship discovered by \citet{scowcroft_2016b} in the Spitzer-IRAC [3.6] and [4.5] filters (i.e, [3.6]-[4.5]; MIR-color, hereafter) using data obtained for the Carnegie Hubble Program \citep[CHP;][]{freedman_2011}. 
The behavior of the MIR-color with metallicity is driven by a CO bandhead in the [4.5] band \citep[see disussion in][]{marengo_2010,monson_2012}, which is confirmed by the specific color variations as a function of the range of temperatures spanned by individual stars of different periods.
Owing to the small photometric uncertainties for MIR photometry (\textless~2\%), the technique permits a more sensitive differential estimate of metallicity than direct probes, though sources of scatter (potentially correlated to C and O abundance differences among the sample) require exploration. 
In \citet{scowcroft_2016b} only 36 of the 200 stars with well sampled MIR light curves had metallicities in the literature; and even these limited measurements were scattered across multiple studies with large systematics. 
Thus, this ancillary program hopes to explore the MIR-color metallicity relationship in anticipation of its future application with JWST and to provide a cross-calibrating sample with the large-scale homogenized sample given in \citet{genovali_2014,genovali_2015} to bring its sample onto the APOGEE metallicity scale.

Targets for this program were selected by cross-matching the General Catalog of Variable Stars \citep[GCVS;][]{samus_2017}, restricted to those stars of normal Cepheid and $\delta$ Cepheid types (CEP and DCEP designations in the GCVS, respectively), to the APOGEE-2N field footprint. 
Given that Cepheid variables experience radial pulsations with periods from $\sim$2 to $\sim$100 days, data obtained over several days (or weeks) will observe the same star at a different temperature and surface gravity. 
As such, abundances derived from co-added data may be suspect. 
Thus, sufficient $S/N$ must be reached in a single visit to guarantee the highest fidelity abundances. 
Thus, the GCVS Cepheid catalog is further restricted to those stars with  6.5 \textless $H$ \textless 11.
We note that while only those stars with periods longer than 7 days are expected to show a MIR-color metallicity relationship \citep{scowcroft_2016b}, Cepheids with periods less than 7 days were intentionally included in the target selection to increase overlap with \citet{genovali_2014,genovali_2015}. 
Including short period sources has the added advantage of permitting a detailed test of the physical cause of the MIR-color metallicity relationship.

%Targets from this program have targeting flag bit apogee2\_target3 = 11 set.
%%%%%%%%%%%%%%%%%%%%%%%%%%%%%%%%%%%%%%%%%%%%%%%%%%%%%%%%%%%%%%%%%%%%%%%%%%%%%%%%%%%%%%%
%%%%%%%%%%%%%%%%%%%%%%%%%%%%%%%%%%%%%%%%%%%%%%%%%%%%%%%%%%%%%%%%%%%%%%%%%%%%%%%%%%%%%%%
\subsection{The Far Disk in Low Extinction Windows} \label{anc:lowextwindow}

%This program takes advantage of low-extinction windows in the Galactic midplane to obtain a sample of luminous giants sampling the far reaches of the Milky Way's disk. Windows are selected based on the \citet{marshall_2006} 3-D extinction map and luminous giants are	targeted using a simple color cut.
%This sample will allow lopsided kinematical and chemical modes to be mapped and provide an improved understanding of the warping and flaring of the Galactic disk.
%One of APOGEE-2's main scientific objectives is a comprehensive study of the chemo-dynamical structure of a large volume of the Galactic disk. 
While the main survey target selection for the disk naturally produces a large sample of stars at distances from 6 to $\sim$9~kpc, the expected number of stars at the largest distances within the mid-plane will only be a few dozen, primarily due to increased extinction at these Galactic latitudes.
With the availability of three-dimensional dust maps covering a large fraction of the sky and a large range of extinctions \citep{marshall_2006, green_2015}, it is now possible to identify low-extinction windows in the inner Milky Way where stars at large distances can be observed at relatively bright optical and infrared magnitudes.
While the dust is highly filamentary on small scales such that much of the area of a typical APOGEE pointing in the mid-plane suffers from high extinction, substantial fractions of a pointing can cover low extinction regions. 
This ancillary program takes advantage of low-extinctions windows in three pointings ($\ell$,$b$): (27$\degs$,0$\degs$), (33$\degs$,0$\degs$), and 65$\degs$,0$\degs$).

A sample of only a few hundred stars a few magnitudes below the tip of the red-giant branch lying in low-extinction windows and probing distances as far as 16~kpc significantly improves APOGEE-2's investigation of the large-scale dynamics and metallicity structure of the disk. 
Because of the low extinction, such stars will have highly precise proper motions from \gaia. 
At large distances proper motions due to Galactic rotation are a few mas yr$^{-1}$ and when these are combined with APOGEE's precise radial velocities, it is possible to study of large-scale lopsided modes in the disk and therefore place more direct constraints on the axisymmetric rotation (the rotation curve) than will be possible from \gaia\ data alone. 
Similarly, a few hundred stars will allow the mean metallicity at otherwise inaccessible regions of the disk to be mapped, leading to much stronger constraints on the azimuthal chemical homogeneity of the disk. 
The $\ell$ = 65$\degs$ field also samples the outer disk in a region that is much less affected by the warp than that at $\ell$ = 180$\degs$, providing for a cleaner study of the outer disk and for stronger constraints on the warp and flaring of the disk by comparison with stars in the $l$ = 180$\degs$ direction.

For each pointing, targets are selected in those regions for which $A_{\rm H}$($D$=7~kpc) \textless~1.4,  as computed from the three-dimensional extinction map of \citet{marshall_2006} using their their native 15$'\times15'$ grid in ($\ell$,$b$) with linear interpolation in distance modulus converted to $A_{\rm H}$ using $A_{\rm H}$/$A_{\rm K_{\rm s}}$ = 0.46/0.31. 
Targets in this region are selected from the 2MASS catalog using the same 2MASS quality cuts and RJCE-dereddening as the APOGEE disk targets in the main red star sample \citepalias{zasowski_2013}. 
All potential targets are obtained by selecting stars with $(J - K_{\rm s})_0$ \textgreater 0.8 and 12 \textless $H$ \textless 13, and of these only a random subset are observed. 
Targets at the top of the list that were already observed as part of the regular disk sample in the $\ell$ = 34$\degs$ and $\ell$ = 64$\degs$ disk fields are flagged, though not re-observed.
%%%%%%%%%%%%%%%%%%%%%%%%%%%%%%%%%%%%%%%%%%%%%%%%%%%%%%%%%%%%%%%%%%%%%%%%%%%%%%%%%%%%%%%
%%%%%%%%%%%%%%%%%%%%%%%%%%%%%%%%%%%%%%%%%%%%%%%%%%%%%%%%%%%%%%%%%%%%%%%%%%%%%%%%%%%%%%%
\subsection{Hot Emission Line Stars} \label{anc:be_stars}
% Drew Chojnowski

This ancillary program seeks multi-epoch APOGEE spectra of a variety of B-type emission line stars, including many of rare subtypes; this project builds on the successful work on emission line stars found through serendipitous targeting in APOGEE-1 \citep[e.g.,][]{Chojnowski_2015}.
As a complement to our dedicated APOGEE-2N field for classical Be star monitoring (in \texttt{FIELD} 135-03; $h$ and $\chi$ Persei), six Be stars from NGC\,7419 were targeted (in \texttt{FIELD} 109+00).  
A small of number of high-interest B-type emission line stars were also targeted in other APOGEE fields, including an unclassified B[e] star (evolutionary status unknown), several Herbig B[e] stars (pre-main sequence), the only known classical Be star + black hole binary, the central classical Be star of NGC\,2023, and an unusual classical Be star (HD\,37115) also observed in APOGEE-1 \citepalias[see][]{zasowski_2013}. 
All of these targets were to be observed multiple times, and observations of the B[e] stars will supplement the ongoing survey of brighter B[e] stars being carried out with the APOGEE instrument attached to the NMSU-1m telescope (\texttt{TELESCOPE} of `apo1m'). 

These targets were selected according to very simple and somewhat subjective criteria. 
First, comprehensive lists of known OBA emission-line stars were obtained via SIMBAD \citep{simbad_2000}, \vizier\ \citep{vizier}, and a search of the literature. 
Next, the stars falling in the field plan were identified and  additional data for them assembled, including spectral types and 2MASS magnitudes \citep{Skrutskie_06_2mass}. 
For stars with suitable $H$ magnitudes, we checked the literature and attempted to sort by sub-type, e.g. classical Be star, B[e] star, unclassified emission star, etc. 
Rare and unusual Be subtypes were targeted first, followed by the 6 members of NGC\,7419.
%Targets from this program have targeting flag bit apogee2\_target3 = 12 set.
%%%%%%%%%%%%%%%%%%%%%%%%%%%%%%%%%%%%%%%%%%%%%%%%%%%%%%%%%%%%%%%%%%%%%%%%%%%%%%%%%%%%%%%%%%%%%%%%
%%%%%%%%%%%%%%%%%%%%%%%%%%%%%%%%%%%%%%%%%%%%%%%%%%%%%%%%%%%%%%%%%%%%%%%%%%%%%%%%%%%%%%%%%%%%%%%%
\subsection{Nearby Young Moving Groups} \label{anc:nymg}
%Juan Jose Downes
%Targets from this program have targeting flag bit apogee2\_target3 = 13 set.
This program observes known stellar members of nearby young moving groups (NYMGs) that fall within APOGEE-2N fields.
These observations of NYMGs will allow for an initial determination of possible differences in the abundances of various elements, which is a relevant issue in the study of the origin of these populations and their relationship with the other components of the solar neighborhood. 
Specifically, the proposed observations will allow us to:
    (1) demonstrate the potential for doing chemical tagging and spectral analysis of stars from NYMGs,
    (2) generate high quality ($S/N\sim$100) stellar spectral templates for relatively young (10-100 Myr) stars of spectral types between F0 and M4.5, which are vital for interpreting spectra of young cluster members because stars belonging to NYMGs are essentially free of the circumstellar material,
    (3) characterize these stars in terms of the abundances of several elements, as well as their radial velocities, \teff, veiling, and \logg, and 
    (4) look for possible differences between the abundances of several elements in different NYMGs and those for other populations of the solar neighborhood. 
These will permit chemical tagging exercises to identify additional members.
%%%%%%%%%%%%%%%%%%%%%%%%%%%%%%%%%%%%%%%%%%%%%%%%%%%%%%%%%%%%%%%%%%%%%%%%%%%%%%%%%%%%%%%%%%%%%%%%
%%%%%%%%%%%%%%%%%%%%%%%%%%%%%%%%%%%%%%%%%%%%%%%%%%%%%%%%%%%%%%%%%%%%%%%%%%%%%%%%%%%%%%%%%%%%%%%%
\subsection{Multiple Populations in NGC6791} \label{anc:multipops}
 % Doug Geisler
 % Targets from this program have targeting flag bit apogee2\_target3 = 15 set.
The old open cluster NGC\,6791 may be the first open cluster known to exhibit the Na-O anti-correlation, which is common among Galactic globular clusters.
Thus, NGC\,6791 may have drastic implications for cluster formation scenarios.  
By observing cluster members, predominantly selected to lie on the cluster horizontal branch where the Na-O anticorrelation is best measured, we will be able to characterize these multiple populations in unprecedented detail.   

Though globular clusters are now known to contain multiple stellar populations, all open clusters until now have shown a homogeneous composition; thus, open clusters are still regarded as simple stellar populations.  
\citet{geisler_2012} obtained high resolution optical spectra of 21 giants in the old, metal-rich, open cluster, NGC\,6791, and found evidence for the intrinsic variations in Na and O abundances that characterize multiple populations in Galactic globular clusters.
While this result is supported by optical and UV photometry \citep{monelli_2013}, previous spectroscopic data, including that from APOGEE-1, did not show evidence for Na-O variations in this cluster \citep{cunha_2015}. 
However, APOGEE-1 primarily targeted red giants and, as a result, excluded the (Na-poor) stars in the \citeauthor{geisler_2012} study. 

For this program, more than 20 confirmed members of NGC\,6791 were targeted, in addition to a significant sample of heretofore unstudied giants selected from 2MASS photometry.
These data will complement existing APOGEE observations for 40 members by prioritizing targets on the red clump, where the Na and O abundances are most easily measured.  
The resulting sample enables a self-consistent analysis of Na, O, Mg, and Al across a statistically meaningful sample.   

All of our targets lie in the existing APOGEE \texttt{FIELD} K21\_071+10.  
Due to $S/N$ requirements, targets were restricted to $H$ \textless 12, and drawn from several sources, in the following descending order of priority: First, all stars observed by \citet{geisler_2012} meeting our S/N requirements were included, except those that already have APOGEE spectra.
Additional candidates were drawn from literature studies which provide membership information \citep{platais_2011,tofflemire_2014,bragaglia_2014} with all confirmed non-members excluded.  
The remaining targets were drawn from 2MASS point sources consistent with the cluster upper RGB or RC in the color-magnitude diagram, and these are sorted in order of increasing radius from the cluster center out to the approximate tidal radius of $\sim$24$'$ \citep{dalessandro_2015}.  
%%%%%%%%%%%%%%%%%%%%%%%%%%%%%%%%%%%%%%%%%%%%%%%%%%%%%%%%%%%%%%%%%%%%%%%%%%%%%%%%%%%%%%%%%%%%%%%%
%%%%%%%%%%%%%%%%%%%%%%%%%%%%%%%%%%%%%%%%%%%%%%%%%%%%%%%%%%%%%%%%%%%%%%%%%%%%%%%%%%%%%%%%%%%%%%%%
\subsection{A Library of Reference Stars} \label{anc:refstars}
 % Melissa Ness 
 % Targets from this program have targeting flag bit apogee2\_target3 = 16 set.
The aim of this ancillary program is to build a comprehensive, state-of-the-art set of reference stars with APOGEE spectra. 
This will, at the most fundamental level, enable the best calibration and test of the APOGEE stellar parameter scale \citep[e.g., for the ASPCAP pipeline][]{garciaperez_2016} while also driving forward new reduction methods for spectra and techniques to derive labels both using data-driven models \citep[e.g., {\it The Cannon};][]{ness_2015,casey_2016}. 
With this science goal, we have the opportunity to place all stellar surveys, both current and future, on a unified stellar parameter and abundance scale, and place APOGEE as a leading and key driver of this critical science goal within the stellar community.

The current APOGEE reference objects only span a limited range of label space and the scale of individual abundances ([X/Fe]) cannot be tied to an external scale given the deficit of reference objects with these labels (e.g., from high resolution optical analysis).
We direct our science targeting goals to procure the best label set of calibration objects for a data-driven model, which is also aligned directly with any global calibration effort.
Our aims are to:
    (1) populate the critical gaps we have identified in the current label space of the set of reference stars (in \teff, \logg, and [Fe/H]),
    (2) extend the current reference objects to cooler stars along the main sequence and to hotter stars at the main sequence turnoff for a broad range in [Fe/H], 
and (3) obtain a set of stars with labels in [X/Fe] from high resolution analysis.
It is essential that these reference objects for calibration have high fidelity labels and be observed at high signal-to-noise ($S/N$ \textgreater 100). 
We also will identify targets that will provide overlap with other large-scale surveys.

%%%%%%%%%%%%%%%%%%%%%%%%%%%%%%%%%%%%%%%%%%%%%%%%%%%%%%%%%%%%%%%%%%%%%%%%%%%%%%%%%%%%%%%%%%%%%%%%
%%%%%%%%%%%%%%%%%%%%%%%%%%%%%%%%%%%%%%%%%%%%%%%%%%%%%%%%%%%%%%%%%%%%%%%%%%%%%%%%%%%%%%%%%%%%%%%%
\subsection{Faint Kepler Giants} \label{anc:faintkepler}
 % Marc Pinnsonneault 
 % Targets from this program have targeting flag bit apogee2\_target3 = 17 set.
The APOKASC survey \citep{apokasc} has successfully characterized targets in \kepler. 
One important limitation has been that the giants that were {\it deliberately} targeted by the \kepler\ mission were both bright and relatively local. 
In particular, this has resulted in a very small sample of astrophysically important metal-poor giants: only 9 out of $\sim$1900 in the first APOKASC sample \citep{apokasc} and only 36 out of $\sim$8200 in the full APOKASC sample \citep{apokasc2}. 

This project is collecting APOGEE observations for a newly discovered population of faint asteroseismic \kepler\ giants that were improperly classified as cool dwarfs in the \kepler\ Input Catalog \citep[KIC][]{brown_2011,huber_2014}, but found to be giants in later analyses \citep{mathur_2016}. 
The stars were identified as giants through the detection of asteroseismic oscillations in the \kepler\ light curves \citep{mathur_2016}. 
Our targets have excellent asteroseismic data with full $\sim$4 years of \kepler\ observations. 
The light curves, even for the faintest targets, can be used to obtain \logg, masses, and radii with precision comparable to that of the primary APOKASC survey \citep{apokasc,apokasc2}.

These stars are important both as tests of stellar physics and as stellar population tracers.
There are 820 targets spread over the 21 \kepler\ tiles. 
This fortuitous sample is not only the single best one for obtaining mass and age estimates for halo stars, but it will arguably remain so for the foreseeable future. 
The \ktwo\ and \tess\ missions are either focused on brighter targets or do not have sufficient photometric precision to detect oscillations for faint giants, so stars with large $R_{G}$ and $Z$ will be rare. 
In addition, their observations will be shorter by a factor of 4-12 than the \kepler\ observations ($\sim$80~days in the case of \ktwo\ and a maximum of $\sim$365~days for \tess). 
We note that there are currently documented problems with asteroseismic mass estimates for halo stars that were first discovered with APOKASC data \citep{epstein_2014}.  
However, there are multiple avenues currently being explored for higher precision and accuracy estimates; more specifically, the availability of \gaia\ parallaxes will provide powerful additional constraints. 

%%%%%%%%%%%%%%%%%%%%%%%%%%%%%%%%%%%%%%%%%%%%%%%%%%%%%%%%%%%%%%%%%%%%%%%%%%%%%%%%%%%%%%%%%%%%%%%%
%%%%%%%%%%%%%%%%%%%%%%%%%%%%%%%%%%%%%%%%%%%%%%%%%%%%%%%%%%%%%%%%%%%%%%%%%%%%%%%%%%%%%%%%%%%%%%%%
\subsection{W3/4/5 Star Forming Complexes} \label{anc:w345}
% Alexandre Roman Lopes
% Targets from this program have targeting flag bit apogee2\_target3 = 18 set.
%This ancillary program performed a spectroscopic study of the three massive star forming complexes W3, W4, and W5 (W3/4/5). 
%It is able to access, at the same time, the massive stellar population together with a significant fraction of the most massive young stellar objects (YSOs) there.
%Indeed, to date very few comprehensive highly homogeneous spectral studies of star forming complexes incorporating a large number of stars have been performed, particularly at high spectral resolution from which radial velocities can be measured. 
%These complexes also provide an ideal laboratory in which to study spatially segregated star formation and the triggering mechanisms, such as the collection and collapse of massive molecular clumps.

The W3/4/5 system in Cassiopeia is located in the Perseus arm of the Galaxy at 2~kpc and stretches in a direction orthogonal to our line-of-sight, making the geometry ideal for the investigation of triggered and sequential star formation. 
Recent studies have traced the distribution of young star populations across the three complexes and have provided clues on the roles played by triggering, local environment, and cloud structure.

The APOGEE-2 survey of W3/4/5 is paramount in understanding the process of massive cluster formation in our Galaxy, a major outstanding puzzle in current star formation research. 
APOGEE data will enable the age progression of stars across the complex to be inferred by observing the whole collection of OB associations in the complexes, the currently forming embedded clusters, the interfaces between Photo-Dissociation Region bubbles and active molecular ridges, sources associated with the bright rimmed clouds (pillar tips), and the surrounding interspersed populations. 
In this context, the spectra will aid in disentangling the ages and dynamical states of the clusters, measuring the kinematics of clusters and subsidiary groups, and comparing these to the main OB populations.
Together, this permits an unprecedentedly comprehensive study of the early evolution of cluster forming complexes.

The massive star sample was selected from the combination of NIR-MIR color CMDs and color-color based selection criteria following \citet{roman-lopes_2016}.
The catalogue was then cross matched with the known OB stars, using the SIMBAD database \citep{simbad_2000} together with the catalogues of spectral classification of stars in the direction of W345.
The YSO sample was selected based on two different approaches. 
The first used {\em Spitzer} colors combined with an X-ray detection that identifies probable young star members. 
These are traditionally subdivided into three categories: Class I sources (deeply embedded protostars with large IR-excesses), Class II sources (those with IR excesses typified by the presence of disks), and Class III sources (those lacking IR excesses, having had their disks dispersed). 
Targets were selected based on these criteria from the catalogs produced by \citet{koenig_2008}, \citet{koenig_2011}, and \citet{chauhan_2011} for W5, and \citet{rivera-ingraham_2011},  \citet{bik_2012}, and \citet{roman-zuniga_2015} for W3/4.
%%%%%%%%%%%%%%%%%%%%%%%%%%%%%%%%%%%%%%%%%%%%%%%%%%%%%%%%%%%%%%%%%%%%%%%%%%%%%%%%%%%%%%%%%%%%%%%%
%%%%%%%%%%%%%%%%%%%%%%%%%%%%%%%%%%%%%%%%%%%%%%%%%%%%%%%%%%%%%%%%%%%%%%%%%%%%%%%%%%%%%%%%%%%%%%%%
\subsection{The Galaxy's Evolved Massive Stars} \label{anc:evolvedstars}
% Targets from this program have targeting flag bit apogee2\_target3 = 19 set. 
This program targets the hottest, most massive and luminous evolved stars, and other rare stellar objects; these targets include, super-giants, Wolf-Rayet stars, transitional WN stars, and O-stars. 
The variability of their rich emission and absorption line spectra can be studied via multiple epoch observations. 
Radial velocities and line profile analysis will be performed, and multiplicity investigated where feasible. 
The high signal-to-noise and the high spectral resolution provided by
APOGEE-2 observations render the cores of some blended strong lines to be partially resolved (e.g., H-He blends), and the weaker lines of HeI and HeII along with higher ionization states of other elements (e.g., NIV and NV lines) to be detected.
Mass loss rates and other physical parameters can be derived by modeling the spectra, allowing the stars complex winds and circumstellar environments to be explored.
These spectral types include supergiant and emission line stars, along with other rare objects. 
We used SIMBAD \citep{simbad_2000} and the online Wolf Rayet star catalog \citep{wolfrayet_cat}\footnote{\url{https://heasarc.gsfc.nasa.gov/W3Browse/all/wrcat.html}} to select targets in APOGEE fields. 

%%%%%%%%%%%%%%%%%%%%%%%%%%%%%%%%%%%%%%%%%%%%%%%%%%%%%%%%%%%%%%%%%%%%%%%%%%%%%%%%%%%%%%%%%%%%%%%%
%%%%%%%%%%%%%%%%%%%%%%%%%%%%%%%%%%%%%%%%%%%%%%%%%%%%%%%%%%%%%%%%%%%%%%%%%%%%%%%%%%%%%%%%%%%%%%%%
\subsection{APOGEE Reddening Survey} \label{anc:redsurvey}
%Targets from this program have targeting flag bit apogee2\_target3 = 20 set. 
The APOGEE Reddening Survey target selection is designed to be as similar to the main APOGEE target selection as possible such that the two surveys will be statistically compatible.
We adopt the exact criteria as for the main red star sample 1-visit designs described by \citetalias{zasowski_2013}.  

We add slightly to the main APOGEE target criteria to improve the selection function for reddening studies.  
Three major additions are made: 
    (1) we prioritize targets in regions of high or interesting reddening,
    (2) we demand that the targets have optical magnitudes from PS1 in at least two of the $griz$ bands, to study the optical reddening law, 
    (3) we demand that the RJCE-based \ejk\ is greater than 0.2~mag, to avoid observations of unreddened stars.
The first criterion is completely independent of the photometry of the stars and only prioritizes stars based on their proximity to stars with $R_{\rm V}$\textgreater\ 4 \citep{fitzpatrick_2007}, or because the star is in a significantly reddened region in a nearby molecular cloud.  
The second criterion is intended to ensure that the targets can be used to infer the shape of the reddening law in those directions.  
The classical parameterization of reddening laws in terms of $R_{\rm V}$ requires measuring the optical slope of the reddening law.  We would ideally require $g-r$ magnitudes for each source, but that constraint limits us to $E(B-V)$ less than about 2.5~mag, preventing us from probing the densest regions.  
Therefore, we require only that two of the $griz$ bands are well measured in the PS1 survey, accepting that only an $i-z$ color may be measured in the densest regions. Finally, the third criterion, that \ejk$_{RJCE}$\textgreater\ 0.2~mag, is made simply to ensure that no stars of very little reddening are observed; these stars are not useful for determining the shape of the reddening law.
%%%%%%%%%%%%%%%%%%%%%%%%%%%%%%%%%%%%%%%%%%%%%%%%%%%%%%%%%%%%%%%%%%%%%%%%%%%%%%%%%%%%%%%%%%%%%%%%
%%%%%%%%%%%%%%%%%%%%%%%%%%%%%%%%%%%%%%%%%%%%%%%%%%%%%%%%%%%%%%%%%%%%%%%%%%%%%%%%%%%%%%%%%%%%%%%%
\subsection{M Dwarf \kepler\ Objects of Interest} \label{anc:mdwarfs_koi}
% Targets from this program have targeting flag bit apogee2\_target3 = 21 set. 

Stellar parameters such as mass ($M_{*}$) and chemical composition play an important role in influencing planetary formation and the nature of the resultant planetary-system architecture.
The abundances of key metals, such as O, Mg, Si, or Fe, affect the nature and structure of the planets themselves. 
Unlike F, G, or K dwarfs, very little is known about the detailed chemical abundance distributions of M-dwarfs in general, and for planet-hosting M-dwarfs in particular, due to the complex nature of their optical spectra.  This situation can be improved greatly by shifting the analysis into the near-infrared where the high-density of molecular lines drops dramatically.  
High-resolution NIR spectra are the preferred dataset for abundance analyses of the cool M-dwarfs, thus APOGEE is well-positioned to contribute to this field.

M-dwarfs are the most numerous stars in the Galaxy and are becoming an increasingly important component in explanet searches using the transit and radial-velocity methods.
Their popularity is to due to the enhanced detectability of small planets owing to their low stellar masses, low luminosities, and small stellar radii. 
The focus of this program is to both derive fundamental stellar parameters and to pioneer a study of the detailed chemical abundance distributions in M-dwarfs with APOGEE. 
The determination of accurate physical parameters for stars that host exoplanets is a crucial step in characterizing the size and nature of the planets themselves; {\it one can only know the planet to the level that the host star is known}.

This project will use APOGEE to observe a sample of M-dwarf \kepler\ systems and \kepler\ Objects of Interest (KOI) to derive both accurate stellar parameters and detailed chemical abundances.
Adding planet-hosting M-dwarfs to APOGEE opens a new window into studying a class of stars that play an increasingly important role in both \ktwo\ and \tess. 
APOGEE is well-positioned to lead the way in chemically  categorizing M-dwarf exoplanetary systems.

The NASA Exoplanet Archive\footnote{\url{https://exoplanetarchive.ipac.caltech.edu/}} was searched for all \kepler\ Objects of Interest having input catalog effective temperatures of \teff\ \textless\ 4300~K. 
The primary emphasis for this project is to push abundance analyses to the M-dwarfs, but some overlap is allowed with late K-dwarfs.  
The exact \teff\ boundary between dwarf spectral types K and M is somewhat fuzzy, but is near \teff$\sim$4000~K (the transition from K7V to M0V).  
A \teff\ cut at 4300~K will include types K6 and K7 and these targets will provide overlap with abundance analyses from optical high-resolution spectra.
%%%%%%%%%%%%%%%%%%%%%%%%%%%%%%%%%%%%%%%%%%%%%%%%%%%%%%%%%%%%%%%%%%%%%%%%%%%%%%%%%%%%%%%%%%%%%%%%
%%%%%%%%%%%%%%%%%%%%%%%%%%%%%%%%%%%%%%%%%%%%%%%%%%%%%%%%%%%%%%%%%%%%%%%%%%%%%%%%%%%%%%%%%%%%%%%%
\subsection{AGB Stars and post-AGB Stars in the Galactic Plane} \label{anc:agbstars}
%Targets from this program have targeting flag bit apogee2\_target3 = 22 set. 
This project aims to study the chemical abundances in a sample of Galactic carbon-rich asymptotic giant branch (AGB) stars, post-AGB stars, and planetary nebulae (PNe). 
The spectral characterization of AGB stars in the $H$ (especially in those stars that are obscured or very faint in the optical range) allows the determination of evolution status and constrain models of stellar nucleosynthesis. 
A second objective is to enable study of how the chemical abundances are affected by circumstellar effects, in a more realistic approximation to the complex atmospheres of AGB stars \citep[see, e.g.,][]{zamora_2014}.
An understanding of the post-AGB evolutionary phase remains elusive, in part due to their scarcity -- only $\sim$300 known objects are very likely Post-AGB stars \citep[][]{szczerba_2007}, with most of the these not having Post-AGB sub-classifications. 
APOGEE spectra of Post-AGB stars will determine the CNO and s-process abundances that signal if ``hot bottom burning'' was activated; this detemination serves to discriminate between the more massive and less massive post-AGBs. 
Finally, we also would like to attempt the challenge of detecting s-process element emission lines for bright PNe to constrain their physical parameters and identify their progenitors.

Most of the stars ($M_{*}$ \textless\ 8 $M_{\odot}$) in the Universe end their lives with a phase of strong mass loss (up to $\sim$10$^{-4}$ to 10$^{-5}$ $M_{\odot}$ year$^{-1}$) on the Asymptotic Giant Branch (AGB), evolving as Post-AGB stars just before becoming Planetary Nebulae (PNe).
AGB stars are among the main contributors to the chemical enrichment of the interstellar medium (ISM) where new stars and planets are born, and thus to the chemical evolution of galaxies. 
More specifically, the more massive ($M_{*}$ $>$ 4-5 $M_{\odot}$) AGB stars form very different isotopes (such as $^{87}$Rb, $^{7}$Li, $^{14}$N) from the isotopes formed by lower mass AGB stars and supernova explosions, as a consequence of different dominant nuclear reaction mechanisms \citep{abia_2001,garcia-hernandez_2007}. 
Stars evolving from the AGB phase to the PNe stage also form complex organic molecules (such as polycyciclic aromatic hydrocarbons or PAHs, fullerenes, and graphene) and inorganic solid-state compounds. 
Thus, the $\sim$10$^{2}$ to $\sim$10$^{4}$ years of evolution following the end of the AGB phase represent a most fascinating laboratory for astrochemistry. 

In AGB stars, the CNO elemental and isotopic abundances, aluminum as well as the abundances of several $\alpha$-elements together with s-process abundances (Rb, Sr, Y, Cs, Nd, Ce) can be measured with APOGEE spectra.
Complementary, circumstellar effects on the chemical abundances in AGB stars will be also investigated and we will attempt the identification of high excitation s-process lines for bright PNe (e.g., [Kr {\sc III}], [Se {\sc IV}], [Rb {\sc IV}]).

The targets were selected using four different sources:
The General Catalogue of Galactic Carbon stars \citep{alksnis_2001}, 
The Strasbourg-ESO Catalogue of Galactic Planetary Nebulae \citep{acker_1992},
The Torun catalogue of Galactic post-AGB and related objects \citep{szczerba_2007} and the PhD thesis of Pedro Garcia-Lario (1992).
Targets were selected with 7.0 \textless\ $H$ \textless\ 13.5.
%%%%%%%%%%%%%%%%%%%%%%%%%%%%%%%%%%%%%%%%%%%%%%%%%%%%%%%%%%%%%%%%%%%%%%%%%%%%%%%%%%%%%%%
%%%%%%%%%%%%%%%%%%%%%%%%%%%%%%%%%%%%%%%%%%%%%%%%%%%%%%%%%%%%%%%%%%%%%%%%%%%%%%%%%%%%%%%
\section{2017 Ancillary Programs} \label{sec:ancillary2017}
% https://trac.sdss.org/wiki/ApprovedAncillaryPrograms2017
Ancillary programs were solicited via an open call across the SDSS-IV collaboration (i.e., not limited to just APOGEE) and were awarded based on a review that focused on technical and scientific feasibility.
The 2017 proposals were due in February 2017, selected by peer review during March/April 2017, and implemented starting in August/September 2017. 
Because the bulk of the Main Survey observations had already been drilled when implementation began, the 2017 Ancillary Science Programs had more limited access to certain parts of the sky.

The timing of the 2017 Ancillary Call and the formulation phases of the Bright Time Extension, while distinct processes, meant that programs were merged during implementation for those cases where the scientific or targeting goals were aligned. 
Such instances have been noted in the main text, but given that these projects were approved based on their scientific merits independent of the Bright Time Extension, we include them here.
%%%%%%%%%%%%%%%%%%%%%%%%%%%%%%%%%%%%%%%%%%%%%%%%%%%%%%%%%%%%%%%%%%%%%%%%%%%%%%%%%%%%%%%
%%%%%%%%%%%%%%%%%%%%%%%%%%%%%%%%%%%%%%%%%%%%%%%%%%%%%%%%%%%%%%%%%%%%%%%%%%%%%%%%%%%%%%%
\subsection{M33 Globular Clusters} \label{anc:m33}
% Targets from this program have targeting flag bit apogee2\_target3 = 23 set.
M33 has both young and old globular clusters (GCs) that span from 1 to 12 Gyr in age \citep{chandar_2006,beasley_2015}. We might expect both old and young GCs to show the CNO and Na/O anomalies -- extragalactic GCs have been inferred to host Na/O anti-correlations since many have high [Na/Fe] as measured from integrated light \citep{colucci_2014, sakari_2015}.
Surprisingly, abundance variations have not been observed in young to intermediate-aged LMC GCs \citep[e.g.,][]{sakari_2017}. 
This motivated a program to collect integrated-light spectra on young and old GCs in the Triangulum Galaxy (M\,33) to test this hypothesis. 

M\,33 is the third most massive galaxy in the Local Group and it is is much less studied than the Milky Way (MW), Andromeda (M31) and the Magellanic Clouds (MCs).
The Pan-Andromeda Archaeological Survey (PANDAS) revealed large-scale substructures of low surface brightness, including arcs, stream and globular clusters, connecting the M\,31 and M\,33 galaxies \citep{huxor_2011}. Substructure in our own Galactic halo reveals its merger history, and certain globular clusters (GCs) appear to be associated with specific accreted satellites (e.g., the Sagittarius dwarf spheroidal galaxy).
Hence, Globular Cluster Systems (GCSs) can be used as tracers of the formation processes and the assembly history of a galaxy. 

To date, integrated-light spectroscopy in the optical has been obtained for only twelve M33 GCs \citep{beasley_2015}; no such data exist in the NIR.
Yet the APOGEE $H$-band wavelength coverage confers some significant advantages for integrated-light  spectroscopy: insensitivity to hot stars, but high sensitivity to red giant branch (RGB) and asymptotic giant branch (AGB) stars, a feature that simplifies integrated-light  analysis \citep{schiavon_2004,sakari_2014}. 
The $H$ also offers access to some chemical features not easily available to optical spectroscopy, in particular, the presence of strong molecular lines of CN, CO and OH, which enable determinations of C, N and O abundances \citep{smith_2013,sakari_2016}Other useful lines for this work are those for Mg, Al, Si, Ca and Ti. 
APOGEE spectra also bring the opportunity to probe multiple populations in GCs, the ability to detect [O/Fe], and to probe directly the Na/O anti-correlation \citep{sakari_2016}. 

We selected our targets from the catalogue with homogeneous $UBVRI$ photometry of 708 M\,33 star clusters and cluster candidates based on archival images from the Local Group Galaxies Survey \citep{fan_2014}.
We use the photometry for the M\,31 GCs \citep[][their table 1]{sakari_2016} to convert from the $V$ to $H$ in the \citet{fan_2014} catalogue. 
We select all the clusters with 12.5 \textless\ $H$ \textless\ 15 for a total of 132~clusters; these also have initial estimates of metallicity, age and mass from \citet{fan_2014}.
%%%%%%%%%%%%%%%%%%%%%%%%%%%%%%%%%%%%%%%%%%%%%%%%%%%%%%%%%%%%%%%%%%%%%%%%%%%%%%%%%%%%%%%
%%%%%%%%%%%%%%%%%%%%%%%%%%%%%%%%%%%%%%%%%%%%%%%%%%%%%%%%%%%%%%%%%%%%%%%%%%%%%%%%%%%%%%%
\subsection{Cepheids Calibrators} \label{anc:cephcalib}
The goal of this program is to provide homogeneous chemical characterization of Galactic Cepheids that have multi-wavelength photometric characterization and will have sub-percent precision trigonometric parallaxes from \gaia. 
Multiple epochs of APOGEE spectra were obtained for each star using the NMSU 1-meter fiber feed. Chemical abundances will be used to calibrate the metallicity effects in the period-luminosity relationship for ten photometric bands and will provide high signal-to-noise templates over multiple phase points for other Cepheid-based programs in APOGEE-2.

For nearly a century, Cepheid variables have been the de facto standard candle buttressing the extra-galactic distance scale. While nearby dwarf galaxies provide excellent probes of the Leavitt law at low metallicity, the overall metallicity sensitivity of the Leavitt Law remains relatively poorly unconstrained due to the lack of high metallicity calibrators (the Galactic field Cepheids) with independent distances. 
The trigonometric parallaxes delivered by \gaia\ are the first to provide these independent distances for a large sample of Cepheids in the Galaxy. 
This NMSU 1-meter fiber extension ancillary project targets a set of the most well characterized Cepheids in the Galaxy that are poised to be the best calibrators in \gaia.
These stars serve not only an important scientific role in themselves, but will support other Cepheid projects in APOGEE-2 and SDSS-V.

The first goal of the program is to calibrate the metallicity dependency of the Leavitt Law and utilize the \citet{fouque_2007} sample of Cepheids with magnitudes in eight optical and near-infrared bandpasses, to which we add newer observations in the near-infrared and mid-infrared for a total of ten bands. 
This sample includes the most nearby Cepheids that will, in turn, have the best \gaia\ parallax measurements. 
The stars will be sampled over a number of epochs to study the stability of the metallicity measurements over phase, which helps to place potential single phase measurements in context. 
A second goal of the project is to calibrate the \citet{scowcroft_2016b} mid-infrared color-metallicity relationship using a homogeneous metallicity characterization spanning 2 dex in [Fe/H] (in combination with other APOGEE-2 Cepheid programs).
This Galactic sample will have the best mid-infrared light curves for Cepheids \citep[][]{monson_2012}. 

We selected the sample from \citet{fouque_2007}, which has homogeneously derived magnitudes in eight bands, to which we add 3.6\micron\ and 4.5\micron\ measurements from the Carnegie Hubble Program \citep{monson_2012}, better NIR sampling \citep{monson_2011}, and revised line-of-sight extinction measurements \citep{madore_2017}. 

The \citet{fouque_2007} multi-wavelength sample contains 60 stars with 2MASS apparent magnitudes in the range 1.8 \textless\ $H$ \textless\ 8.2 (note that for bright stars the 2MASS magnitude uncertainties are 50\% of the pulsation amplitude of the star). Restricting to those objects with $\delta$ \textgreater\ $-30\degs$, there are 34 remaining sources that range in magnitude over 1.8 \textless\ $H$ \textless\ 7.7. We prioritized the sources by putting those sources with TGAS parallaxes as top priority and those without TGAS parallaxes as lower priority \citep{gaia_dr1}. 
Due to saturation limits with \gaia, this distinction amounts to down-weighting the brightest sources (in the range  1.8 \textless\ $H$ \textless\ 5). 

%%%%%%%%%%%%%%%%%%%%%%%%%%%%%%%%%%%%%%%%%%%%%%%%%%%%%%%%%%%%%%%%%%%%%%%%%%%%%%%%%%%%%%%
%%%%%%%%%%%%%%%%%%%%%%%%%%%%%%%%%%%%%%%%%%%%%%%%%%%%%%%%%%%%%%%%%%%%%%%%%%%%%%%%%%%%%%%
\subsection{Brown Dwarfs} \label{anc:browndwarf}
% Targets from this program have targeting flag bit apogee2\_target3 = 24 set.
This program aims to build a library of late-M and L dwarf APOGEE spectra for the purpose of (1) measuring spatial and rotational kinematics for the nearby population and (2) extending spectral modeling for abundance analysis to \teff\ \textless\ 2700~K. Sources with spectral types later than M7 and 11 \textless\ $H$ \textless\ 14.5  were selected. The desired scientific outcome is improved spectral models across the hydrogen-burning limit, with an eye toward improving characterization of potential low-mass terrestrial exoplanet host systems.

The transition between the M dwarf and L dwarf spectral classes for very low mass, $M_{*}$ $\leq$ 0.1 $M_{\odot}$ \citep[VLM;][]{kirkpatrick_2005}, stars and brown dwarfs is a critical benchmark in studies of Galactic populations, substellar evolution, stellar magnetic field generation, angular momentum evolution, star formation processes and history, and exoplanet atmospheric chemistry and habitability. 
This transition spans the temperature range for atmospheric condensate formation, a focus of current star and exoplanet spectral modeling work \citep{helling_2008,marley_2010}; and the decoupling of atmospheres from internally-generated magnetic fields \citep{mohanty_2002}, which results in sharp declines in the incidence and strength of H$\alpha$ and X-ray non-thermal emission \citep{west_2011} -- albeit with dramatic exceptions \citep[e.g.,][]{schmidt_2014}, and reduced angular momentum transport resulting in rotation periods as short as 1-3 hr \citep{konopacky_2010,irwin_2011}. 
Late-M and L dwarfs also span the hydrogen-burning mass limit ($M_{*}=0.07~M_{\odot}$) and are the densest hydrogen-rich bodies known, probing a minimum in the mass-radius relationship and potentially exotic states of matter \citep{burrows_2001}. 
The long lifetimes of these sources (trillions of years) and their limited fusion and full convective mixing make them ideal time capsules for Galactic star formation and chemical evolution history.
Finally, their small radii ($R_{*} \sim$0.1 $R_{\odot}$), close-in habitable zones ($d R_{*}^{-1} \sim$ 10-30), and apparent preference for forming Earth-sized planets \citep{dressing_2013} make VLM dwarfs ideal targets for probing Galactic habitability through the transit method  \citep[e.g., Trappist-1,][]{gillon_2017}.

Characterizing the physical and populative properties of local VLM dwarfs is optimally accomplished with high-resolution infrared spectroscopy. 
APOGEE's resolution and sensitivity are well matched to the infrared magnitudes and rapid rotation of these objects (up to 80 \kms), while its broad spectral coverage is critical for measuring atmospheric abundances, probing both bulk composition and atmospheric chemical dynamics. 
However, late-M and L dwarfs fall below current APOGEE stellar modeling temperature limits \citep[][]{garciaperez_2016,schmidt_2016,souto_2017}. 
DR14 included $\sim$25 VLM dwarfs \citep{holtzman_2018} and this provided only a narrow scope for examining population properties. 
This program aims to increase the observe sample 20-fold by targeting up to 448 M6-L6 dwarfs. 
These data will provide radial velocities, which, combined with PanSTARRS/\gaia\ proper motions, will: yield precise kinematics (\textless 1-2 \kms); measure rotational velocities down to 5 \kms, to examine angular momentum evolution and magnetic activity trends in conjunction with ancillary optical spectral observations; enable searches for close-separation companions down to planetary masses through RV variability measurements; improve spectral modeling in the low-temperature regime; and characterize planet-host candidates targeted by MEarth \citep{nutzman_2008}, SPECULOOS \citep{gillon_2013}, and \tess\ \citep{ricker2015}, among others.

Targets were selected from compilations of known VLM dwarfs, including 
    Dwarf Archives\footnote{\url{DwarfArchives.org}}; 
    Pan-STARRS DR1 \citep{best_2017}; 
    BOSS Ultracool Dwarfs \citep{schmidt_2015}, 
    BASS \citep{gagne_2015}, 
    LaTE-MoVeRS \citep{theissen_2017}, 
    MEarth \citep{newton_2017}, 
    SPECULOOS \citep{gillon_2013}, and 
    the \tess\ Input Catalog \citep{stassun_2018}. 
Targets were selected to have reported spectral types later than M7 (\teff\ \textless 2700~K) and 11 \textless $H$ \textless 14.5. 
%%%%%%%%%%%%%%%%%%%%%%%%%%%%%%%%%%%%%%%%%%%%%%%%%%%%%%%%%%%%%%%%%%%%%%%%%%%%%%%%%%%%%%%
%%%%%%%%%%%%%%%%%%%%%%%%%%%%%%%%%%%%%%%%%%%%%%%%%%%%%%%%%%%%%%%%%%%%%%%%%%%%%%%%%%%%%%%
\subsection{Distant Halo Giants} \label{anc:distanthalo}
%PI: Paul Harding 
%Targets from this program have targeting flag bit apogee2\_target3 = 25 set. 
As discussed in Section \autoref{sec:halo}, the initial targeting strategy of APOGEE unfortunately yielded relatively few spectra for distant halo stars ($D$\textgreater10~kpc).
On the other hand, the SEGUE
catalog consists of 6000 confirmed K~giants \citep[][]{xue_2014}, many of which are at large heliocentric distances.
The 0.5 \textless\ ($g$-$r$) \textless\ 1.3 color range is within the temperature range of the APOGEE pipeline. 
Observations that deliberately target known SEGUE giants will dramatically increase the number of distant stars observed by APOGEE and, in particular, those in the distant halo. 
By comparison there were approximately 60 halo giants observed by APOGEE in DR14 \citep[based on metallicity, velocity and distance;][]{holtzman_2018,dr14} from approximately 100 halo or stream plates observed by APOGEE to that point in the survey \citepalias[][]{zasowski_2013,zasowski_2017}.
The halo is lumpy and so the number of giants within the APOGEE magnitude varies greatly, but by targeting known stars we will confidently boost this sample of stars and ensure that APOGEE obtains a chemical fingerprint of this Galactic component.
This program was subsumed, fully, into the BTX Halo targeting described in \autoref{sec:new_halo_targeting}. 
%%%%%%%%%%%%%%%%%%%%%%%%%%%%%%%%%%%%%%%%%%%%%%%%%%%%%%%%%%%%%%%%%%%%%%%%%%%%%%%%%%%%%%%
%%%%%%%%%%%%%%%%%%%%%%%%%%%%%%%%%%%%%%%%%%%%%%%%%%%%%%%%%%%%%%%%%%%%%%%%%%%%%%%%%%%%%%%
\subsection{The Young Galaxy} \label{anc:younggal}
%PI Inno
%Targets from this program have targeting flag bit apogee2\_target3 = 26 set. 
\gaia\ DR2 has increased our understanding of the Galaxy. 
However, a complete description of the Galactic thin disk remains challenging, because even \gaia\ has severe limitations in the Galactic plane, where dust extinction dims stars below the detection limit at distances greater than 5~kpc. 
We target faint/distant young Cepheids that were identified in the PanSTARRs \citep[PS1;][]{panstars_dr1} multi-epoch catalog to sample the farthest obscured reaches of the Galactic disk.

Cepheids are powerful probes of both the structure and the recent history of the Milky Way: 
they are luminous and can be seen to great distances, even through substantial dust extinction;
their individual ages and distances can be precisely determined from their periods; and, with ages of 20-150 Myrs, they are young stars but they are relatively cool and hence their spectra show a rich metal-line absorption spectrum, from which many element abundances can be determined. 
The new PanSTARRs (PS1) catalog of variable stars is currently still under construction.

We selected a list of targets determined on the basis of the following steps:
[1] identification of Cepheids-like variables in the entire PS1 database, by using the variability parameters by \citet{Hernitschek_2016} and colors defined on the basis of magnitudes in the $grizy$ (PS1) and near-infrared bands from 2MASS \citep{Skrutskie_06_2mass} and ALLWISE \citep{cutri_2014}; 
[2] multi-band template fitting of the sparsely sampled PS1 light curves for all the 200,000 sources selected and determination of pulsation period and other parameters (including distance modulus and extinction);
[3] use of machine-learning techniques to extract the variables classified as Cepheids with highly significant probability  (i.e., higher classification score).
The targets existing in APOGEE fields will be targeted for multi-epoch observations as is feasible.
%%%%%%%%%%%%%%%%%%%%%%%%%%%%%%%%%%%%%%%%%%%%%%%%%%%%%%%%%%%%%%%%%%%%%%%%%%%%%%%%%%%%%%%
%%%%%%%%%%%%%%%%%%%%%%%%%%%%%%%%%%%%%%%%%%%%%%%%%%%%%%%%%%%%%%%%%%%%%%%%%%%%%%%%%%%%%%%
\subsection{Kapteyn Selected Area 57} \label{anc:sa57}
%PI: Steve Majewski 
%Targets from this program have targeting flag bit apogee2\_target3 = 27 set.

Selected Area 57 (SA57) is the nearest of Kapteyn's Selected Areas (an ambitious program to systmatically study the Milky Way first organized by Jacobus Kapteyn  in the year 1906) to the North Galactic Pole. 
SA57, therefore, lies in a direction relatively free of reddening and where ``in situ'' halo stars can most easily be accessed. 
SA57 has traditionally played a significant role for a variety of astronomical studies, from probing the vertical density laws of the Galactic stellar populations to deep studies of galaxies and quasars, the latter because of the minimal stellar foreground.    
As a result of more than a century of interest in this direction of the sky, SA57 has received extensive attention by surveys of photometry, astrometry and spectroscopy --- but never, to this point, high-resolution spectroscopy.  

The APOGEE observations here are intended to tie state-of-the-art chemistry of giant stars in SA57 to the rich legacy of previous observations in this field.  
Stars were selected following the criteria of main red star sample \citepalias{zasowski_2013,zasowski_2017}.
Four overlapping plate centers (N, S, E, W) were targeted yielding three cohorts in a ``wedding-cake'' arrangement that follows the magnitude limits of the main red star sample, but also takes into account the overlap between the plates to reach the full $S/N$.
%%%%%%%%%%%%%%%%%%%%%%%%%%%%%%%%%%%%%%%%%%%%%%%%%%%%%%%%%%%%%%%%%%%%%%%%%%%%%%%%%%%%%%%
%%%%%%%%%%%%%%%%%%%%%%%%%%%%%%%%%%%%%%%%%%%%%%%%%%%%%%%%%%%%%%%%%%%%%%%%%%%%%%%%%%%%%%%
\subsection{Tidally Synchronized Binaries} \label{anc:tlockbinary}
%PI: Greg Simonian 
%Targets from this program have targeting flag bit apogee2\_target2 = 12 set. 
It is now common knowledge that more stars have binary companions than are truly isolated stars \citep{Raghavan_2010}. 
While the formation of single-star systems is still under active study, the formation and evolution of binaries demonstrates rich behavior and, yet, is even more poorly understood. 
Binary formation is governed by the interplay between fragmentation and the circumprotobinary disk
\citep{Artymowicz_1991,Artymowicz_1994,Bonnell1994,Bate_2000,gunther_2002}, while the subsequent evolution occurs due to dynamical interactions in the cluster \citep{Bodenheimer_2001,clarke_2001}.
Unfortunately, obtaining data on the field binary population to compare to evolutionary models is notoriously difficult.
The selection function for field binaries is so heterogeneous and biased that statistical analyses do not provide strong constraints. 
Binaries in clusters are more easily interpretable; however, binaries currently in clusters may not be representative of the escaped field population.

The aim of this project is to collect a verified sample of tidally-synchronized binaries selected from the rapid rotators in the \kepler\ field. 
The modulation amplitude in tidally-synchronized binaries should be easily measured because rapid rotation enhances starspot activity \citep{Basri_1987}.
Applying solar neighborhood binary properties \citep{Raghavan_2010} to the \kepler\ field predicts around 300 non-eclipsing tidally-synchronized binaries showing detectable rotational modulation, compatible with the actual number of rapid-rotators found.
%%%%%%%%%%%%%%%%%%%%%%%%%%%%%%%%%%%%%%%%%%%%%%%%%%%%%%%%%%%%%%%%%%%%%%%%%%%%%%%%%%%%%%%
%%%%%%%%%%%%%%%%%%%%%%%%%%%%%%%%%%%%%%%%%%%%%%%%%%%%%%%%%%%%%%%%%%%%%%%%%%%%%%%%%%%%%%%
\subsection{SubStellar Companions} \label{anc:substellar}
%PI: Nick Troup 
%Targets from this program have targeting flag bit apogee2\_target3 = 29 set. 
With the ever-growing sample of companions (both stellar and substellar) being discovered around stars from ever more diverse environments, population studies comparing host environments are now possible.
For example, recent work suggests the hot Jupiter occurrence rate in M67 is higher than in the field \citep{Brucalassi_2016}, although it is not clear whether this is the norm for open clusters generally.
We aim to expand upon the substellar companion science program in APOGEE-2 \citepalias{zasowski_2017} to include stars in a variety of cluster environments with a wide range of ages, metallicities, and densities.
These observations will help constrain the role  environment has to play in the formation and evolution of stellar systems containing companions of various masses. 
Critically, this program will both extend the time baseline for targets observed in APOGEE-1 and increase the number of RV measurements to enable robust orbital fits in the final dataset. 
The \texttt{FIELD} names from APOGEE-1 are: IC348\_RV, M67\_RV, and M3\_RV. 

%%%%%%%%%%%%%%%%%%%%%%%%%%%%%%%%%%%%%%%%%%%%%%%%%%%%%%%%%%%%%%%%%%%%%%%%%%%%%%%%%%%%%%%
%%%%%%%%%%%%%%%%%%%%%%%%%%%%%%%%%%%%%%%%%%%%%%%%%%%%%%%%%%%%%%%%%%%%%%%%%%%%%%%%%%%%%%%
\subsection{M dwarfs in \ktwo} \label{anc:mdwarfk2}
%PI: Verne Smith
% Targets from this program have targeting flag bit apogee2\_target3 = 28 set. 
M-dwarf stars are increasingly important objects in both the fields of exoplanet searches, due to the emphasis on cool dwarfs in the target lists of \ktwo\ and \tess, as well as studies of Milky Way stellar populations, as these are the most numerous stars in the Galaxy. 
M-dwarfs are notoriously difficult to analyze via optical spectroscopy, due to intense molecular line absorption; this difficulty is alleviated greatly at infrared wavelengths. 
Our team is using APOGEE spectra of M dwarfs to pioneer the derivation of fundamental stellar parameters and detailed chemical compositions for these types of stars \citep{souto_2017,Souto_2020}. 
M dwarfs are a particularly important component in both transit and radial-velocity searches for exoplanets, thanks to the enhanced detectability of small planets due to low stellar masses, low luminosities, and small stellar radii. 
This work opens a new window into a class of planet-hosting stars that will play an increasingly important role in ongoing and future planet surveys and missions, such as \ktwo\ and \tess. 
Targets from this program are shown in \autoref{fig:k2}a.
%%%%%%%%%%%%%%%%%%%%%%%%%%%%%%%%%%%%%%%%%%%%%%%%%%%%%%%%%%%%%%%%%%%%%%%%%%%%%%%%%%%%%%%
%%%%%%%%%%%%%%%%%%%%%%%%%%%%%%%%%%%%%%%%%%%%%%%%%%%%%%%%%%%%%%%%%%%%%%%%%%%%%%%%%%%%%%%
%\subsection{MaStar} \label{anc:mastar}
% Never implemented.
%%%%%%%%%%%%%%%%%%%%%%%%%%%%%%%%%%%%%%%%%%%%%%%%%%%%%%%%%%%%%%%%%%%%%%%%%%%%%%%%%%%%%%%
%%%%%%%%%%%%%%%%%%%%%%%%%%%%%%%%%%%%%%%%%%%%%%%%%%%%%%%%%%%%%%%%%%%%%%%%%%%%%%%%%%%%%%%
\subsection{Local Group Stellar Populations in Integrated Light} \label{anc:m31}
% Targets from this program have targeting flag bit apogee2\_target3 = 30 set. 

The MW provides us with a unique opportunity to constrain galaxy evolution using individual interstellar clouds and single stars, both of which serve as the fundamental building blocks of galaxies. 
The results from these stellar measurements in the Galaxy have guided our understanding of the chemo-dynamical processes that impact galaxies on all spatial scales.
This understanding, then, relies on the assumption that the processes required to produce these spatial and chemical distributions are universally available and commonplace for galaxies generally. M\,31 provides one opportunity for testing this assumption.

With 3\arcsec\ fibers at APO, APOGEE-2 cannot resolve individual stars in M\,31, so integrated light observations are required to identify multiple stellar populations with distinct chemistry and dynamics.
APOGEE collected integrated light spectra of the inner disk of M\,31, spaced in a grid of points within $R_{\rm M31} < 5$~kpc, and of the centers of M\,32 and M\,110.  
All fiber positions were visually checked against optical and near-IR imaging and shifted by up to 10\arcsec\ to ensure that no bright stars, clusters, or emission line regions fell within the aperture.  
Fibers that could not be fit onto these targets due to fiber packing and plugging limitations were placed on positions observed by the MaNGA survey and on Luminous Blue Variable stars in the field of view.

%%%%%%%%%%%%%%%%%%%%%%%%%%%%%%%%%%%%%%%%%%%%%%%%%%%%%%%%%%%%%%%%%%%%%%%%%%%%%%%%%%%%%%%
\section{Glossary} \label{sec:glossary}
%%%%%%%%%%%%%%%%%%%%%%%%%%%%%%%%%%%%%%%%%%%%%%%%%%%%%%%%%%%%%%%%%%%%%%%%%%%%%%%%%%%%%%%
%%
% Synced this up with the south paper on 04 Dec 2020 -- NO FURTHER EDITS ARE ALLOWED.
%%
This Glossary contains SDSS- and APOGEE-specific terminology appearing in this paper and throughout the data documentation. 

\begin{description} \itemsep -2pt
\item[1-Meter Target] Target observed with the NMSU 1-m telescope (\texttt{TELESCOPE} tag of `apo1m'), which has a single fiber connection to the APOGEE-2N instrument. The NMSU 1-m telescope is described in \citet{holtzman_2010} with the reduction specific to its connection to the APOGEE-N instrument given in \citet{holtzman_2015}.
\item[Ancillary Target] Target observed as part of an approved Ancillary Science Program. Ancillary science Programs from APOGEE-1 are described in \citetalias{zasowski_2013} and from APOGEE-2 in this work (\autoref{sec:ancillary2015} and \autoref{sec:ancillary2017}). 
\item[APO] Apache Point Observatory; site of the Sloan Foundation 2.5-m telescope \citep{gunn_2006} on which the APOGEE-N spectrograph operates. 
\item[ASPCAP] The APOGEE Stellar Parameters and Chemical Abundances Pipeline; the analysis software that calculates basic stellar parameters (T$_{\rm eff}$, $\log{g}$, [Fe/H], [$\alpha$/Fe], [C/Fe], [N/Fe]) and elemental abundances \citep{holtzman_2015,garciaperez_2016}.
\item[BTX] The Bright Time Extension, an APOGEE-2N program executed in the last 1.5 years of the APOGEE-2 Survey.
\item[CIS] The Carnegie Institution for Science or CIS is an SDSS-IV partner and operates the Las Campanas Observatory in Chile. 
\item[Cohort] Set of targets in the same field that are observed together on all of their visits.  A given plate may have multiple cohorts on it.
\item[Contributed Program] Term for programs allocated to Principal Investigators by the CIS or CNTAC but whose data are contributed to the APOGEE-2 survey. These data appear in SDSS data releases, but their targeting was performed by the PI. These programs are described in \inprep{(F.~Santana et al.~in prep.)}.
\item[CNTAC] The Chilean National Telescope Allocation Committee, which allocates observing resources to the Chilean community. 
\item[Design] Set of targets drilled together on a plate, consisting of up to one each of short, medium, and long cohorts.  A design is identified by an integer Design ID. Changing a single target on a design results in a new design.
\item[Design ID] Unique integer assigned to each design.
\item[Drill Angle] Hour angle (distance from the meridian) at which a plate is drilled to be observed.  This places the fiber holes in a way that accounts for differential refraction across the FOV.
\item[External Program] General term for programs and targets observed during the APOGEE-2S time allocated by the Carnegie Observatories (OCIS) or the Chilean Time Allocation Committee (CNTAC). These targets will not be included in the SDSS dataset.
\item[Fiber Collision] A situation in which two targets, separated by less than the protective ferrule around the fibers, are included in the same design.  The higher-priority target is drilled on the plate(s); the lower-priority target is removed.
\item[Fiber ID] Integer (1--300) corresponding to the rank-ordered spectrum on the detector. Fiber IDs can vary from visit to visit for a given star.
\item[Field] Location on the sky, defined by central coordinates and a plate radius.
\item[GAP] The \ktwo\ Galactic Astrophysics Program, which is described in \citet{Stello_2017}.
\item[LCO] Las Campanas Observatory, site of the Ir\'en\'ee~du~Pont $2.5$-${\rm m}$ telescope \citep{Bowen_1973} on which the APOGEE-S spectrograph operates. 
\item[Location ID] Unique integer assigned to each field on the sky.
\item[Main Red Star Sample] The sample drawn from a simple selection function defined by magnitude and color that comprises the bulk of the APOGEE program. This program is explained in \citetalias{zasowski_2013} and \citetalias{zasowski_2017}.
%
%\item[Normal Targets]
%
\item[MaNGA] Mapping Nearby Galaxies at Apache Point Observatory;  A SDSS-IV program described in \citet{bundy_2015}.
\item[MaStar] The MaNGA Stellar Program; a program within the MaNGA Survey with the objective of constructing a high-fidelity stellar library. An overview of the project and its first data release is described in \citet{yan_2019}.
\item[POI] Photometric Object of Interest; an umbrella term for stars targeted due to their \kepler, \ktwo, or \tess\ light curves.
\item[Plate] Piece of aluminum with a design drilled into it.  Note that while ``plate'' is often used interchangeably with ``design'', multiple plates may exist for the same design -- e.g., plates with a common design but drilled for different hour angles.
\item[Plate ID] Unique integer assigned to each plate.
\item[RJCE] The Rayleigh-Jeans Color Excess method, a technique used to estimate the line-of-sight reddening to a star \citep{majewski_2011}.  APOGEE-2 uses this method to estimate intrinsic colors for many potential targets \citepalias[for details see][]{zasowski_2013,zasowski_2017}.
\item[Sky Targets] Empty regions of sky on which a fiber is placed to collect a spectrum 
used to remove the atmospheric airglow lines and sky background from the target spectra observed simultaneously with the same plate.
\item[Special Targets] General term for targets selected with criteria other than the color and magnitude criteria of the main red giant sample.  For example, special targets include ancillary science program targets and calibration cluster members.
\item[Targeting Flag and Bits] A targeting ``flag'' refers to one of the three long integers assigned to every target in a design, each made up of 31 ``bits'' that correspond to particular selection or assignment criteria.  APOGEE-2's flags are named \texttt{APOGEE2\_TARGET1}, \texttt{APOGEE2\_TARGET2}, \texttt{APOGEE2\_TARGET3}, and \texttt{APOGEE2\_TARGET4}; see Table~\ref{tab:targeting_bits} for a list of the bits as of this publication. 
\item[Telluric Standards] Hot blue stars observed on a plate to derive corrections for the telluric absorption lines.
\item[Visit] The base unit of observation, equivalent to approximately one hour of on-sky integration (but this can vary, as discussed in \autoref{sec:obsoverview}) and comprising a single epoch.  Repeated visits are used to both build up signal and provide a measure of spectral and RV stability.
\item[Washington\textit{+DDO51}] Also ``W+D photometry''; adopted abbreviation for the combination of Washington $M$ and $T_2$ photometry \citep{Canterna_1976} with $DDO51$ photometry \citep{McClure_1973}, used in the photometric classification of dwarf/giant stars \citep{majewski_2000}.
\end{description}
\end{appendix} 

\end{document}